\documentclass[fleqn,usenatbib,useAMS,usedcolumn]{mnras}

\usepackage{newtxtext,newtxmath}
\usepackage[T1]{fontenc}
\usepackage{ae,aecompl}

\usepackage{graphicx}	% Including figure files
\usepackage{amsmath}	% Advanced maths commands
\usepackage{amssymb}	% Extra maths symbols
\usepackage{siunitx}
\title[Photoevaporation in Carbon Depleted Discs]{Radiation\hbox{-}Hydrodynamical\,Models\,of\,X\hbox{-}ray\,Photoevaporation in\,Carbon\,Depleted\,Circumstellar\,Discs}
\author[W\"olfer et al.]{Lisa W\"olfer$^{1,2}$\thanks{E-mail: woelfer@mpe.mpg.de},
Giovanni Picogna$^{2}$,
Barbara Ercolano$^{2,3}$,
Ewine F. van Dishoeck$^{1,4}$
\\
$^{1}$Max-Planck-Institut f\"ur extraterrestrische Physik, Gie\ss enbachstr. 1 , 85748 Garching bei M\"unchen, Germany\\
$^{2}$Universit\"ats-Sternwarte M\"unchen, Scheinerstr. 1, 81679 M\"unchen, Germany\\
$^{3}$Excellence Cluster Origin and Structure of the Universe, Boltzmannstr. 2, 85748 Garching bei M\"unchen, Germany \\
$^{4}$Leiden Observatory, Leiden University, P.O. Box 9513, NL-2300 RA Leiden, The Netherlands
}

\date{Accepted 2019 October 3. Received 2019 October 2; in original form 2019 April 30}

\pubyear{2019}
\volume{{\rm accepted}}

\begin{document}
%\unboldmath
%\lsstyle
\label{firstpage}
\pagerange{\pageref{firstpage}--\pageref{lastpage}}
\maketitle

\begin{abstract}
The so-called transition discs provide an important tool to probe various mechanisms that might influence the evolution of protoplanetary discs and therefore the formation of planetary systems. One of these mechanisms is photoevaporation due to energetic radiation from the central star, which can in principal explain the occurrence of discs with inner cavities like transition discs. Current models, however, fail to reproduce a subset of the observed transition discs, namely objects with large measured cavities and vigorous accretion. For these objects the presence of (multiple) giant planets is often invoked to explain the observations. In our work we explore the possibility of X-ray photoevaporation operating in discs with different gas-phase depletion of carbon and show that the influence of photoevaporation can be extended in such low-metallicity discs. As carbon is one of the main contributors to the X-ray opacity, its depletion leads to larger penetration depths of X-rays in the disc and results in higher gas temperatures and stronger photoevaporative winds. We present radiation-hydrodynamical models of discs irradiated by internal X-ray+EUV radiation assuming Carbon gas-phase depletions by factors of 3,10 and 100 and derive realistic mass-loss rates and profiles. Our analysis yields robust temperature prescriptions as well as photoevaporative mass-loss rates and profiles which may be able to explain a larger fraction of the observed diversity of transition discs.
\end{abstract}


%%%%%%%%%%%%%%%%%%%%%%%%%%%%%%%%%%%%%%%%%%%%%%%%%%

%%%%%%%%%%%%%%%%% BODY OF PAPER %%%%%%%%%%%%%%%%%%
%
\section{Introduction}\label{intro}
The nurseries of planets, circumstellar discs, are dense remnants of the star formation process, enclosing all the gas and dust material crucial for the formation of planetary systems. Far from being static, they evolve and ultimately disperse while they give birth to planets, moons and minor bodies. As the disc dispersal proceeds on timescales which are of the same order as the planet formation timescales (e.g. \citealp{Helled2014}), the disc evolution and planet formation processes are directly linked and occur as a highly coupled and complex problem. 

In this regard the so-called transition discs (TD's) are of particular interest, as they show evidence for inner dust (and possibly gas) depleted regions (e.g. \citealp{Strom1989}) and are therefore often treated as being on the verge of dispersal. These cavities can reach various sizes from sub-au to several tens of au, with many transition discs simultaneously showing evidence for gas accretion onto the central star. Understanding the occurrence and underlying physics of transition discs, may enable to probe various mechanisms that could play a role during disc evolution and influence the planet formation and migration processes. 

Many different mechanisms have been proposed so far to explain the observed diversity of transition discs (e.g. photoevaporation, planet-disc interactions, MHD processes), none of which however is able to explain the whole database of observations (e.g. \citealp{Espaillat2014, Alexander2014}). One promising mechanism is internal photoevaporation, which describes the formation of inner holes or gaps as a result of the interaction of high-energy stellar radiation with the disc material, naturally producing transition discs. It was however assumed for a long time that photoevaporation can only account for very few of the observed objects. Especially those discs which were found to have cavities at large disc radii and simultaneously vigorous gas accretion onto the central star (of order $\textrm{10}^{-8}\,\textrm{M}_{\odot}\,\textrm{yr}^{-1}$) are not explained by current photoevaporation models \citep{Owen2011b, ErcolanoPas2017, Picogna2018}. These discs are therefore often suggested as being an indicator for the presence of (multiple) giant planets, which are in principle able to dynamically carve significant gaps into a disc.

Recent studies have however shown that the range of photoevaporative influence can be extended in discs of reduced metallicity compared to the solar elemental abundances \citep{ErcolanoWeber2018}. Indeed, several observations of gas-phase depletion of volatile carbon and oxygen in outer disc regions have been reported in the last years \citep{Ansdell2016,Du2017,Favre2013,Hogerheide2011,Kama2016,Miotello2017}. Carbon and oxygen represent the main contributors to the X-ray opacity, thus a disc depleted in these elements experiences stronger (X-ray) photoevaporative winds and enhanced mass-loss rates, as the X-ray radiation can penetrate further into the disc and heat the gas in deeper disc layers. 

In this paper, we investigate the effects of X-ray photoevaporation in such metal depleted discs, adopting different degrees of carbon depletion and performing detailed radiation-hydrodynamical simulations, following the approach of \citet{Picogna2018}. FUV photoevaporation is not included in this work, yet it can play a role at larger disc radii (e.g. \citealp{GortiDullemond2009}). Thus the presented mass-loss rates are a lower limit to the actual mass-loss rates. We describe the numerical methods and setups we used in \autoref{sec:Methods} whereas we present our main results in \autoref{sec:Results}. A conclusion of our analysis and an outlook for future research are given in \autoref{sec:Conclusions}. 
\section{Methods}\label{sec:Methods}
\begin{figure}
\centering
\includegraphics[width=\columnwidth]{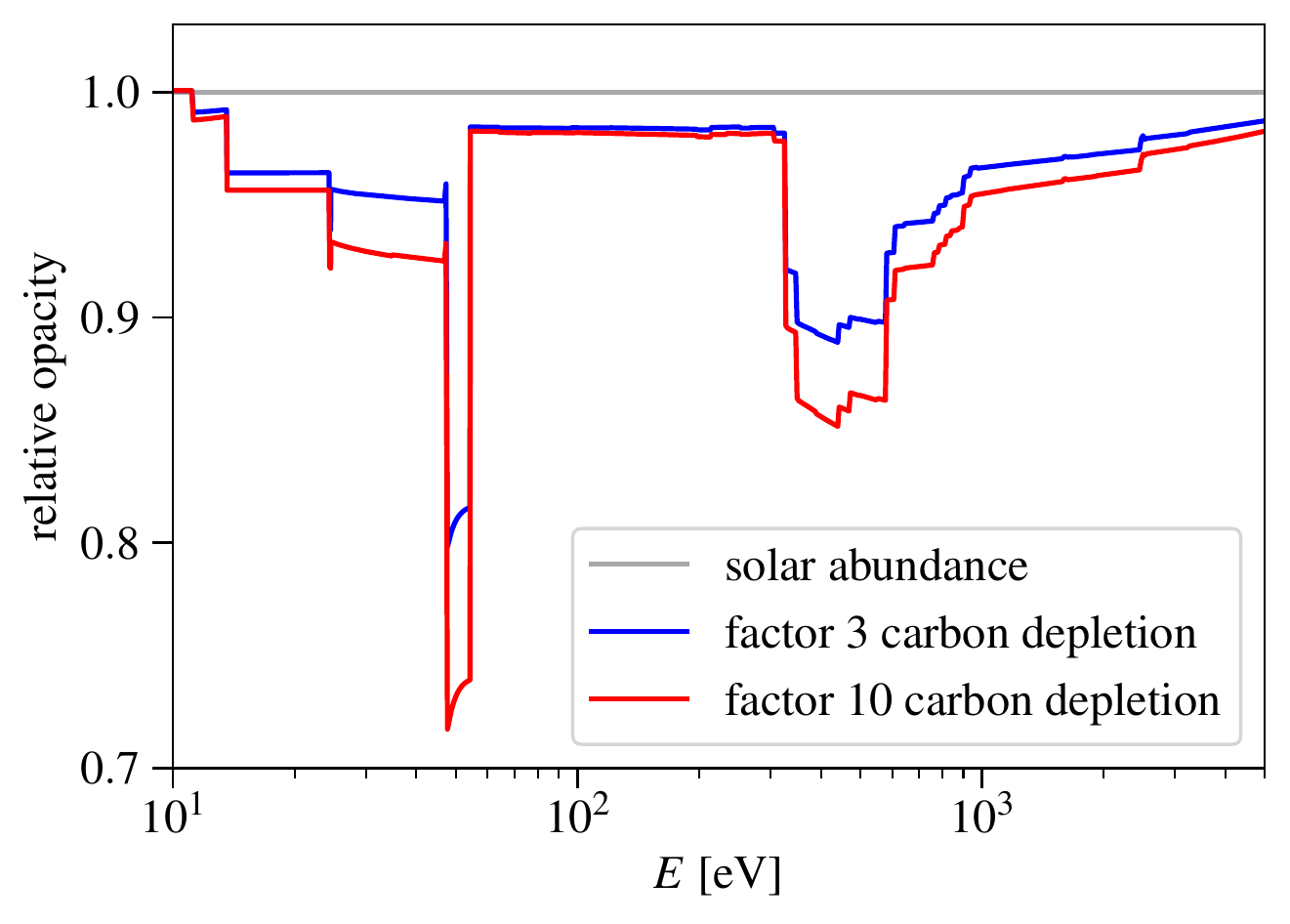}
\caption{Relative opacity of carbon depletion (with respect to the undepleted case) by a factor of 3 (blue) and 10 (red).}
\label{fig:Opacity}
\end{figure}
\begin{table}
\caption{Coefficients of the temperature parametrisation for the different carbon depletions by factors of 3, 10 and 100 and all 10 column density bins up to 2.5 $\times$ 10$^{22}$\,pp\,cm$^{-2}$.}
%\vspace{0.1cm}
\label{tab:Parameters}
\centering
\begin{tabular}{lcccc}
\hline
\hline
\multicolumn{5}{l}{\textit{carbon depletion by a factor of 3}} \\
\hline
$\textit{N}_{\textrm{H}}$ & \textit{b} & \textit{c} & \textit{d} & \textit{m}\\
1 $\times$ 10$^{20}$\,pp\,cm$^{-2}$ & & & &\\
\hline
0 - 25 & -49.6442 & -7.0423 & 3.9952 & 0.1008 \\
25 - 50 & -15.6516 & -5.7592 & 3.9144 & 0.3904 \\
50 - 75 & -13.5273 & -5.2914 & 3.8841 & 0.5038\\
75 - 100 & -13.8039 & -5.1523 & 3.8620 & 0.4904\\
100 - 125 & -20.0278 & -5.2913 & 3.8378 & 0.3184\\
125 - 150 & -18.2243 & -5.1041 & 3.8208 & 0.4003 \\
150 - 175 & -19.2923 & -5.3050 & 3.8429 & 0.2354 \\
175 - 200 & -23.5695 & -5.3299 & 3.8464 & 0.1839 \\
200 - 225 & -16.7558 & -4.9177 & 3.8138 & 0.3483\\
225 - 250 & -22.9758 & -5.0689 & 3.8247 & 0.2440\\
\hline
\multicolumn{5}{l}{\textit{carbon depletion by a factor of 10}} \\
\hline
$\textit{N}_{\textrm{H}}$ & \textit{b} & \textit{c} & \textit{d} & \textit{m}\\
1 $\times$ 10$^{20}$\,pp\,cm$^{-2}$ & & & &\\
\hline
0 - 25 & -21.1849 & -7.7162 & 4.0001 & 0.2214 \\
25 - 50 & -15.1575 & -6.4422 & 3.9176 & 0.3672 \\
50 - 75 & -14.1757 & -6.2253 & 3.8915 & 0.3679 \\
75 - 100 & -10.8864 & -5.8325 & 3.8743 & 0.4958 \\
100 - 125 & -11.1109 & -5.6791 & 3.8418 & 0.4705\\
125 - 150 & -11.2723 & -5.5136 & 3.8344 & 0.4798\\
150 - 175 & -17.3954 & -5.7711 & 3.8030 & 0.2998\\
175 - 200 & -13.5226 & -5.3788 & 3.8126 & 0.4469\\
200 - 225 & -13.9993 & -5.4703 & 3.7953 & 0.4657\\
225 - 250 & -19.0899 & -5.5465 & 3.7807 & 0.3046\\
\hline
\multicolumn{5}{l}{\textit{carbon depletion by a factor of 100}} \\
\hline
$\textit{N}_{\textrm{H}}$ & \textit{b} & \textit{c} & \textit{d} & \textit{m}\\
1 $\times$ 10$^{20}$\,pp\,cm$^{-2}$ & & & &\\
\hline
0 - 25 & -11.3726 & -8.2547 & 4.0024 & 0.3494\\
25 - 50 & -7.3249 & -6.7159 & 3.9200 & 0.6860\\
50 - 75 & -6.9106 & -5.9662 & 3.8872 & 0.8848 \\
75 - 100 & -6.3211 & -5.6836 & 3.8557 & 0.9324 \\
100 - 125 & -5.6213 & -5.3946 & 3.8461 & 1.1009\\
125 - 150 & -4.7809 & -4.7992 & 3.8218 & 1.5653\\
150 - 175 & -5.5289 & -5.0542 & 3.8155 & 1.1728\\
175 - 200 & -5.1865 & -4.5065 & 3.7945 & 1.7157\\
200 - 225 & -5.5705 & 5.0308 & 3.7948 & 1.1407\\
225 - 250 & -5.0972 & -4.1973 & 3.7693 & 2.2123\\
\hline
\end{tabular}
\end{table}
\begin{figure*}
	\includegraphics[width=\columnwidth]{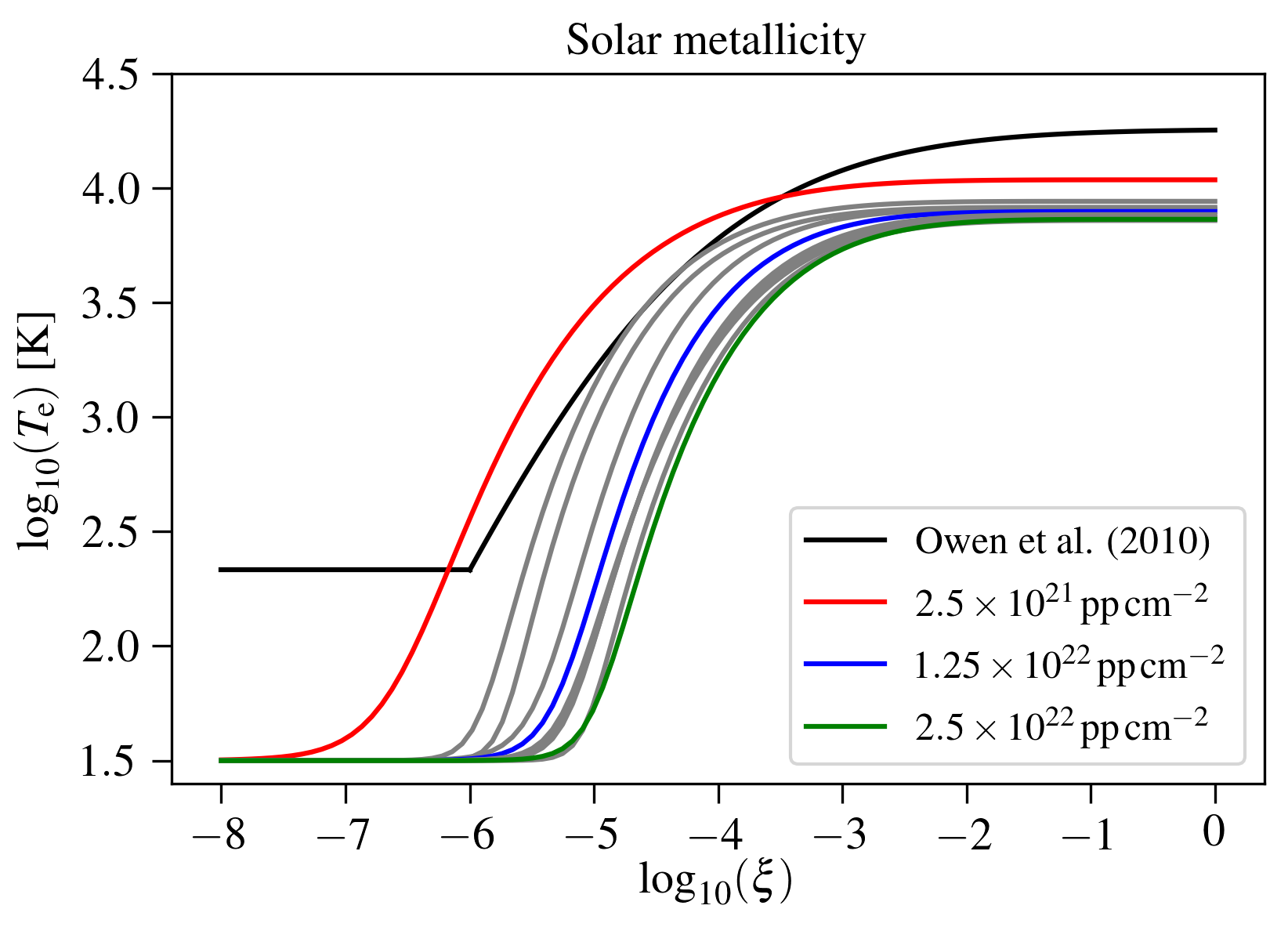}
	\includegraphics[width=\columnwidth]{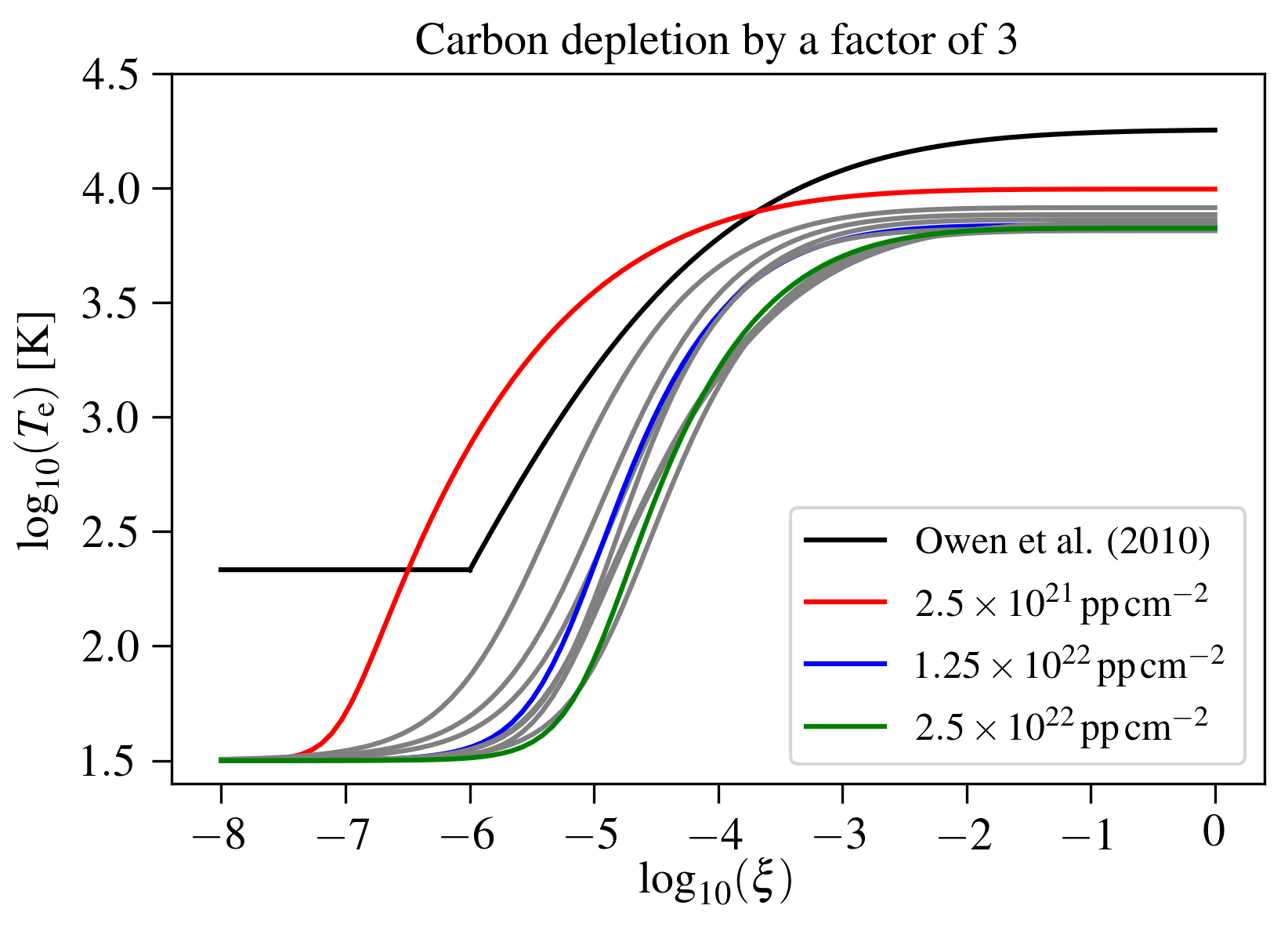}
	\includegraphics[width=\columnwidth]{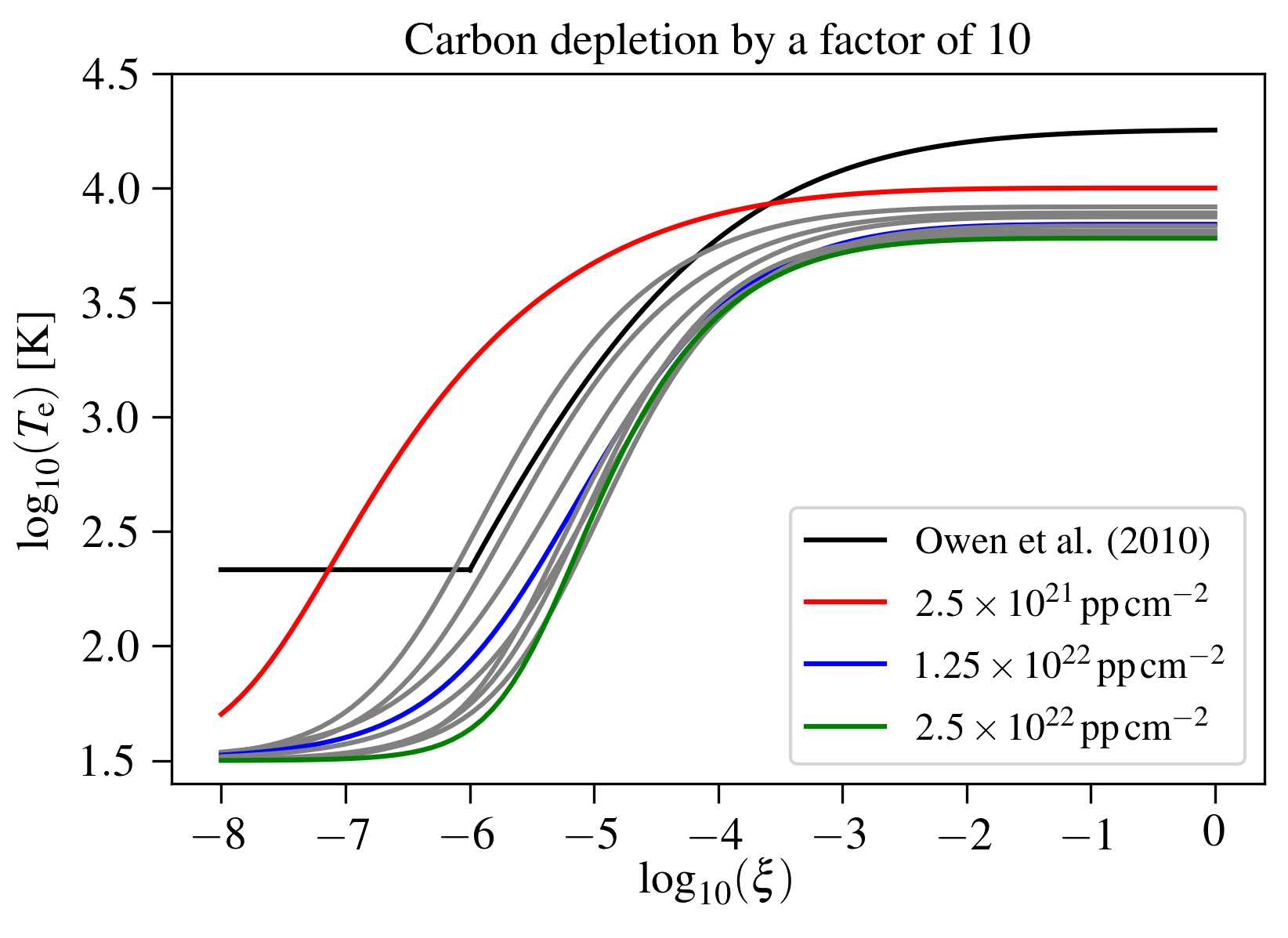}
	\includegraphics[width=\columnwidth]{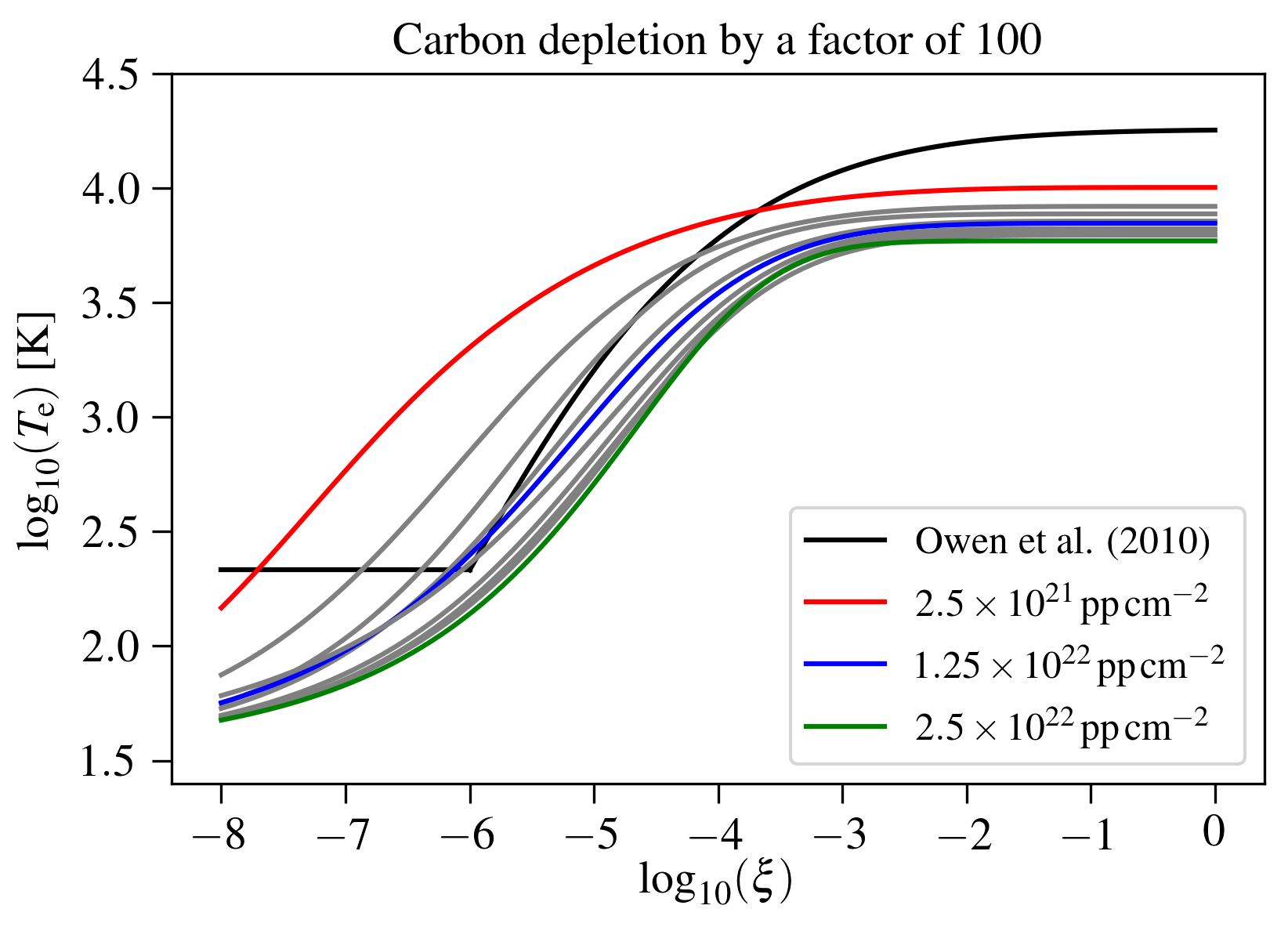}
    \caption{Temperature parametrisation for the three different carbon depletions by a factor of 3 (top right), 10 (bottom left) and 100 (bottom right). The scheme for solar metallicity is included as a reference in the top left panel of the plot \citep{Picogna2018}. In each panel the lowest column density curves are highlighted in red, the medium ones in blue and the highest ones in green while the black curve represents the parametrisation by \citet{Owen2010}. The four different carbon abundance sets clearly differ from each other, showing higher gas temperatures for stronger depletion.}
    \label{fig:TempCarbon}
\end{figure*}
\subsection{Thermal Calculations}\label{Thermal}
We have used the gas and dust radiative transfer code {\large \textsc{mocassin}} \citep{Ercolano2003,Ercolano2005,Ercolano2008a} to model gas temperatures of circumstellar discs with different carbon abundances that are irradiated by an X-ray+EUV spectrum (presented in \citealp{Ercolano2008b,Ercolano2009}, unscreened spectrum of Figure 3 in \citealp{Ercolano2009}) of a $\textrm{0.7}\,\textrm{M}_{\odot}$ star. In total we set up three simulations with mostly standard solar abundances but varying degrees of carbon depletion. Our standard interstellar gas-phase abundances are taken from \citet{Savage1996} (C:~$\textrm{1.4} \times \textrm{10}^{-4}$;~O:~$\textrm{3.2} \times \textrm{10}^{-4}$). These values take into account that some fraction of the solar abundances \citep{Asplund2005} are locked up in refractory material. Subsequently, we have depleted the gas-phase carbon abundance relative to the interstellar value by factors of 3, 10 and 100. This will have a strong impact on the opacity as visible in \autoref{fig:Opacity} where the relative opacity of the carbon depletion by a factor of 3 and 10 to the undepleted case is shown, respectively. The curves are presented for a column density of $\approx \textrm{5} \times \textrm{10}^{20}\,\textrm{pp}\,\textrm{cm}^{-2}$ and an ionisation parameter $\xi = \frac{\textit{L}_{\textrm{X}}}{\textit{nr}^2}$ \citep{Tarter1969} of $\textrm{log(}\xi\textrm{)} = -\textrm{2}$, where $\textit{L}_{\textrm{X}}$ is the X-ray luminosity, \textit{r} the distance from the star and \textit{n} the electron number density. The adopted synthetic thermal spectrum was created with the plasma code PINTofALE \citep{Kashyap2000} in order to match \textit{Chandra} spectra of T Tauri stars observed by \cite{Maggio2007}. 

All simulations were run for column densities up to $\textrm{2.5} \times \textrm{10}^{22}\,\textrm{pp}\,\textrm{cm}^{-2}$ and for in total 20 ionisation parameters between $\textrm{log(}\xi\textrm{)} = -\textrm{8}$ and $\textrm{log(}\xi\textrm{)} = -\textrm{2}$. From the output of the simulations we obtained the equilibrium gas temperature at the upper disc layers as a function of the ionisation parameter. 
We furthermore divided the disc into 10 sections of size $\textrm{2.5} \times \textrm{10}^{21}\,\textrm{pp}\,\textrm{cm}^{-2}$ giving a temperature prescription for each column density bin. For higher column densities than $\textrm{2.5} \times \textrm{10}^{22}\,\textrm{pp}\,\textrm{cm}^{-2}$ we assume that the gas and dust are thermally coupled and use the dust temperatures from the models of \citet{Alessio2001}, mapped to our models. \footnote{The radiation-hydrodynamical calculations were actually performed using temperature parametrisations which extended to columns of $\textrm{5} \times \textrm{10}^{22}\,\textrm{pp}\,\textrm{cm}^{-2}$. We however found a posteriori, that the high column density curves (> $\textrm{2.5} \times \textrm{10}^{22}\,\textrm{pp}\,\textrm{cm}^{-2}$) are severely affected by Monte Carlo noise and as a consequence carry large errors on the temperatures. We have thus decided not to include these high column parametrisation in this work. We further note that the errors on the high column parametrisation do not affect the hydrodynamical simulations presented here, since the region of parameter space affected represents only a very small percentage of our simulation domain, well below the wind launching region.}
\\
\\
\\
In order to fit the modelled data we used the following ad-hoc relation 
\begin{equation}\label{eq:fit}
\log_{10}\left(T(\xi)\right) = d + \frac{1.5 - d}{\left[1.0 + (\log_{10}(\xi)/c)^b\right]^m}
\end{equation}
with the resulting curves being shown in \autoref{fig:TempCarbon} and the corresponding coefficients being listed in \autoref{tab:Parameters}.
In \autoref{fig:TempCarbon} we also include a parametrisation for a solar metallicity disc as a reference (the underlying data were taken from \citealp{Picogna2018}). The lowest, medium and highest column density are highlighted with color. \autoref{fig:TempCarbon} shows that the three different carbon depletion sets clearly vary from each other and from the solar metallicity set and that the temperatures increase as expected with increasing degree of depletion. In addition, the curves become flatter and are distributed more narrowly over the whole column density range for higher depletion. This results from models with stronger depletion having a lower gas opacity in the X-ray regime.

Our parametrisation schemes include the column density independent curve for solar metallicity used by \citet{Owen2010,Owen2011,Owen2012}. In this respect we would like to point out that the inclusion of a column density parameter helps to model the temperatures more accurately at different disc locations. Similar to \citet{Picogna2018} we find the temperature error to be reduced to less than 1\,\% for all simulations (compare Appendix \ref{appendix:TempError}). Furthermore, our calculations extend to lower $\xi$ values ($\textrm{log(}\xi\textrm{)} = -\textrm{8}$ instead of $\textrm{log(}\xi\textrm{)} = -\textrm{6}$), which allows us to simulate the outer disc regions that are important for studying the evolution of transition discs more extensively. The prescription of \citet{Owen2010} reaches a higher maximum temperature due to integration over a finer grid. This in principle allows to resolve a region of low density that is heated by EUV radiation; however this region does not contribute to the total mass-loss rate and is therefore not relevant for the purpose of this work. A detailed description and discussion of the new temperature prescriptions for solar abundance discs and their impact on photoevaporative mass-loss rates and profiles can be found in \citet{Picogna2018}. 

To test the reliability of our temperature prescriptions we performed additional Monte Carlo simulations with higher resolution and furthermore applied different binnings, with both tests however yielding the same results as presented in \autoref{fig:TempCarbon}. In terms of the microphysics, which are relatively well known, the {\large \textsc{mocassin}} code has been thoroughly benchmarked (see \citealp{Ercolano2003,Ercolano2005,Ercolano2008a}), which together with the small temperature error confirms the robustness of our parametrisation.
\subsection{Hydrodynamics}\label{sec:Hydro}
We have used the open source hydro-code {\large \textsc{pluto}} \citep{Mignone2007} to model different carbon depleted as well as solar metallicity protoplanetary discs until a "steady-state" is reached, in order to find reliable photoevaporative mass-loss rates $\dot{M}$ and $\dot{\mathit{\Sigma}}$ profiles. We therefore performed several simulations with {\large \textsc{pluto}}, adopting a two-dimensional, spherical coordinate system centred around a 0.7\,M$_{\odot}$ star in the \textit{r}-$\theta$ plane, since the problem we address is symmetric along the $\phi$ dimension. We furthermore implemented the temperature prescriptions described in \autoref{Thermal} and interpolated from the curves for the whole column density range directly in {\large \textsc{pluto}}. Outside of this range, we set the lowest column density of 2.5 $\times$ 10$^{21}$\,pp\,cm$^{-2}$ as a limit and used the assumption described in the previous subsection for higher column densities than 2.5 $\times$ 10$^{22}$\,pp\,cm$^{-2}$. In terms of the $\textrm{log(}\xi\textrm{)}$ range, we assume $\textit{T} = \textit{T}_{\textrm{dust}}$ for values smaller than $\textrm{log(}\xi\textrm{)} = -\textrm{8}$ and apply the maximum temperature we found in our temperature parametrisation for values larger than $\textrm{log(}\xi\textrm{)} = -\textrm{2}$. As an initial density and temperature structure of the discs, we took the results of \citet{Ercolano2008b,Ercolano2009}, which were obtained from hydrostatic equilibrium models. 

To avoid numerical issues in the low density regions near the pole and at larger radii, we defined a logarithmic grid scaling in both directions. Being positive in the radial and negative in the polar direction this leads to a finer grid close to the star. Another issue that needs to be considered is the outer boundary of the domain. Here, unwanted oscillations can occur (observed also in \citealp{Picogna2018} and \citealp{Nakatani2018a}) and affect the inner disc regions and therefore the final results. To deal with this, we adopted an outer boundary inside the computational domain at 980\,au, after which the gas is not evolved in time. Due to this sort of damping region, unrealistic oscillations and reflections could successfully be prevented. 

All simulations described in the upcoming sections were run for 300$-$500 orbits at 10\,au. In this context, a good compromise needs to be found for the total number of orbits: If too few orbits are performed, a steady state value of $\dot{M}$ cannot be reached. As the disc is however continuously losing mass, a real equilibrium cannot be found and the mass-loss rate will change over time due to the disc's evolution. We therefore have to find a time span in which first of all, the change of the total disc mass $\textit{M}_{\textrm{disc}}$ is stable and not too rapid and secondly, the disc has not evolved significantly yet. Above a certain number of orbits, depending on the discs properties (e.g. the mass), no steady state is established and $\textit{M}_{\textrm{disc}}$ will decrease rapidly due to the wind, resulting in a rapid change in the mass-loss rates.
\subsubsection{Primordial Discs}
With the purpose of investigating the effects of carbon abundance in various protoplanetary discs, we set up six types of primordial disc (i.e., full disc without a hole) simulations for four disc masses in a range between $\textit{M}_{\textrm{disc}} = \textrm{0.005}\,\textit{M}_{*}$ and $\textit{M}_{\textrm{disc}} = \textrm{0.1}\,\textit{M}_{*}$. Besides a solar metallicity simulation, these simulation types included three simulations with a homogeneous carbon depletion by a factor of 3, 10 and 100 throughout the whole disc and two additional inhomogeneous simulations where we assumed solar abundances within 15\,au distance from the star and carbon depletion factors of 3 and 10, respectively, outside of this radius. No self-gravity is included in our simulations, which may play a role for the highest-mass disc of our sample ($\textit{M}_{\textrm{disc}} = \textrm{0.1}\,\textit{M}_{*}$). The parameter space of all primordial disc simulations is shown in \autoref{tab:parSpacePluto}.
\begin{table}
\caption{Parameter space for the primordial disc simulations with {\large \textsc{pluto}}.}\label{tab:parSpacePluto}
\centering
\begin{tabular}{l c}
\hline
\hline
\bf{variable} & \bf{value}\\
\hline
\textit{disc extent} & \\
\hline
radial [au] & 0.33$-$1000 [log spaced]\\
polar [rad] & 0.005 $- \uppi$/2 [log spaced]\\
\hline
\textit{grid resolution} & \\
\hline
radial & 412 \\
polar & 160 \\
\hline
\textit{physical properties} & \\
\hline
$\textit{M}_{\textrm{disc}}$ [$\textit{M}_*$] & 0.005, 0.01, 0.05, 0.1 \\
luminosity $\textit{L}_{\textrm{X}}$ [erg/s]& 2 $\times$ 10$^{30}$ \\
luminosity $\textit{L}_{\textrm{EUV}}$ & 1.26\,$\textit{L}_{\textrm{X}}$ \\ 
viscosity parameter $\alpha$ & 0.001\\
mean molecular weight $\mu$ & 1.37125\\
\hline
\end{tabular}
\end{table} 
\begin{figure*}
    \includegraphics[width=0.93\textwidth]{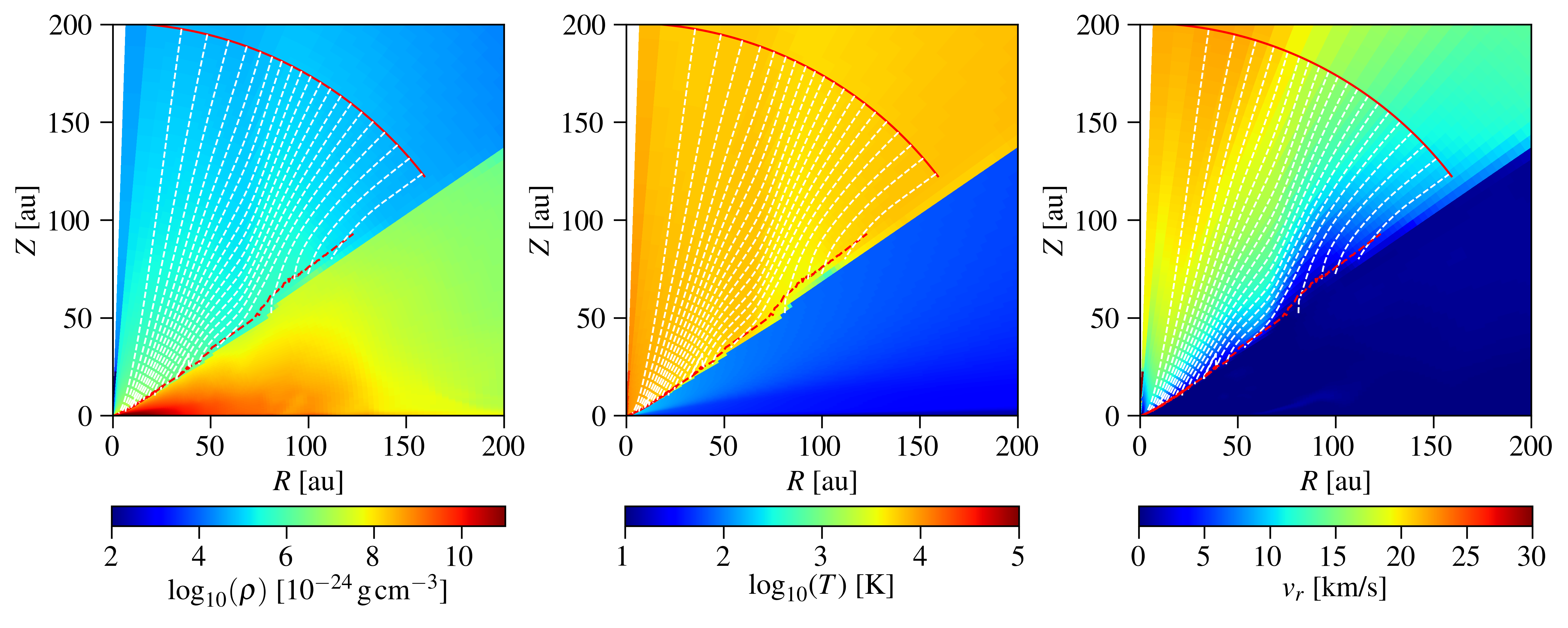}
    \includegraphics[width=0.93\textwidth]{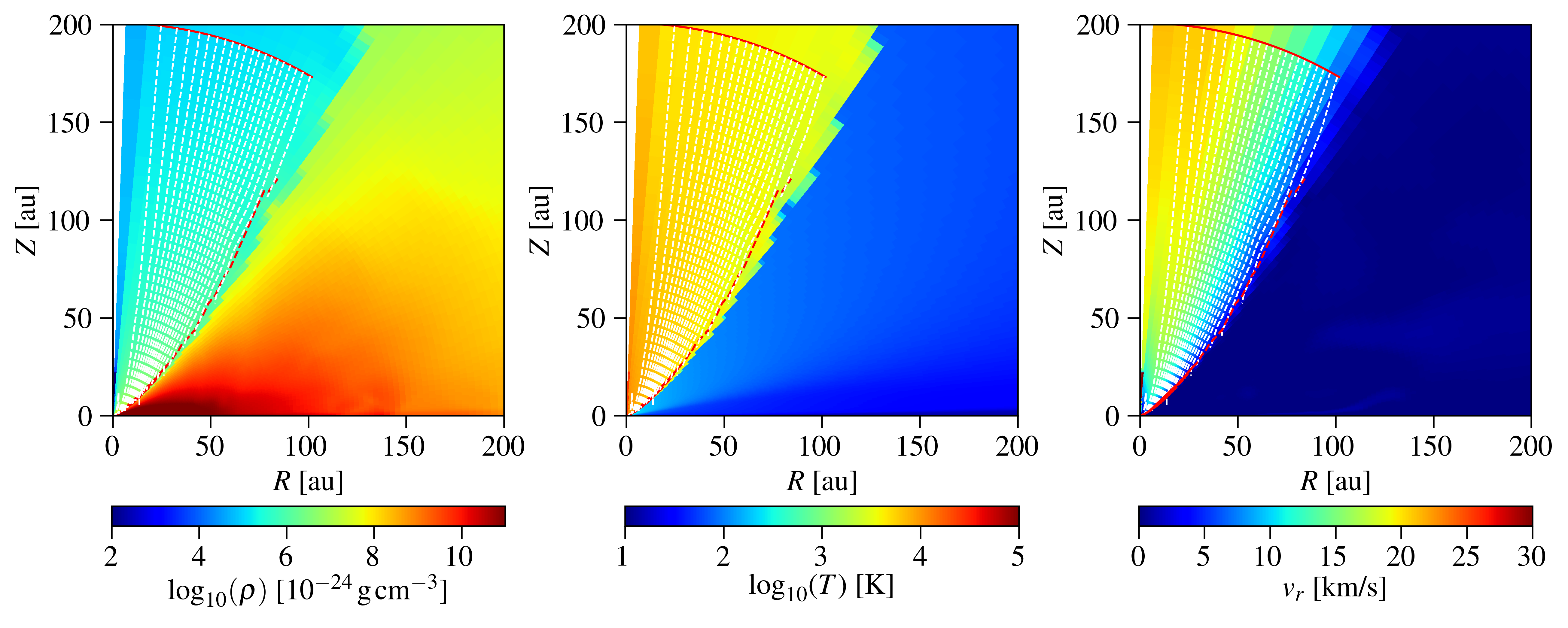}
    \caption{Disc structure for the lowest-mass (0.005\,$\textit{M}_{*}$, top panels) and highest-mass (0.1\,$\textit{M}_{*}$, bottom panels) primordial discs at the end of a simulation with carbon depletion by a factor of 3. Depicted are the mass density (left panels), temperature (middle panels) and radial velocity (right panels). The wind streamlines are overlaid as white dashed lines at 5\,\% intervals of the integrated mass-loss rate. The radius of the streamlines calculation and sonic surface are plotted with solid and dashed red lines respectively.}
    \label{fig:DiscStructureP}
\end{figure*}
\begin{figure*}
    \includegraphics[width=0.93\textwidth]{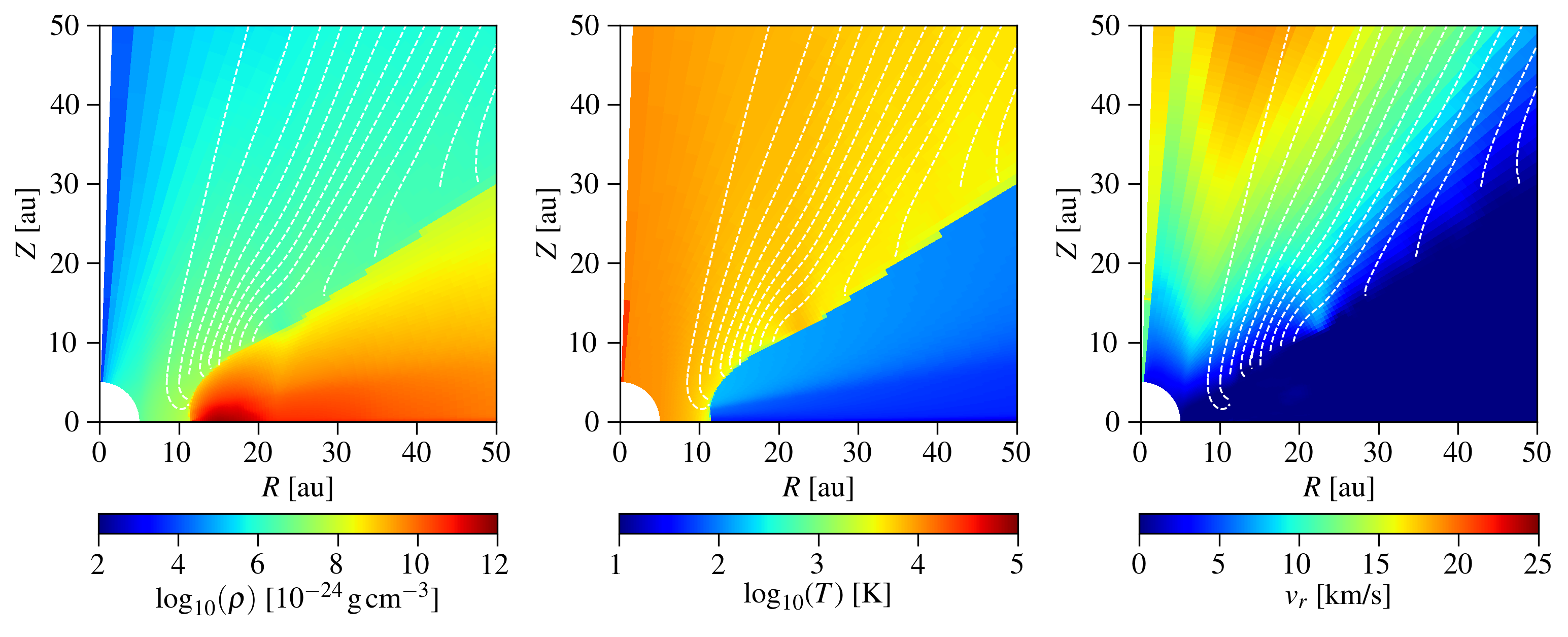}
    \includegraphics[width=0.93\textwidth]{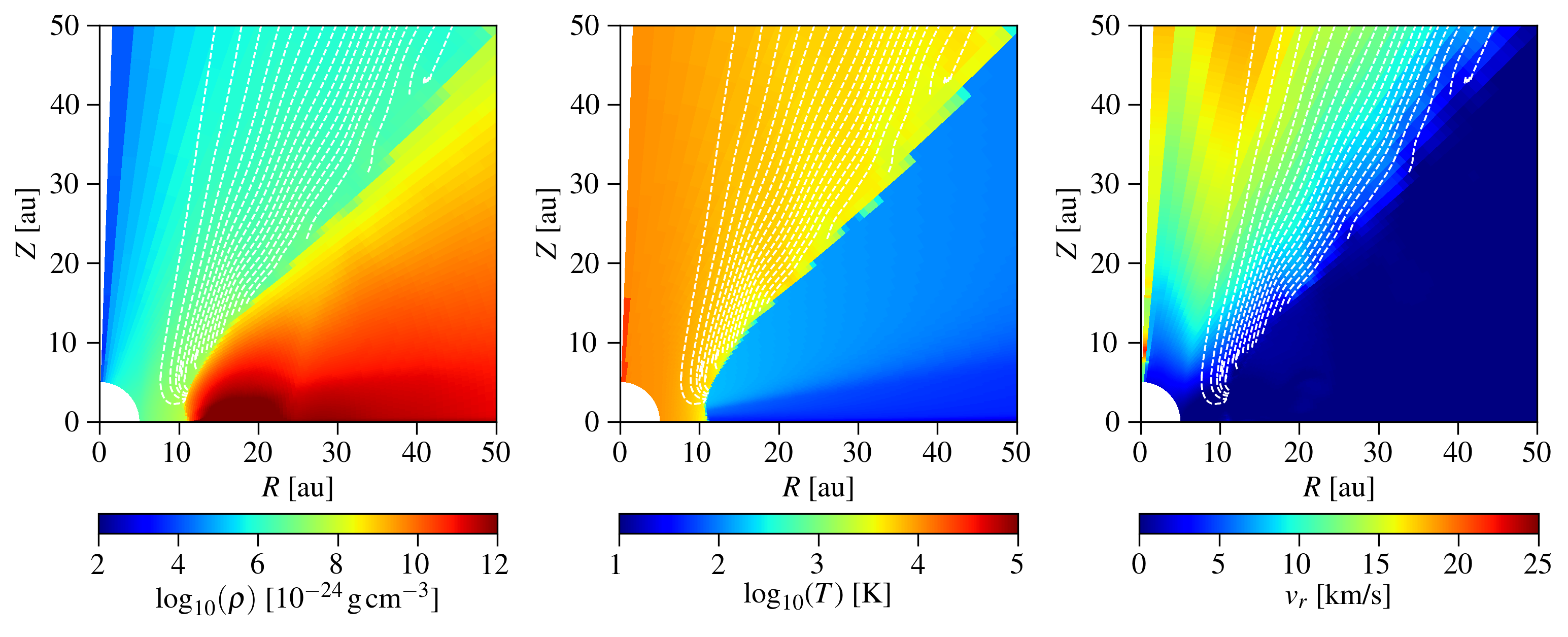}
    \caption{Disc structure for the lowest-mass (0.005\,$\textit{M}_{*}$, top panels) and highest-mass (0.1\,$\textit{M}_{*}$, bottom panels) transition discs (factor 3 depletion) displayed for a hole radius of $\textit{R}_{\textrm{H}} \approx \textrm{11\,au}$. Depicted are the mass density (left panels), temperature (middle panels) and radial velocity (right panels). The wind streamlines are overlaid as white dashed lines at 5\,\% intervals of the integrated mass-loss rate.}
    \label{fig:DiscStructureT}
\end{figure*}
\subsubsection{Transition Discs}
Alongside the primordial discs, we also modelled several transition discs for different initial hole radii and corresponding to all primordial disc simulations. In this context we choose a similar setup as before, increasing however the inner radial boundary, depending on the hole radius, and adjusted the number of radial grid cells in order to have the same resolution in the modelled region as for the primordial disc simulations. To set up a realistic gap without an abrupt density change, we added an exponential decay of the density close to the defined gap radius. Again, we used the hydrostatic models of \citet{Ercolano2008b,Ercolano2009} as initial conditions.  
\\
\\
Similar to \citet{Picogna2018} and \citet{Owen2010} we find that adiabatic cooling can be negelected in our calculations. We thus conclude that the gas should be in thermal equilibrium, which we prove in Appendix \ref{appendix:equilib} by directly comparing the advection and recombination timescales throughout the computational domain. Here we find that the advection timescale is significantly exceeding the timescale for the recombination processes. This result stands in contrast to \citet{Wang2017} who found adiabatic cooling to play an important role for the thermal balance of their models. There are however a number of important differences in the model setup and assumptions which may contribute to these discrepancies. This is discussed in more detail in \citet{Picogna2018}.
%
%\subsection{Calculation of the mass-loss rates and $\dot{\Sigma}$ profiles}
\subsection{Calculation of the mass-loss rates and \texorpdfstring{$\dot{\mathit{\Sigma}}$}{} profiles}
In order to derive the mass-loss rates and $\dot{\mathit{\Sigma}}$ profiles, we adopted the approach used by \citet{Picogna2018}, which is similar to the methods followed by \citet{Owen2010}. In this context, we first remapped the grid onto a Cartesian grid of 4000 x 4000 and defined a radius in the disc from which we followed the streamlines of the gas to the base of the flow. Here the location of the flow base is characterized by the local maximum of the derivative of the temperature profile at each cylindrical radius. We checked that this definition is consistent with the Bernoulli parameter. 

While the domain of our calculations extends to 1000\,au we choose to calculate mass-loss rates out to 200\,au. The reasons for this choice are discussed in detail in Appendix \ref{appendix:Radius}. From the streamline calculations we derived the mass-loss as a function of the cylindrical radius and a value for the total mass-loss rate. We furthermore applied a fit 
\begin{equation}\label{eq:mdot}
\dot{M}(R) = 10^{a \lg^6(R) + b \lg^5(R) + c \lg^4(R) + d \lg^3(R) + e \lg^2(R) + f \lg(R) + g}
\end{equation}\noindent
for the mass-loss rates from which we were able to calculate the $\dot{\mathit{\Sigma}}$ profiles via 
\begin{alignat}{2}\label{eq:array}
\dot{\mathit{\Sigma}} &= \ln(10) \bigg(\frac{6 a \ln^5(R)}{R \ln^6(10)} + 
\frac{5 b \ln^4(R)}{R \ln^5(10)} + 
\frac{4 c \ln^3(R)}{R \ln^4(10)} +  
\frac{3 d \ln^2(R)}{R \ln^3(10)} + \nonumber 
\\
&\hspace{1.3cm}\frac{2 e \ln(R)}{R \ln^2(10)} +
\frac{f}{R \ln(10)}\bigg)
\frac{\dot{M}(a,b,c,d,e,f,g,R)}{2\pi R}\, .
\end{alignat}
\begin{figure*}
\includegraphics[width=\columnwidth]{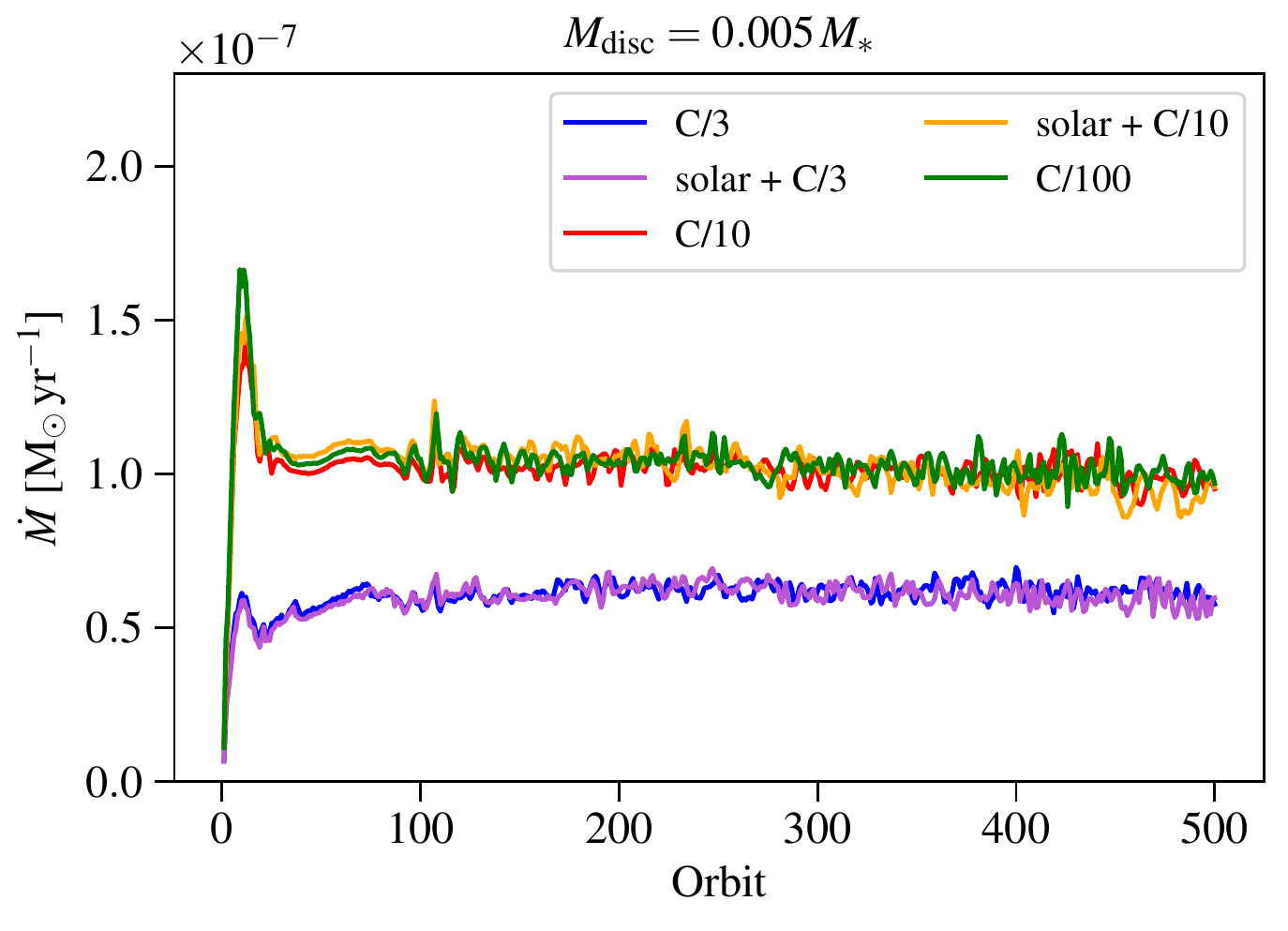}
\includegraphics[width=\columnwidth]{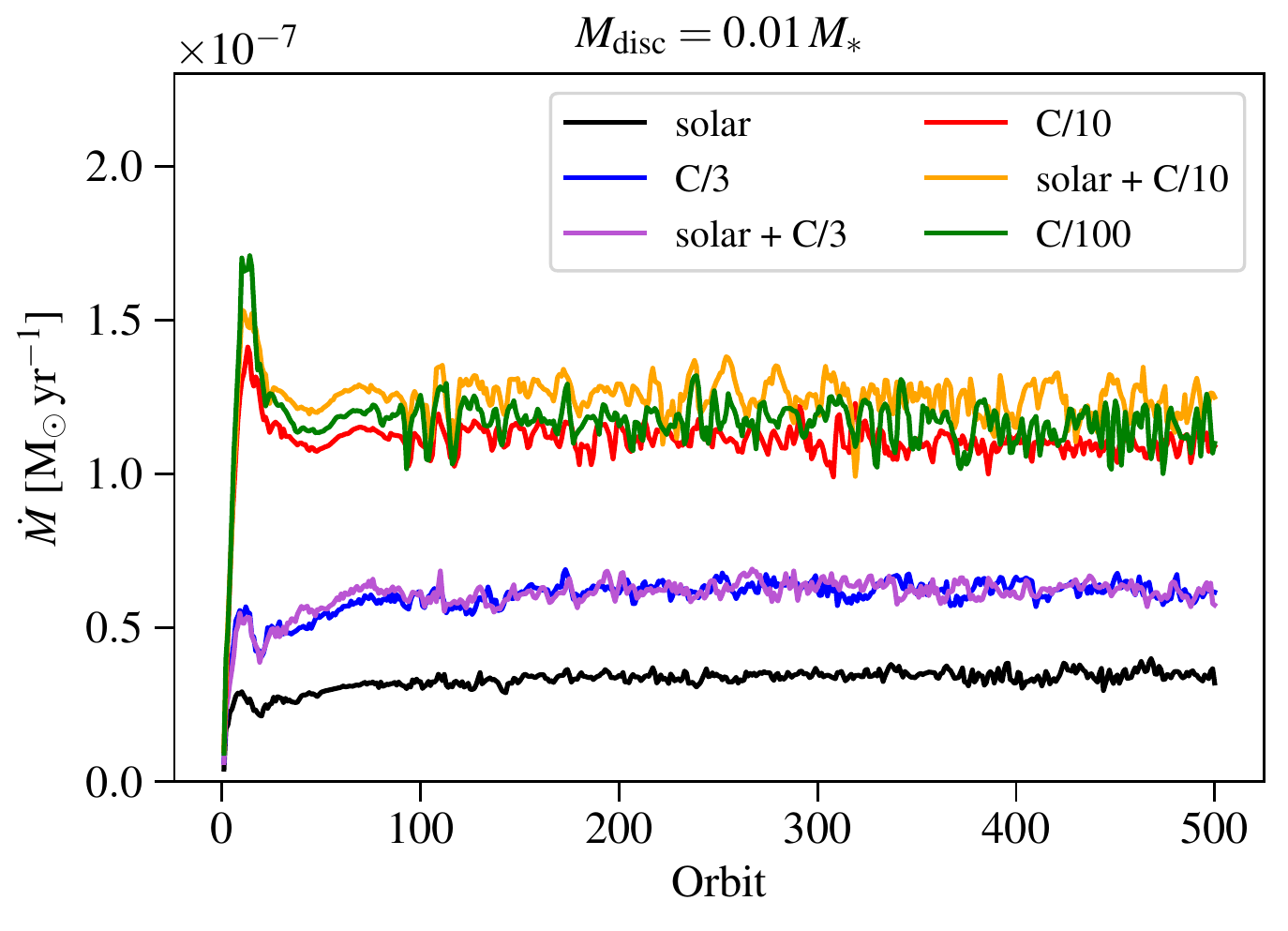}
\includegraphics[width=\columnwidth]{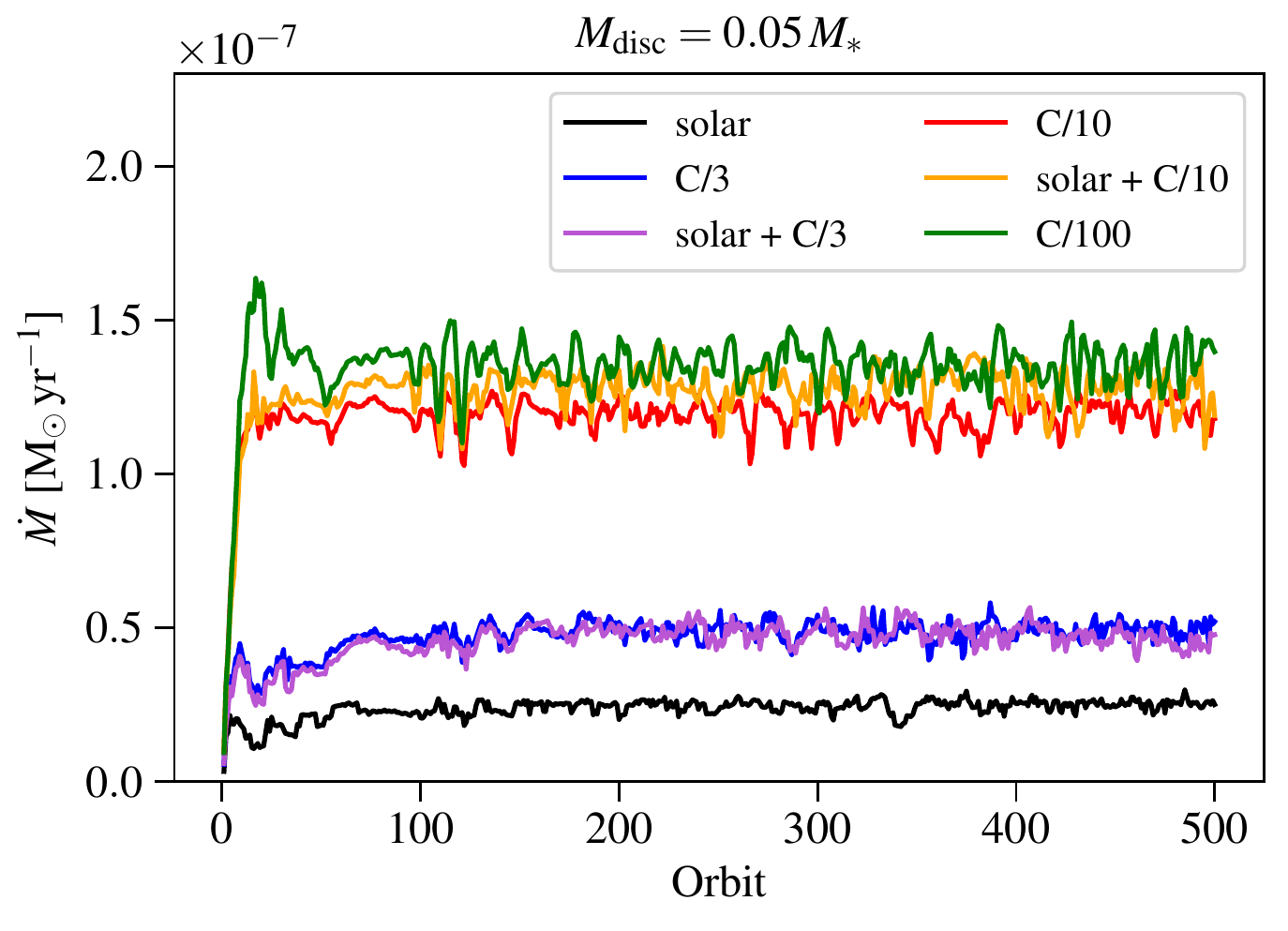}
\includegraphics[width=\columnwidth]{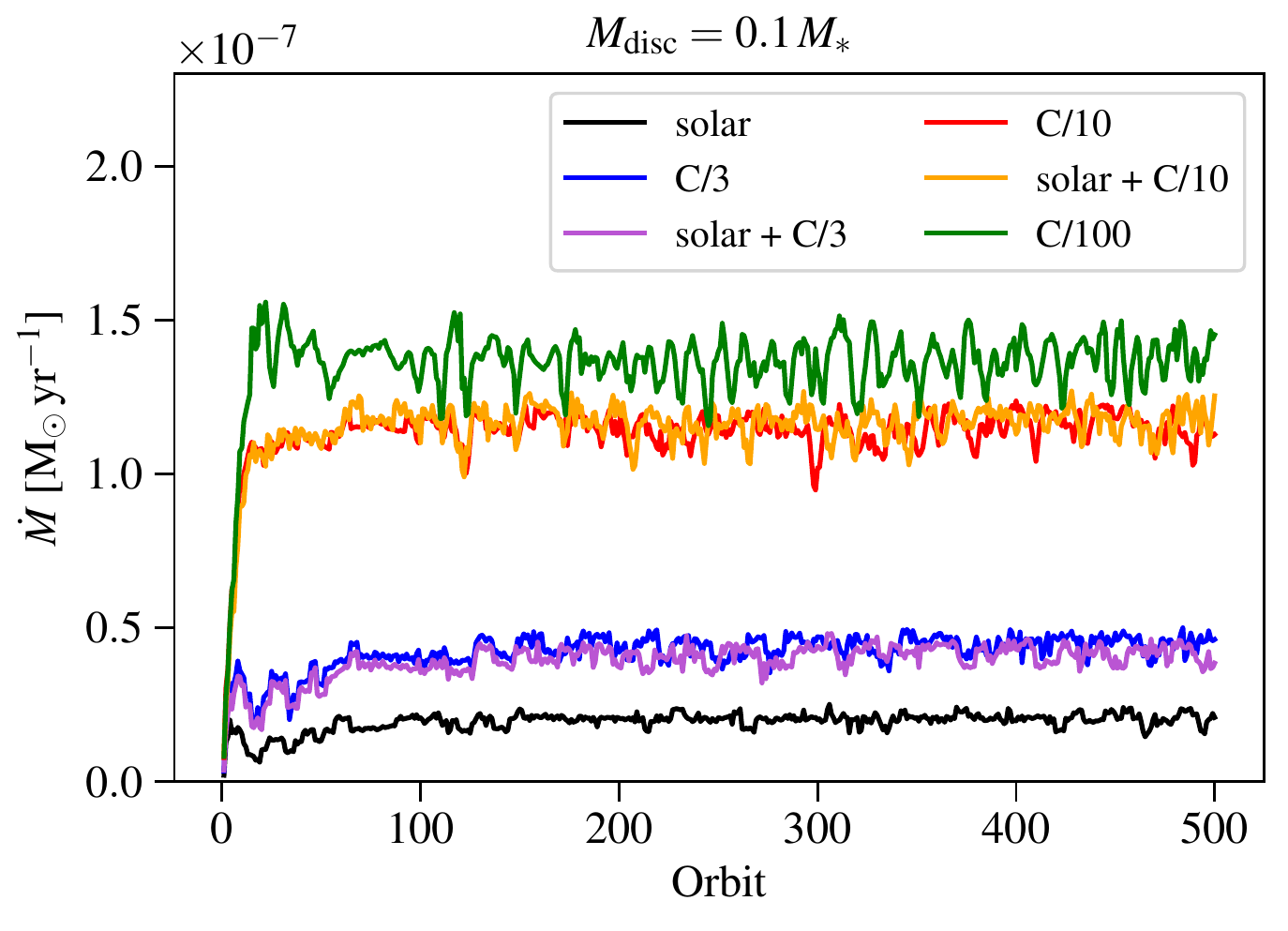}
\caption{Mass-loss rate as a function of orbits for the different carbon depletion setups. Shown are the results of the 0.005\,$\textit{M}_{*}$ (top left panel), 0.01\,$\textit{M}_{*}$ (top right panel), 0.05\,$\textit{M}_{*}$ (bottom left panel) and 0.1\,$\textit{M}_{*}$ (bottom right panel) primordial disc simulations. Besides a small scatter, the mass-loss rates behave stable after $\approx$ 100 orbits.}\label{fig:Mverlauf}
\end{figure*}
\section{Results}\label{sec:Results}
\autoref{fig:DiscStructureP} and \autoref{fig:DiscStructureT} display an example of the density, temperature and radial velocity structure of the primordial and transition discs, respectively. In each case, an example for the lowest-mass disc of 0.005\,$\textit{M}_{*}$ (top panels) and the highest-mass disc of 0.1\,$\textit{M}_{*}$ (bottom panels) is shown at the end of a simulation with carbon depletion by a factor of 3. The transition discs in \autoref{fig:DiscStructureT} have cavities with radius $\textit{R}_{\textrm{H}} \approx \textrm{11\,au}$. Furthermore, we overlaid the disc structure with the streamlines of the photoevaporative wind flow (white dashed lines), plotting a streamline for every interval of 5\,\% of the integrated mass-loss. The radius of 200\,au, from which the streamline calculation starts, is marked by a solid red line while the dashed red line indicates the sonic surface. For the primordial discs, we find that the streamlines mostly originate from a radius inside of 50\,au, whereas the percentage of these lines drops with decreasing carbon abundance. In general, the fraction is comparable for the various disc masses but we still notice a slight drop of the percentage of streamlines inside of 50\,au with decreasing mass as well.

In total, all primordial disc simulations behave in a stable manner over the whole range of orbits after a small adjustment time. Quite in contrast to that, the transition disc simulations evolve relatively fast within a few hundred orbits, showing two sorts of behaviour: First, the inner hole radius moves outwards about 1$-$5\,au within 100 orbits ($\sim$ 3800\,yrs), depending on the disc mass, degree of depletion and initial hole radius, which indicates some sort of inside-out clearing. Secondly, for a disc mass of 0.005\,$\textit{M}_{*}$ (and partly 0.01$\,\textit{M}_{*}$), the disc quickly starts to thin out for depletion factors above 3, whereby this effect is more pronounced for a larger initial cavity. Such a behaviour indicates some kind of rapid clearing of (lower-mass) transition discs that are harbouring a very extended hole (see \autoref{sec:sweeping}). 
\subsection{Mass-loss rates for the primordial disc simulations} \label{sec:MassLoss}
The evolution of the mass-loss rate of the primordial disc models is presented in \autoref{fig:Mverlauf} for all five (six) simulations of each disc mass. First, it becomes clear that the mass-loss rate is, apart from a small scatter, relatively stable beyond 100 orbits. Moreover, the mass-loss rates of the homogeneously and the inhomogeneously depleted discs lie relatively close to each other, implying that the overall mass-loss is mostly dominated by outer disc regions, with the solar abundances inside of 15\,au causing no significant effect. We note however that despite the small mass-loss rate variation, the $\dot{\mathit{\Sigma}}$ profiles can be noticeably influenced by the different depletion architectures and differ from each other significantly (see \autoref{sec:profiles}). As expected, the mass-loss rates increase with carbon depletion, whereas the difference between the carbon depletion by a factor of 10 and 100 becomes more pronounced with higher disc mass.    

\begin{figure}
\includegraphics[width=\columnwidth]{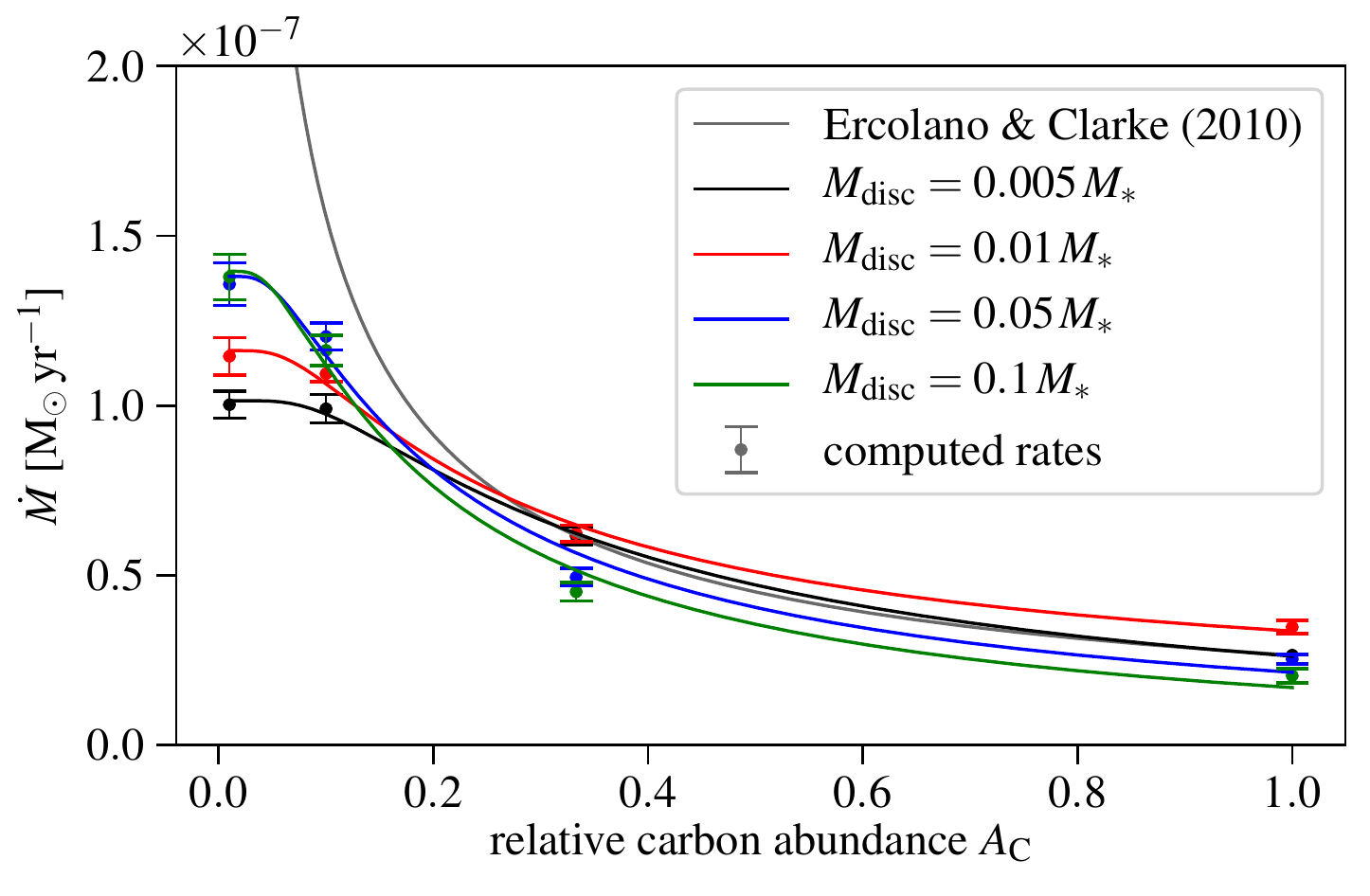}
\caption{Mass-loss rate as a function of the relative carbon abundance $\textit{A}_{\textrm{C}}$. Shown are the data and fits according to \autoref{eq:relationZ} for the four disc masses. The metallicity relation of \citet{ErcolanoClarke2010} is included as a reference. In contrast to their results, our relations predict a less extreme increase of the mass-loss rate with decreasing carbon abundance (metallicity).}\label{fig:MZrel}
\end{figure}

\vspace{0.5cm} In \autoref{fig:MZrel} we fit the mass-loss rate as a function of the relative carbon abundance $\textit{A}_{\textrm{C}}$ (compared to the solar carbon abundance value) for all four disc masses. Here, the average mass-loss rates were calculated from the last 100 orbits, the solar abundance value for the lowest-mass disc was adopted from \citet{Picogna2018}. In order to fit the data, we applied the following relation
\begin{equation}\label{eq:relationZ}
\dot{M}(A_{\mathrm{C}}) = a \cdot e^{-\frac{b}{A_{\mathrm{C}}}} + c
\end{equation}\noindent
\noindent
finding 
\begin{equation}\label{eq:05M}
\dot{M}(A_{\mathrm{C}}) = (-9.33 \times 10^{-8})\,\frac{\mathrm{M}_{\odot}}{yr} \cdot e^{-\frac{0.29}{A_{\mathrm{C}}}} + (1.02 \times 10^{-7})\,\frac{\mathrm{M}_{\odot}}{yr}   
\end{equation}
\noindent
for the 0.005\,$\textit{M}_{*}$ disc, 
\begin{equation}\label{eq:1M}
\dot{M}(A_{\mathrm{C}}) = (-1.05 \times 10^{-7})\,\frac{\mathrm{M}_{\odot}}{yr} \cdot e^{-\frac{0.24}{A_{\mathrm{C}}}} + (1.16 \times 10^{-7})\,\frac{\mathrm{M}_{\odot}}{yr}   
\end{equation}
\noindent
for the 0.01\,$\textit{M}_{*}$ disc, 
\begin{equation}\label{eq:5M}
\dot{M}(A_{\mathrm{C}}) = (-1.4 \times 10^{-7})\,\frac{\mathrm{M}_{\odot}}{yr} \cdot e^{-\frac{0.18}{A_{\mathrm{C}}}} + (1.38 \times 10^{-7})\,\frac{\mathrm{M}_{\odot}}{yr}   
\end{equation}
\noindent
for the 0.05\,$\textit{M}_{*}$ disc and 
\begin{equation}\label{eq:highM}
\dot{M}(A_{\mathrm{C}}) =( -1.45 \times 10^{-7})\,\frac{\mathrm{M}_{\odot}}{yr} \cdot e^{-\frac{0.17}{A_{\mathrm{C}}}} + (1.4 \times 10^{-7})\,\frac{\mathrm{M}_{\odot}}{yr}   
\end{equation}
\noindent
for the 0.1\,$\textit{M}_{*}$ disc. Beside these four relations, we also included the metallicity relation 
\begin{equation}
    \dot{M}_{\mathrm{w}} \propto Z^{-0.77}
\end{equation}
found by \citet{ErcolanoClarke2010} as a reference in \autoref{fig:MZrel}. 
While there are fundamental differences between the approach used here and that of \citet{ErcolanoClarke2010}, as also discussed below, a comparison is still interesting as previous work used this relation to investigate the effect of carbon depletion on transition disc populations \citep{ErcolanoWeber2018}. We show here that there are important differences, particularly at low values of carbon abundance, highlighting the need of further work on population synthesis of transition discs using our current results. In contrast to their result, our simulations predict a flatter and somewhat saturating increase of the mass-loss rate with carbon abundance (metallicity). In \autoref{fig:MZrel} we are only showing the relation of \citet{ErcolanoClarke2010} for the lowest-mass disc, using the mass-loss rate for solar metallicity found by \citet{Picogna2018} as $\dot{M}_0$. Comparing our new and the old relation for each disc mass individually we find that the two curves follow (except for the disc mass of 0.01\,$\textit{M}_{*}$) a very similar slope down to a carbon abundance of 0.2$-$0.3 but differ significantly for smaller carbon abundances. 

The comparison of our model to the model of \citet{ErcolanoClarke2010} is mostly for illustrative purposes, as the two models have substantial differences. Rather than performing hydrodynamical calculations to extract mass-loss rates, \citet{ErcolanoClarke2010} perform thermal calculations and look for a hydrostatic solution. The mass-loss rates are then calculated assuming that at each radius the surface mass-loss rate $\dot{\mathit{\Sigma}}$ is the product of the density and the sound speed at the base of the flow. Here the base of the flow at each radius is identified as the first height starting from the midplane, where the temperature of the gas becomes equal to the local escape temperature. This simplified method carries large uncertainties (see discussion in \citealp{Owen2010}). In contrast, this work performs detailed hydrodynamical calculations to extract the wind mass-loss rates and profiles. Furthermore \citet{ErcolanoClarke2010} lower the abundance of all elements by the same amount to investigate the metallicity dependency, since their work aimed at studying disc lifetimes in regions of lower metallicity (e.g. the extreme outer Galaxy) and their effect on planet formation. The goal of this work is different as we want to investigate the effects of the observationally determined gas-phase depletion of carbon in discs. Therefore we only lower the abundance of carbon. It is thus not surprising that the resulting effect on the mass-loss rate is lower, since the opacity suppression is not as high as in \citet{ErcolanoClarke2010}. 
\begin{figure}
\includegraphics[width=\columnwidth]{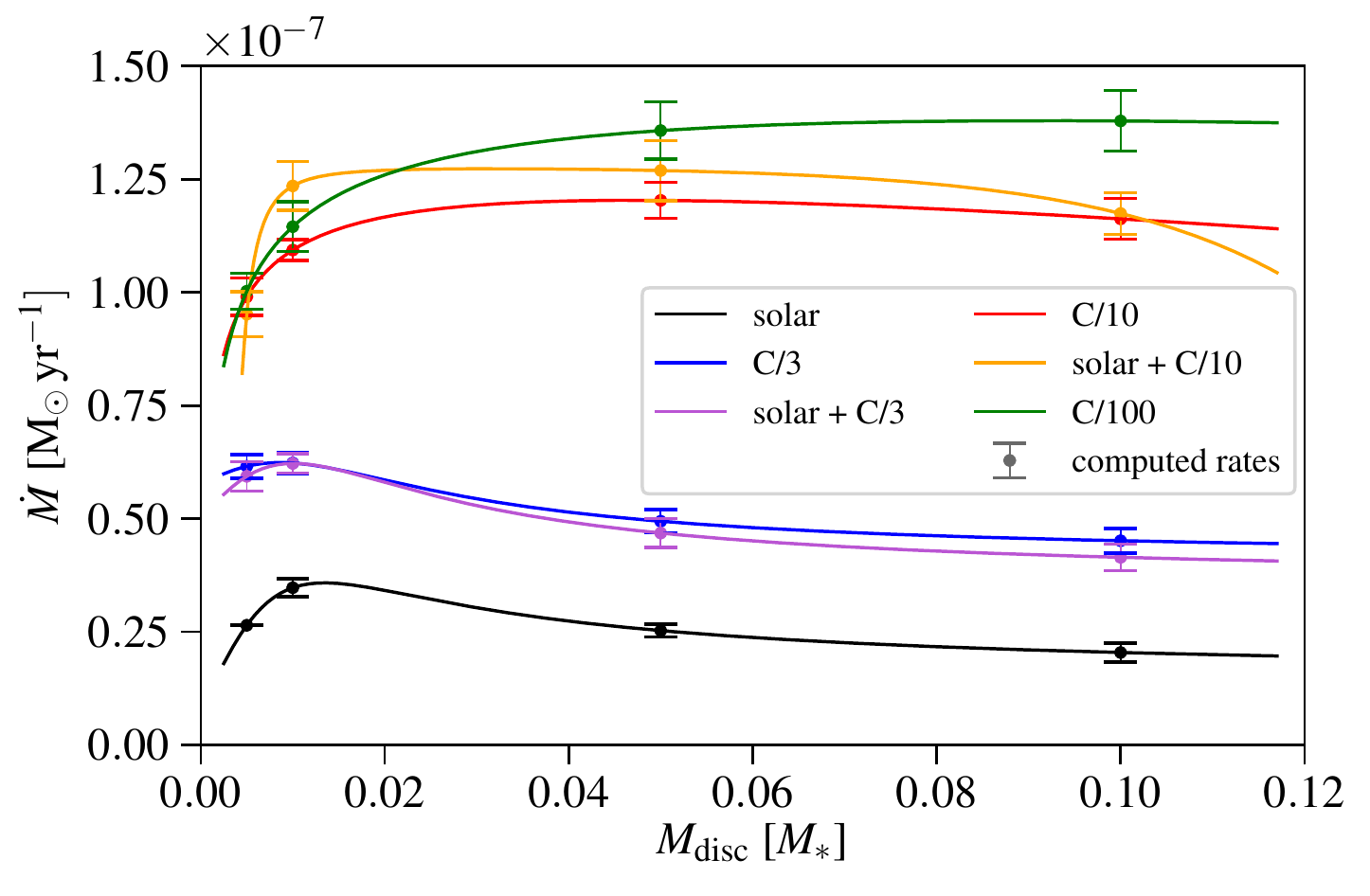}
\caption{Mass-loss rate as a function of disc mass, shown for the six different carbon abundance cases. While for higher carbon abundances the mass-loss is overall going down with disc mass, it increases when the carbon abundance is low.}\label{fig:DiscM}
\end{figure}

Comparing the four disc masses, we notice a reversing behaviour, as the mass-loss rates are decreasing with disc mass for larger carbon abundances, but increasing with disc mass for smaller carbon abundances. Being comparable for the lower-mass discs, a significant rise in the mass-loss rate from factor 10 to factor 100 carbon depletion can be distinguished for the higher-mass discs. The reason behind the various effects connected to the disc mass, is that depending on the carbon abundance, photoevaporation is efficient in distinct regions of the disc. While for high carbon abundances ($\textit{A}_{\textrm{C}}$ $\gtrsim$ 0.3) the total mass-loss is mainly dominated by the inner disc, the disc becomes more transparent to X\hbox{-}ray radiation for lower carbon abundances, which can then drive a significant flow from the outer disc regions. Now two effects have to be considered: One is that radiation can penetrate radially further into a lower-mass disc, whereas more mass can in principle be removed if a larger reservoir is hit. In this context, the low-mass disc experiences stronger winds when the depletion is moderate because the effect of reaching larger radii dominates over the effect of the larger mass content, which is anyway small near the star. For strong depletion however, the radiation can heat the large amount of mass present in the outer disc, which is why the radius of the layers reached by the radiation becomes less important. To conclude, we would like to note that even though clear variations can be distinguished between the four disc masses, these differences are in fact remarkably small, keeping in mind that the discs span a wide realistic mass range. 
\begin{table}
\caption{Average mass-loss rates of the primordial disc simulations calculated from the last 100 orbits.}\label{tab:massLoss}
\centering
\begin{tabular}{l c}
\hline
\hline
\bf{simulation} & \bf{$\dot{M}$ $[\mathrm{M}_{\odot} \mathrm{yr}^{-1}]$}\\
\hline
\textit{disc mass 0.005\,M$_*$} &  \\
\hline
solar \citep{Picogna2018}& $2.644 \times 10^{-8}$\\
C/3 & $(6.16 \pm  0.26)\times 10^{-8}$ \\
solar + C/3 & $(5.94 \pm  0.32)\times 10^{-8}$ \\
C/10 &  $(9.91 \pm  0.41)\times 10^{-8}$ \\
solar + C/10  & $(9.52 \pm  0.50)\times 10^{-8}$ \\
C/100 &  $(1.0 \pm  0.04)\times 10^{-7}$ \\
\hline
\textit{disc mass 0.01\,M$_*$} &  \\
\hline
solar & $(3.47 \pm  0.2) \times 10^{-8}$\\
C/3 & $(6.23 \pm  0.24)\times 10^{-8}$ \\
solar + C/3 & $(6.22 \pm  0.21)\times 10^{-8}$ \\
C/10 &  $(1.09 \pm  0.02)\times 10^{-7}$ \\
solar + C/10  & $(1.24 \pm  0.05)\times 10^{-7}$ \\
C/100 &  $(1.14 \pm  0.06)\times 10^{-7}$ \\
\hline
\textit{disc mass 0.05\,M$_*$} &  \\
\hline
solar & $(2.53 \pm  0.14) \times 10^{-8}$\\
C/3 & $(4.94 \pm  0.26)\times 10^{-8}$ \\
solar + C/3 & $(4.68 \pm  0.32)\times 10^{-8}$ \\
C/10 &  $(1.2 \pm  0.04)\times 10^{-7}$ \\
solar + C/10  & $(1.27 \pm  0.07)\times 10^{-7}$ \\
C/100 &  $(1.36 \pm  0.06)\times 10^{-7}$ \\
\hline
\textit{disc mass 0.1\,M$_*$} & \\
\hline
solar & $(2.04 \pm 0.21) \times 10^{-8}$\\
C/3 & $(4.51 \pm  0.28)\times 10^{-8}$ \\
solar + C/3 & $(4.15 \pm  0.29)\times 10^{-8}$ \\
C/10 &  $(1.16 \pm  0.04)\times 10^{-7}$ \\
solar + C/10  & $(1.17 \pm  0.05)\times 10^{-7}$ \\
C/100 &  $(1.38 \pm  0.07)\times 10^{-7}$ \\
\hline
\end{tabular}
\end{table}

In \autoref{fig:DiscM}, we display the dependency of the total mass-loss rate on the disc mass for each individual carbon abundance, applying the following ad-hoc functions
\begin{equation}\label{eq:relation1}
\dot{M}(M_{\mathrm{disc}}) = \frac{a + M_{\mathrm{disc}}}{b + c \cdot M_{\mathrm{disc}}^2} + d
\end{equation}
for the higher carbon abundance and
\begin{equation}\label{eq:relation2}
\dot{M}(M_{\mathrm{disc}}) = a \cdot M_{\mathrm{disc}}^{\left(b \cdot M_{\mathrm{disc}}^c\right)} + d
\end{equation}\noindent
for the lower carbon abundance cases. For no or moderate depletion (black, purple and blue curve) the mass-loss rate is overall decreasing with increasing disc mass due to the fact that the radiation can not reach the radially further disc layers. As the radiation can however hit a larger mass reservoir if more material is present, the mass-loss rate does not follow a steep, but rather flat slope after a short increase. If on the other hand the carbon abundance is low (red, orange and green curve), the mass-loss rate is in general increasing with disc mass. Similar to the high carbon abundance cases these curves are marked by a flat rise and are then slightly decreasing when the disc mass becomes too high for the radiation to penetrate far enough into the disc layers. 

All average mass-loss rates for the primordial disc simulations, calculated from the last 100 orbits, are listed in \autoref{tab:massLoss}. The corresponding uncertainties are calculated from the standard deviation. 
\subsection{Hole radius dependency} \label{sec:HoleM}
\begin{figure*}
\centering
\includegraphics[width=17.5cm]{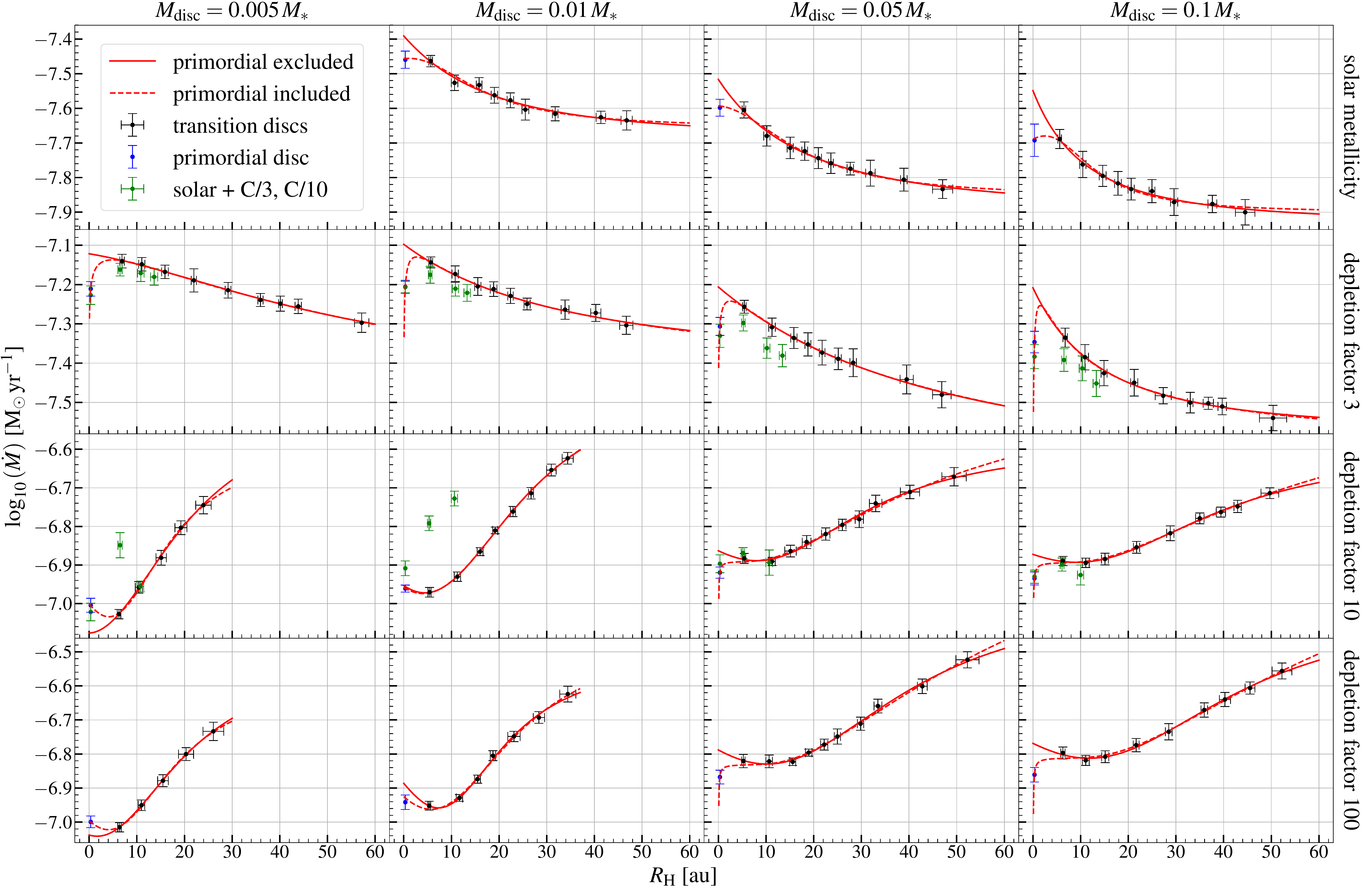}
\caption{Mass-loss rate as a function of the hole radius for the four different disc masses and carbon abundances. The black and blue dots represent the computed mass-loss rates for the transition and primordial discs respectively. The solid red lines display a fit for the transition discs only, while the primordial disc simulations are taken into account for the fit shown by the red dashed lines. With green dots, the mass-loss rates for the inhomogeneously depleted discs are included.} \label{fig:MRh}
\end{figure*}
\begin{figure*}
\centering
	\includegraphics[width=\columnwidth]{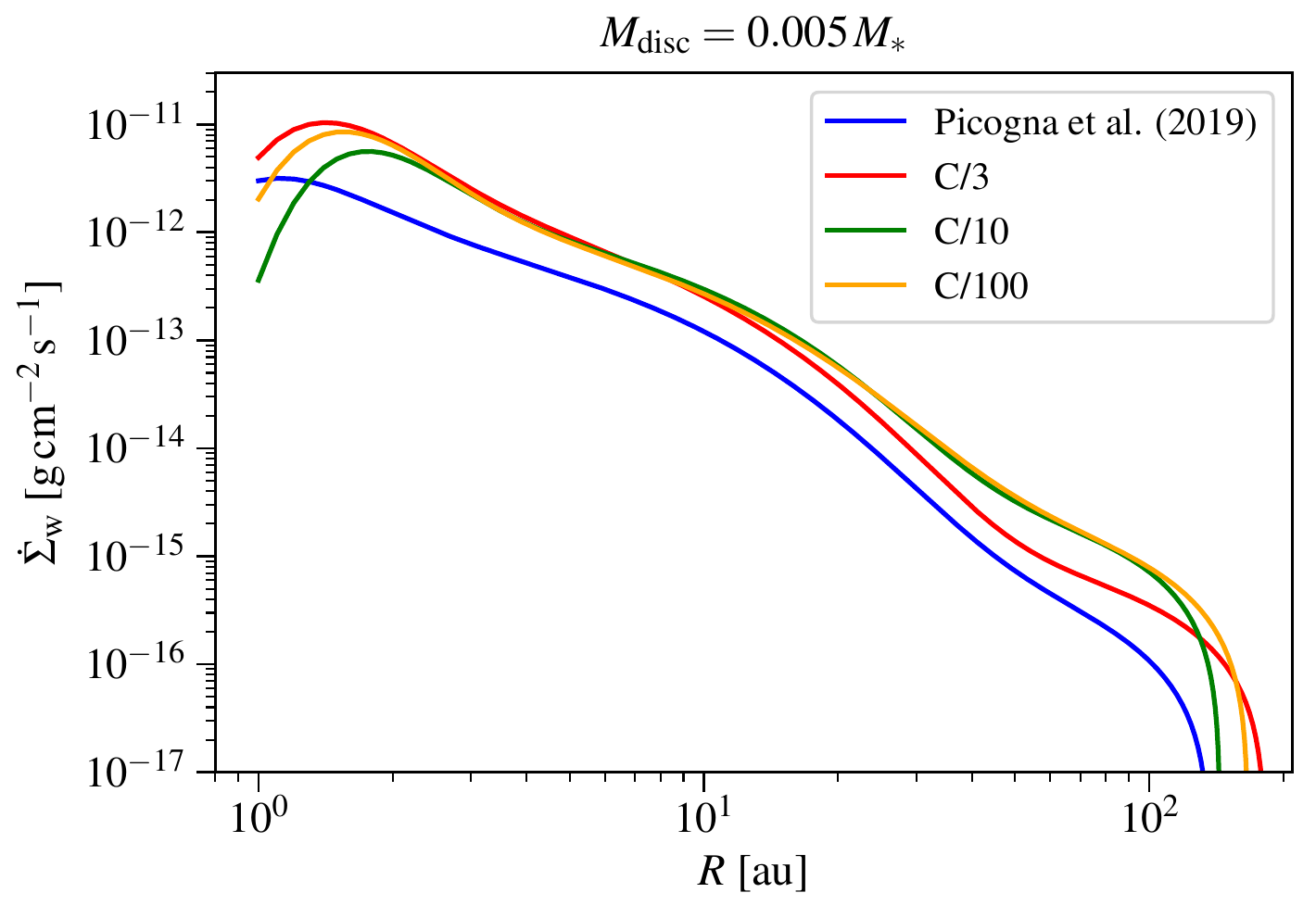}
	\includegraphics[width=\columnwidth]{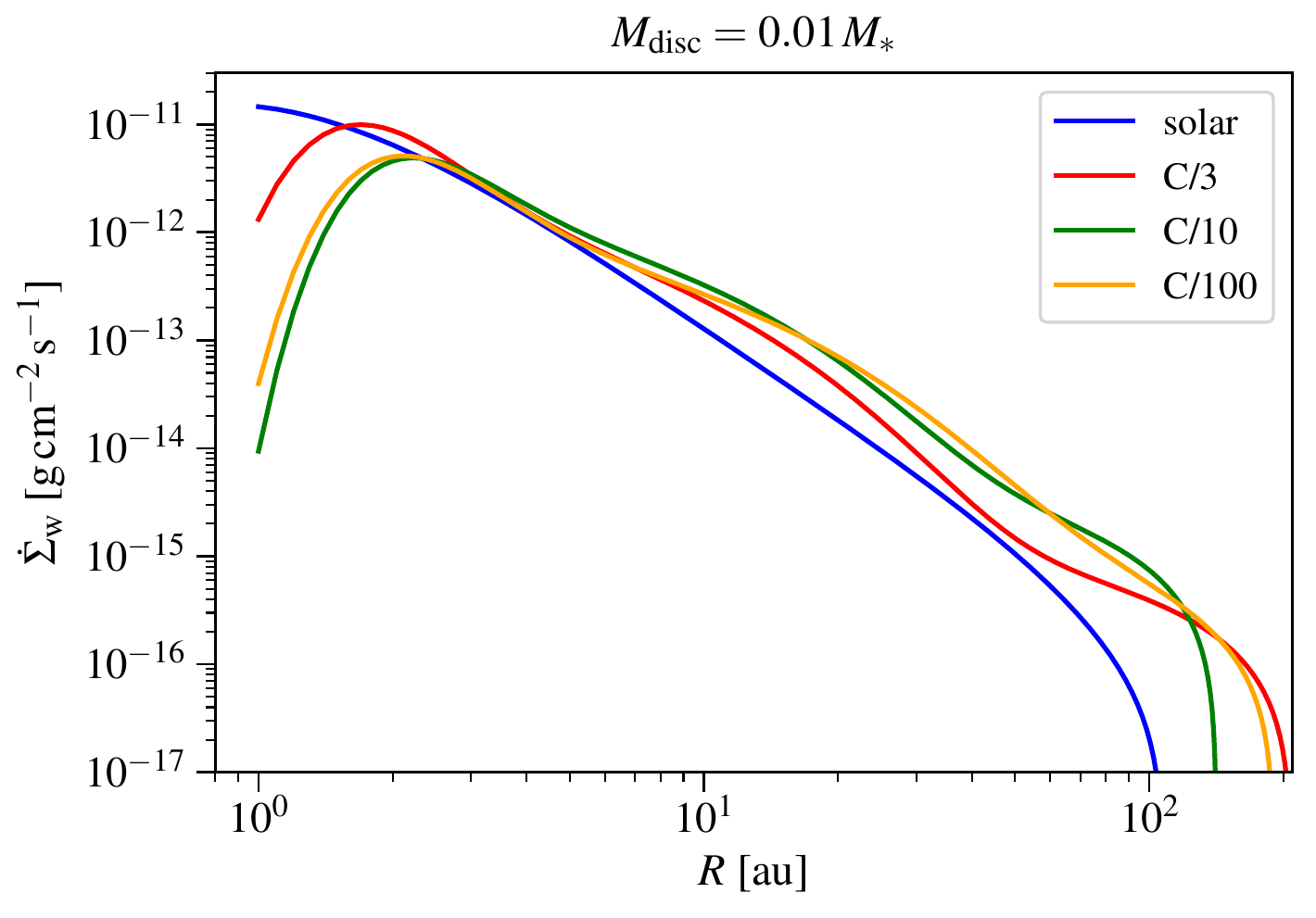}
	\includegraphics[width=\columnwidth]{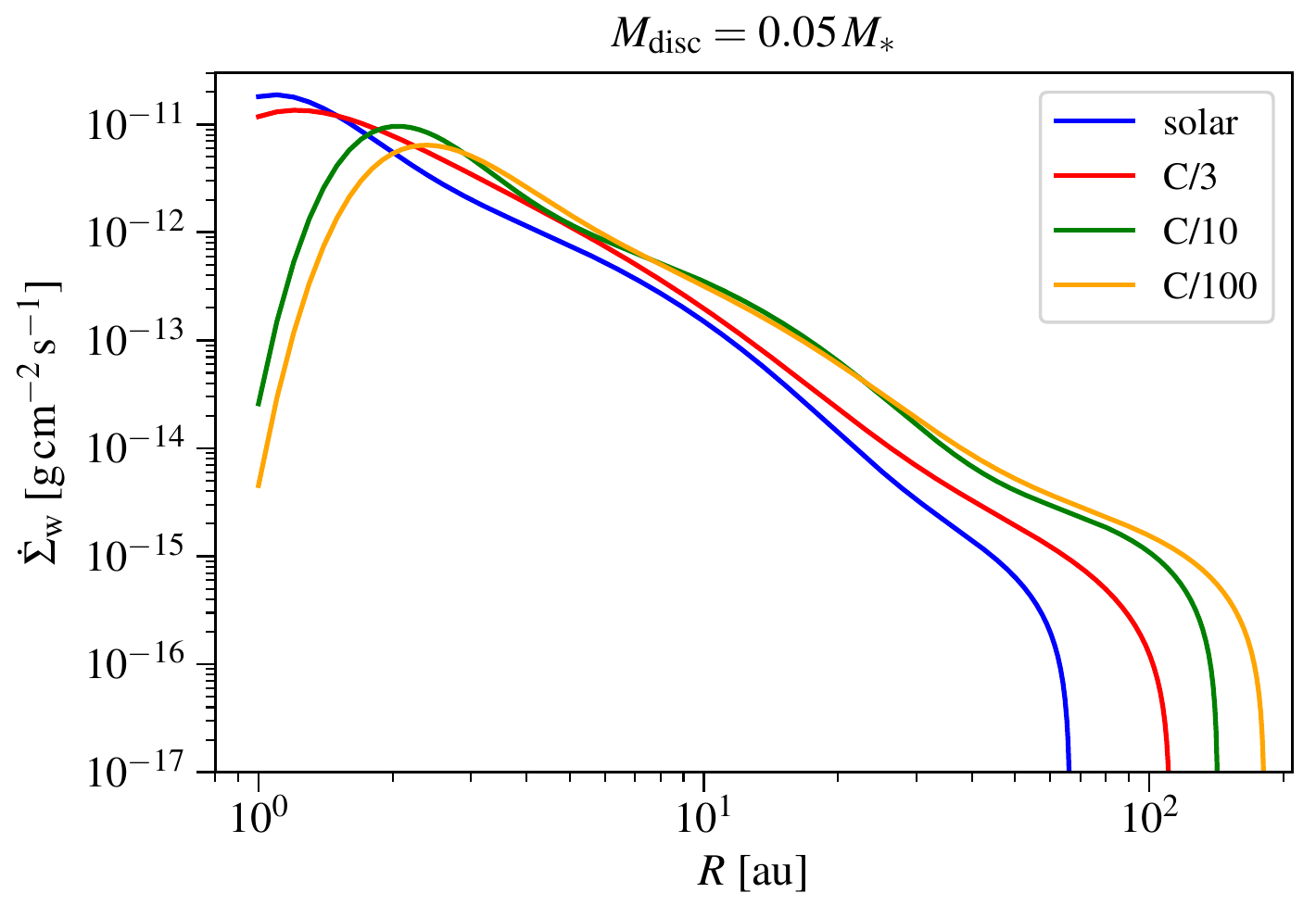}
	\includegraphics[width=\columnwidth]{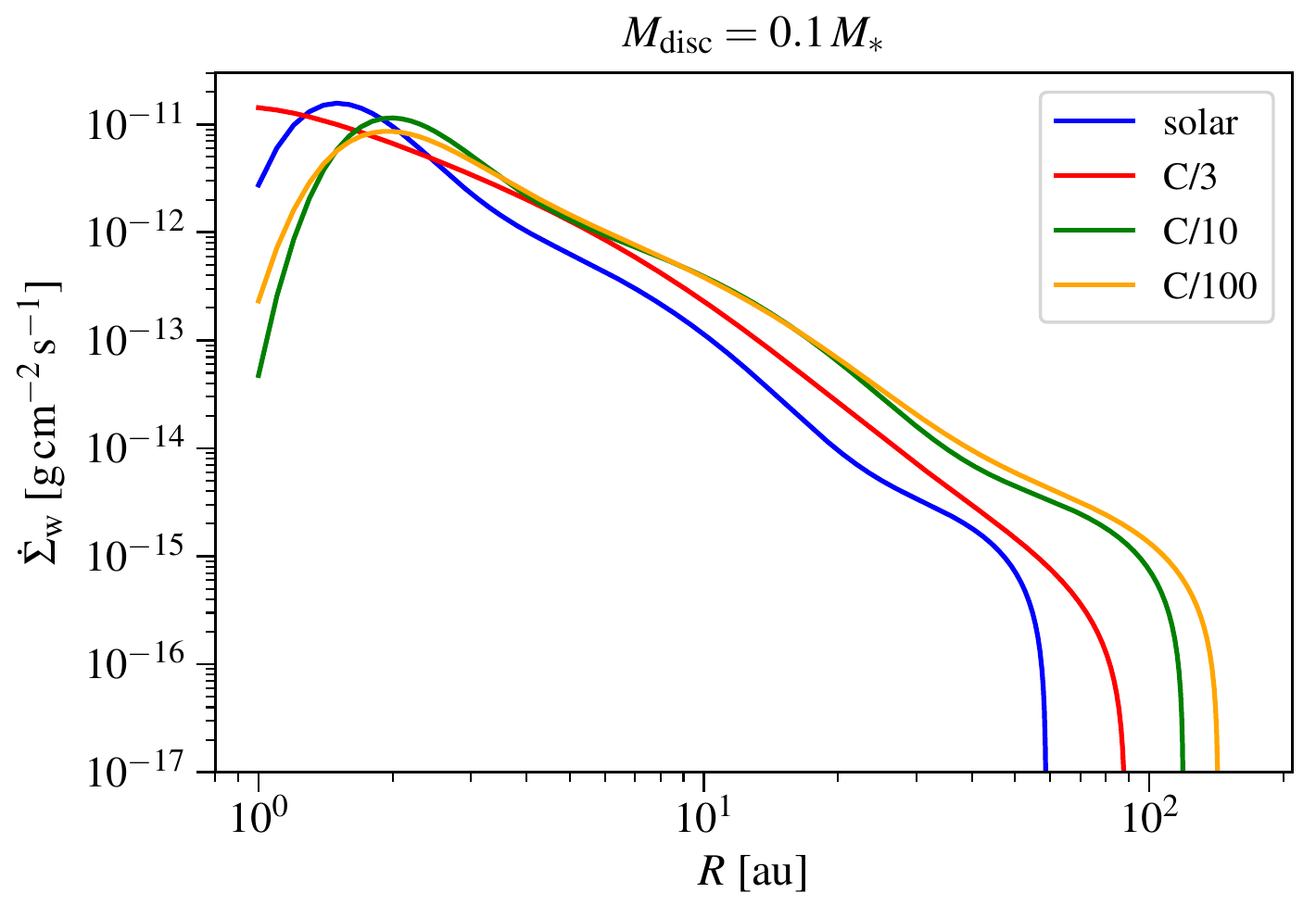}
    \caption{Mass-loss profiles $\dot{\mathit{\Sigma}}$ of the primordial discs, shown for the four different disc masses and homogeneous carbon abundances. With increasing depletion, the profiles extend to larger disc radii.}
    \label{fig:SigmaDotPrim1}
\end{figure*}
\begin{figure*}
\centering
	\includegraphics[width=\columnwidth]{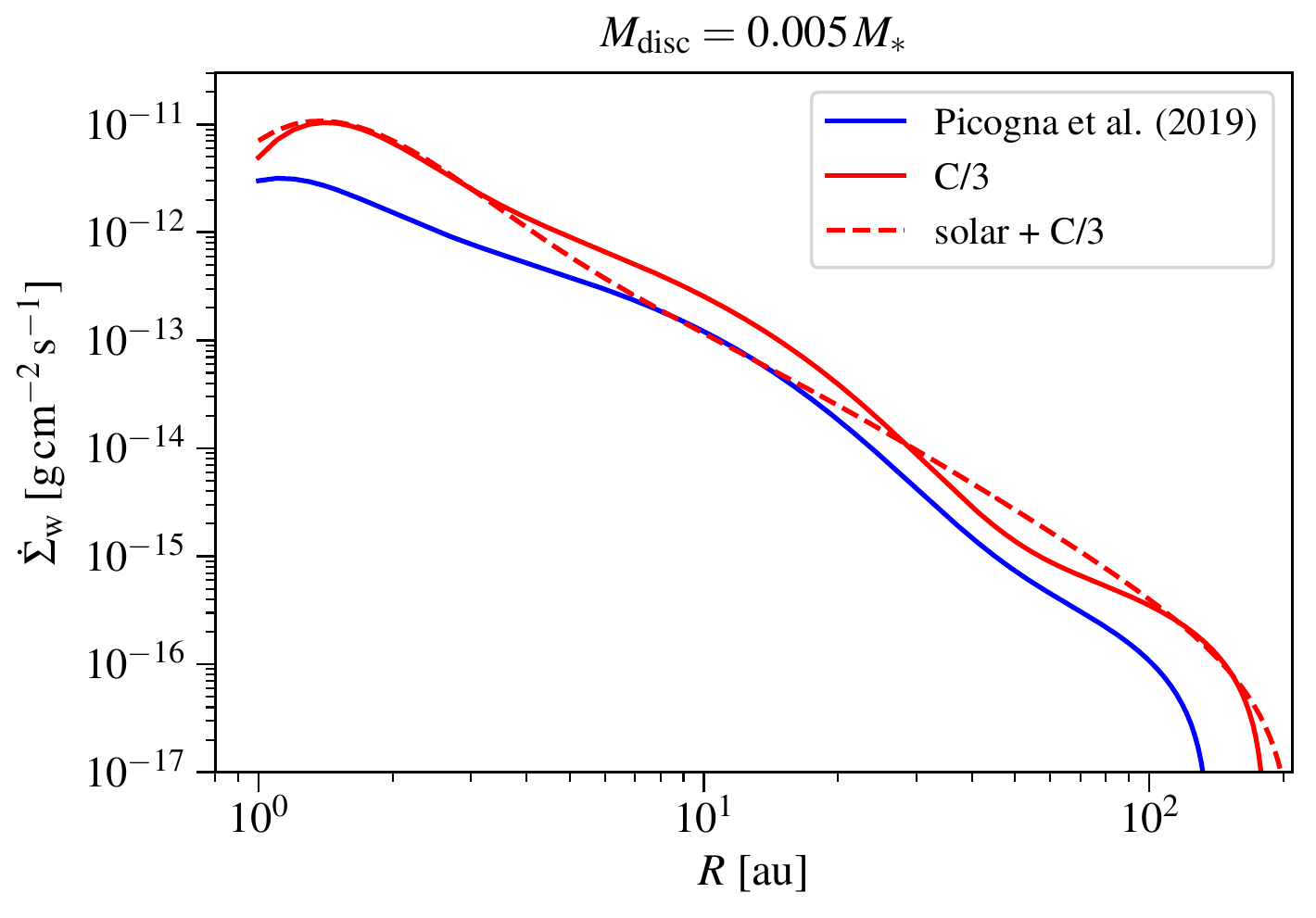}
	\includegraphics[width=\columnwidth]{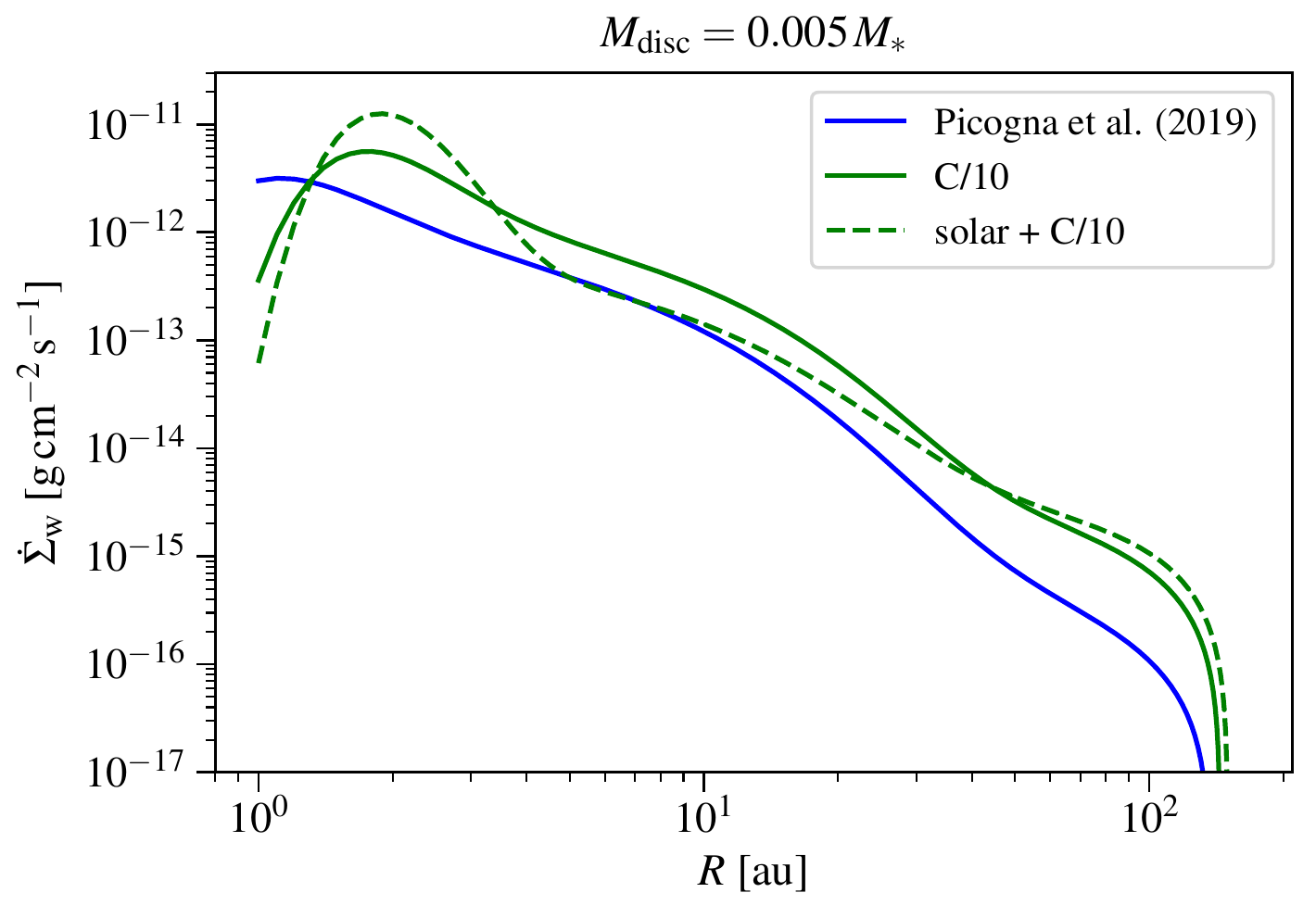}
	\includegraphics[width=\columnwidth]{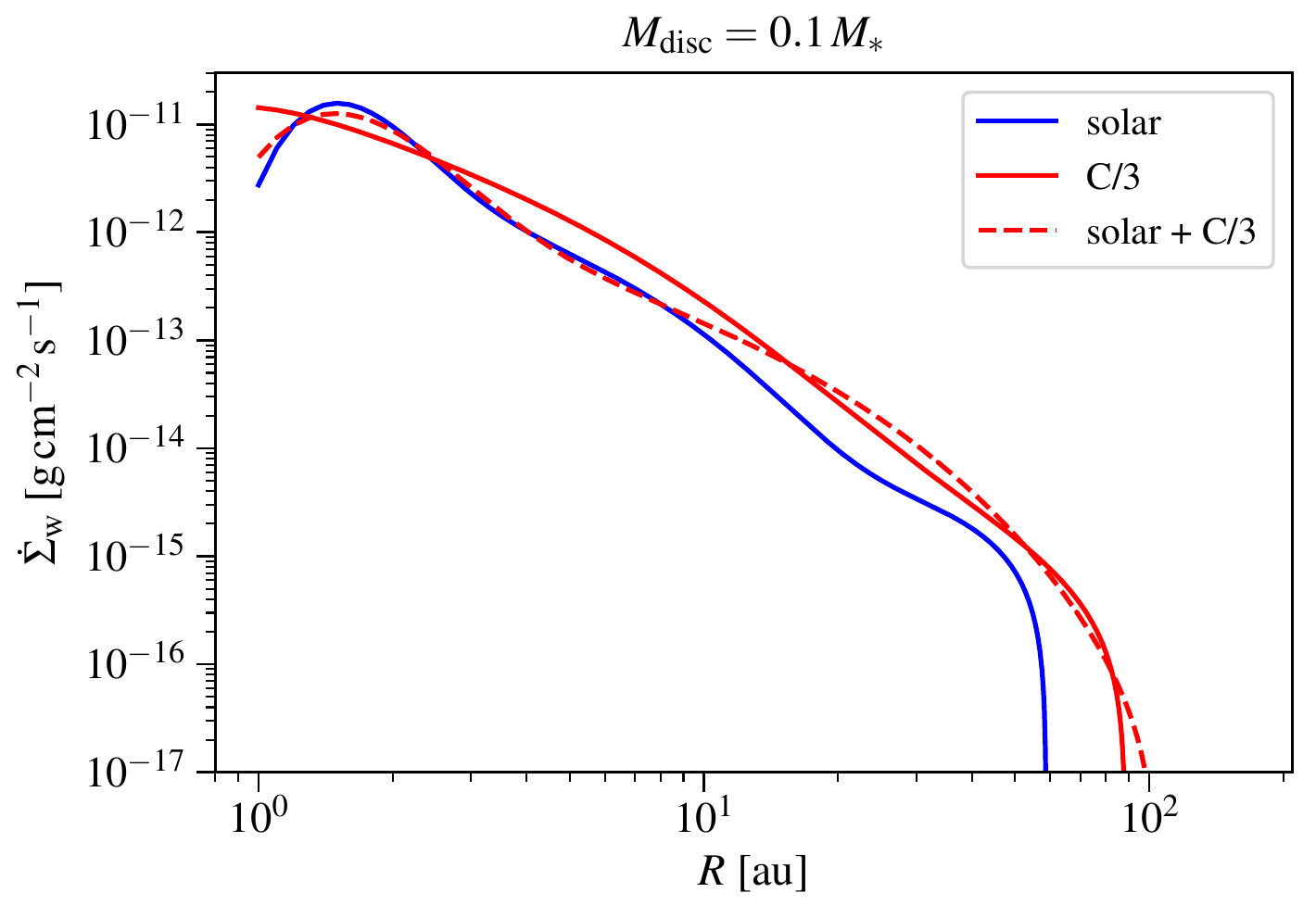}
	\includegraphics[width=\columnwidth]{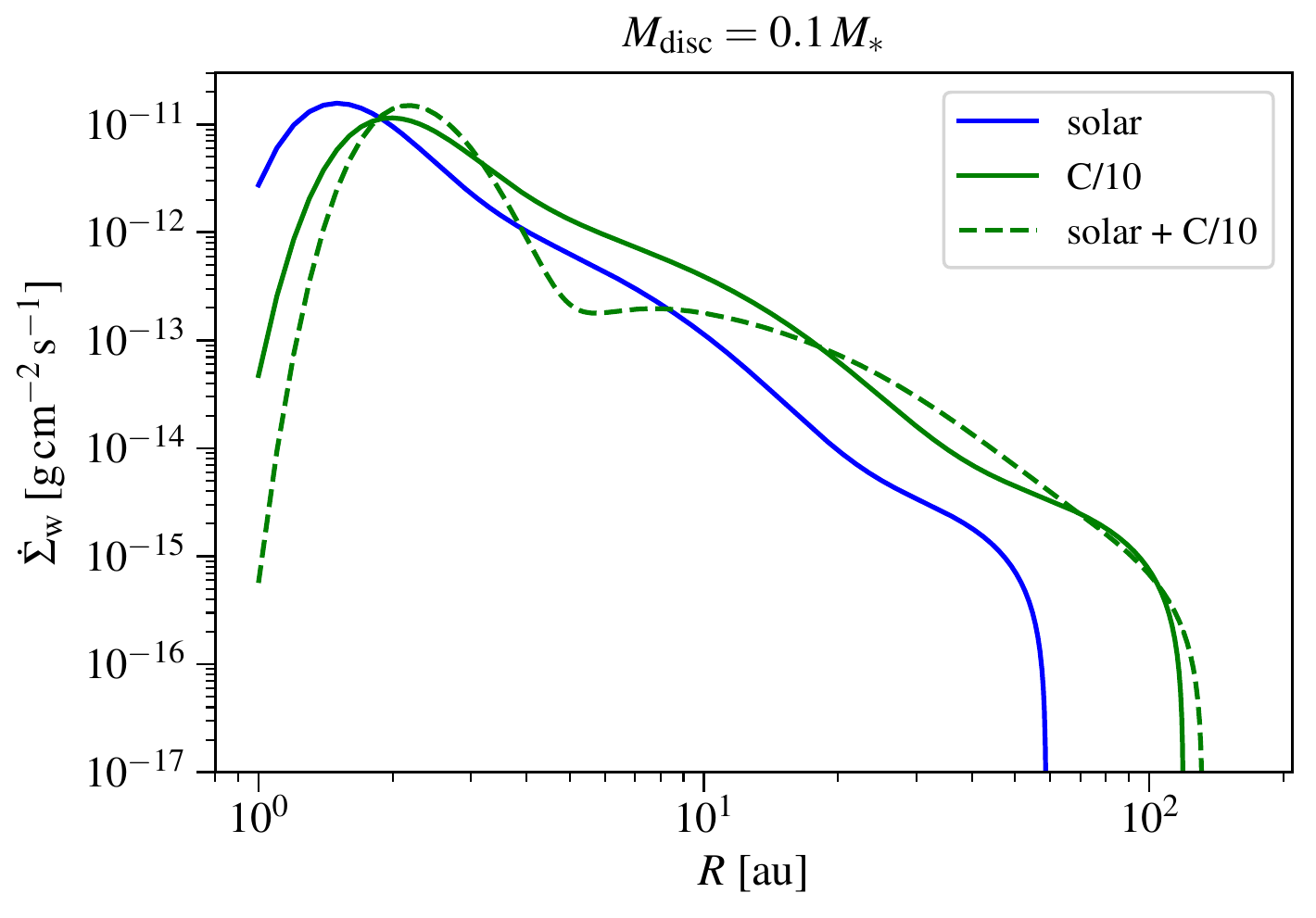}
    \caption{Mass-loss profiles $\dot{\mathit{\Sigma}}$ for the inhomogeneously depleted discs, shown for the lowest-mass disc of 0.005\,$\textit{M}_{*}$ (top plots) and the highest-mass disc of 0.1\,$\textit{M}_{*}$ (bottom plots). Compared to the homogeneously depleted discs, the mass-loss is slightly higher close to and far from the star and lower in the mid disc regions.}
    \label{fig:SigmaDotPrim2}
\end{figure*}
\begin{figure*}
\includegraphics[width=4.3cm]{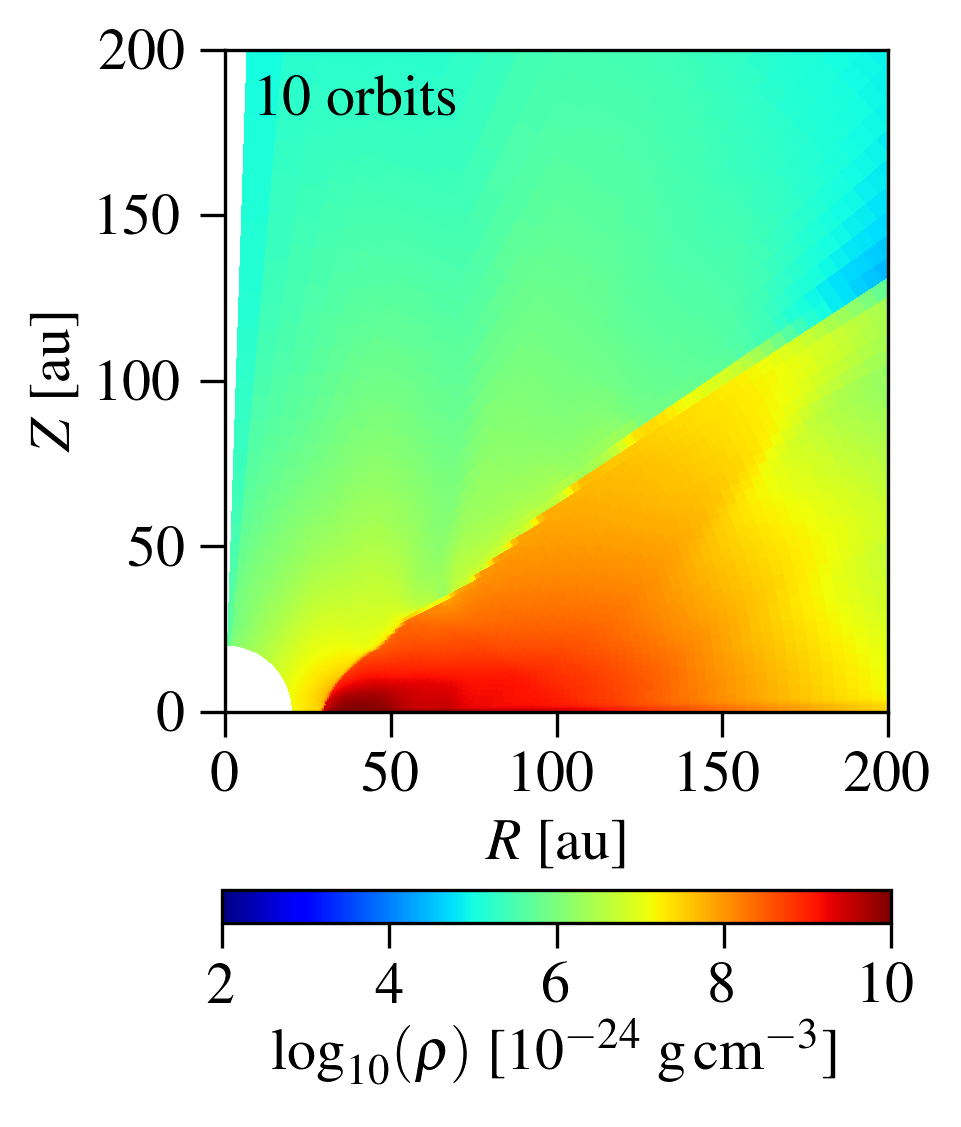}
\includegraphics[width=4.3cm]{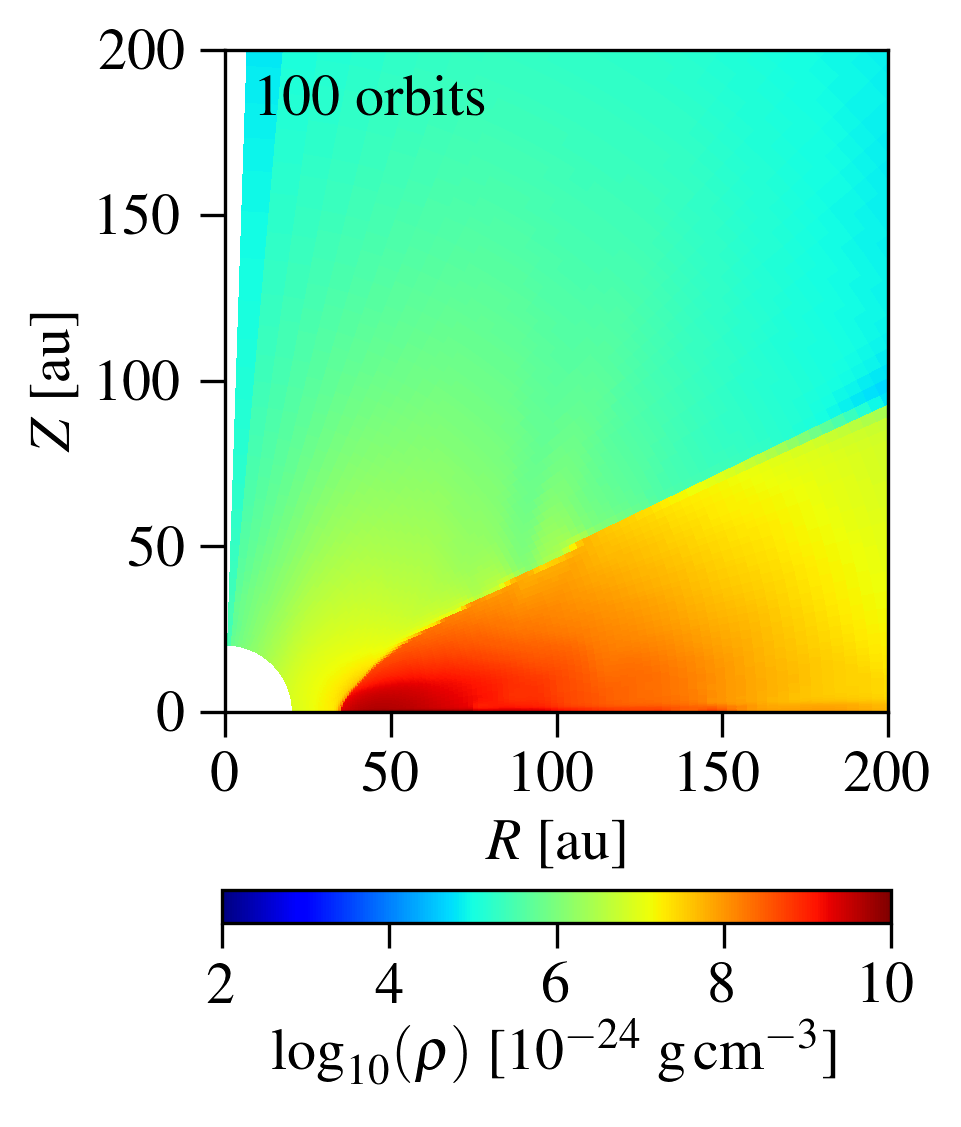}
\includegraphics[width=4.3cm]{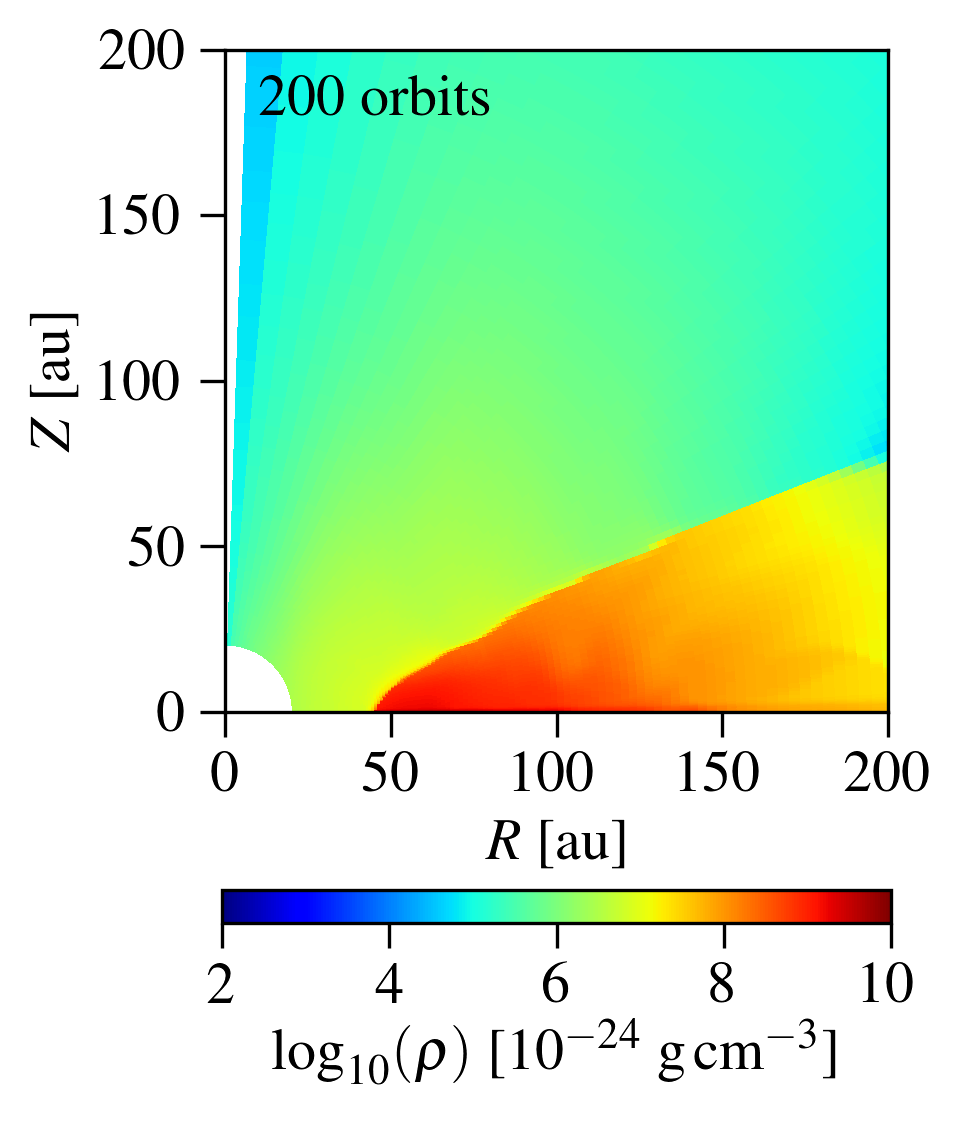}
\includegraphics[width=4.3cm]{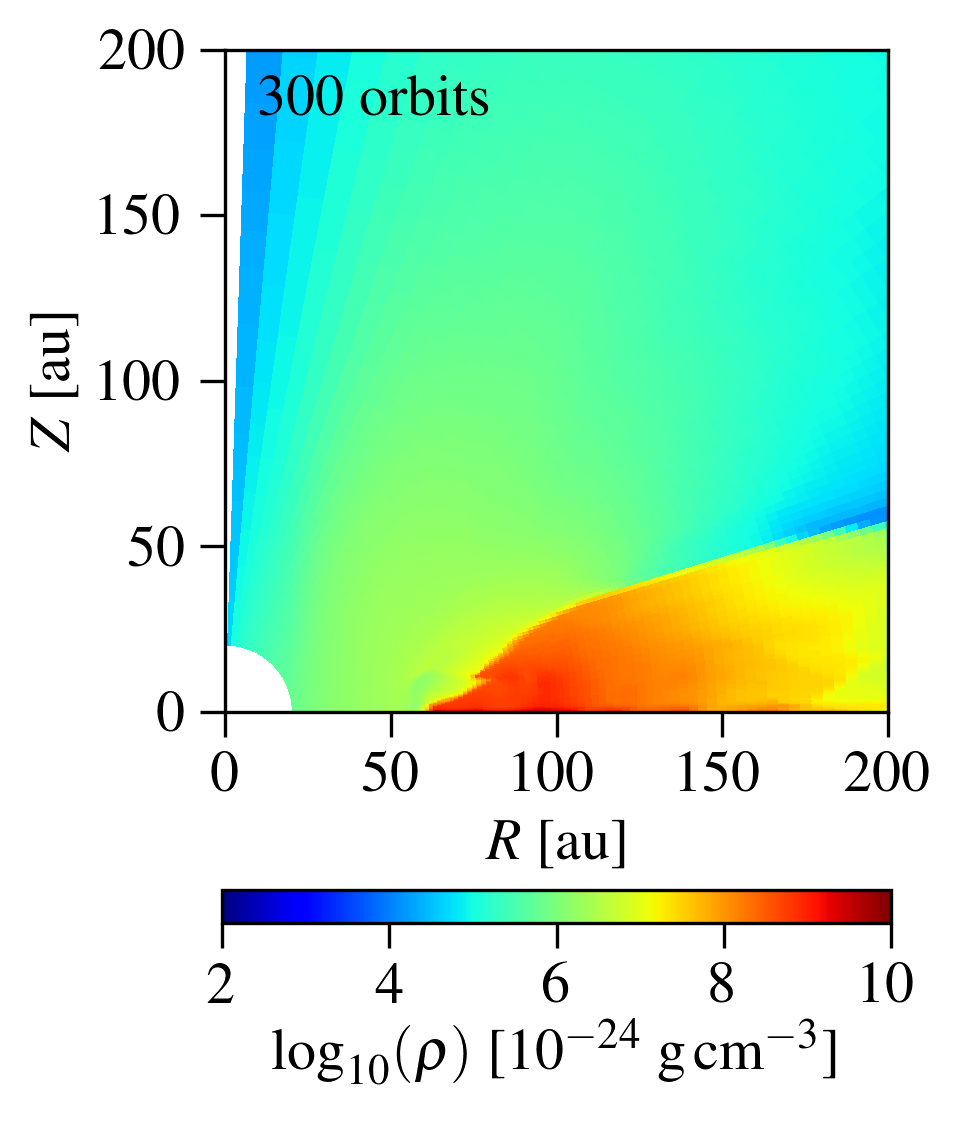}
\\
\includegraphics[width=4.3cm]{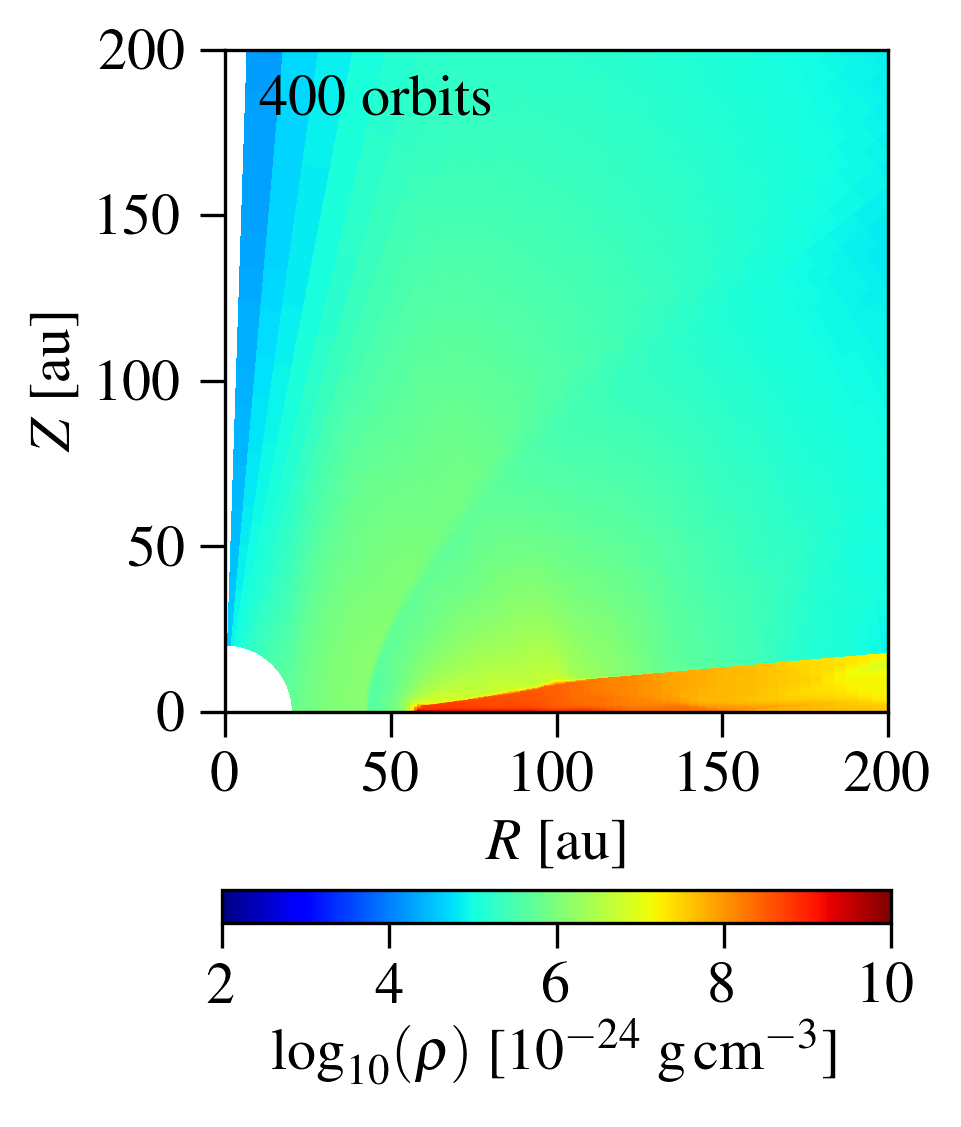}
\includegraphics[width=4.3cm]{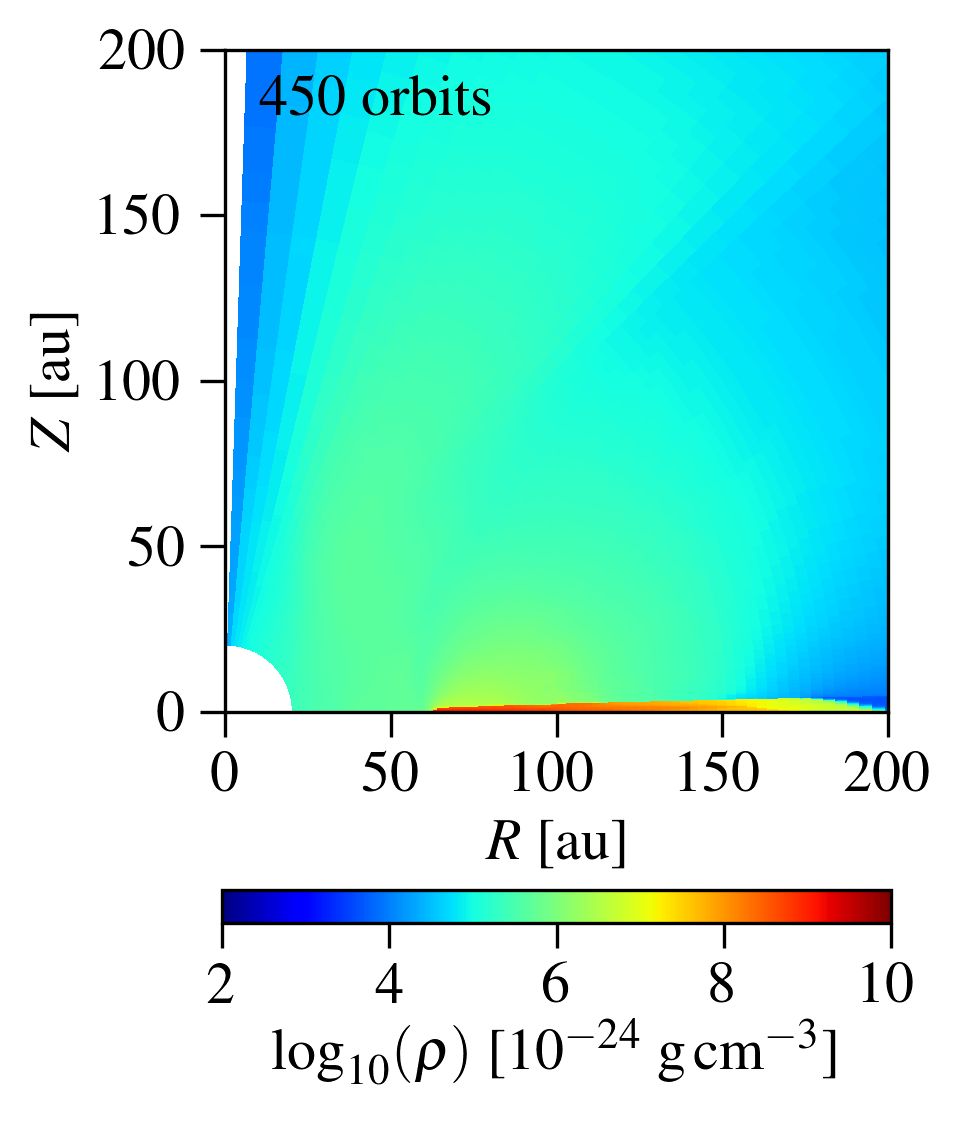}
\includegraphics[width=4.3cm]{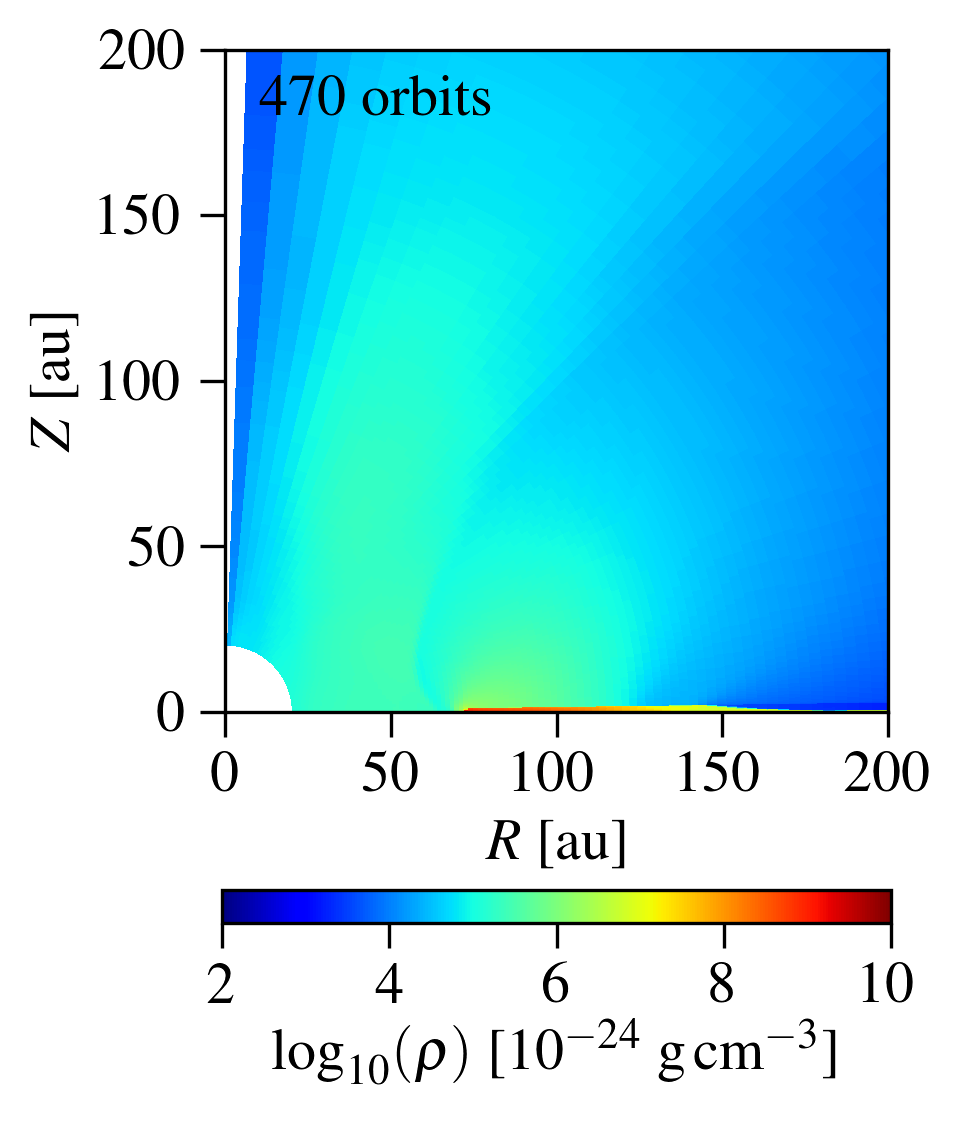}
\includegraphics[width=4.3cm]{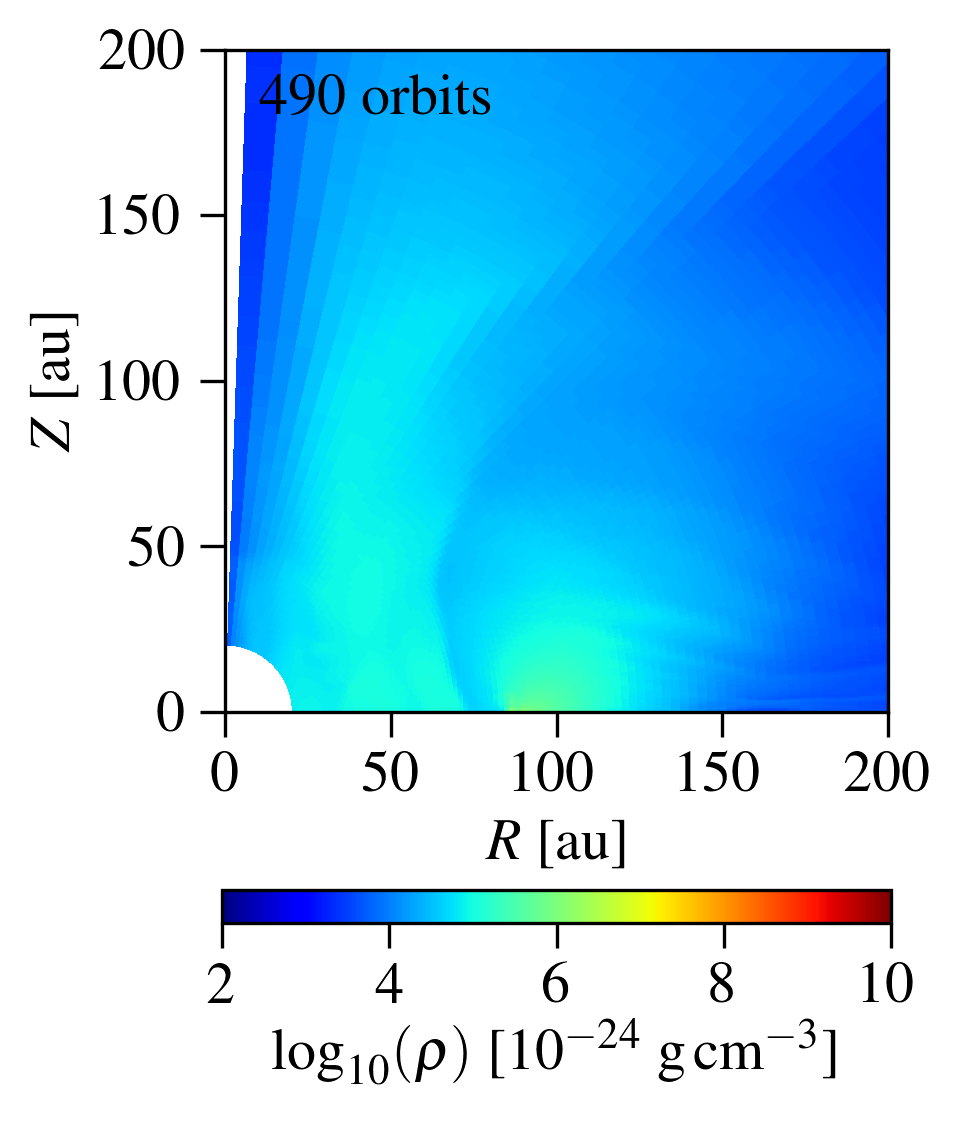}
\caption{Rapid disc clearing of a low-mass transition disc (0.005\,$\textit{M}_{*}$) with an initial hole radius of $\textit{R}_{\textrm{H}} \approx \textrm{30\,au}$ and a carbon depletion by a factor of 10. The disc is fully dispersed within 500 orbits, corresponding to $\approx$ 19000\,yrs.} \label{fig:Sweeping}
\end{figure*}
As mentioned before, the transition discs are evolving relatively fast during our simulations. It was therefore more challenging to find stable mass-loss rates, and thus profiles, because the full range of orbits could not be taken into account. We therefore decided to use a suitable range of 100 orbits (and not necessarily the last orbits), for which we calculated the average hole radius and mass-loss rate. In this context, we considered several factors in order to find the best possible time span. First, we tried to find a range for which the mass-loss rate was relatively stable. Furthermore, we checked if the evolution of the disc mass was moderate and not too rapid in this range. In addition, we only chose orbits for which significant thinning of the disc had not begun yet. In general, it was easier to match these three conditions (simultaneously) for the higher-mass disc simulations. In case of the larger depletions (factor 10 and 100) no stable mass-loss rates for hole radii above $\textit{R}_{\textrm{H}} \approx \textrm{25\,au}$ could be found for the 0.005\,$\textit{M}_{*}$ disc. Similarly no stable mass-loss rates were established at these depletion factors for the 0.01\,$\textit{M}_{*}$ disc simulations above $\textit{R}_{\textrm{H}} \approx  \textrm{35\,au}$. These discs are evolving extremely fast and are (almost) completely dispersed during the simulation. We discuss the implications of this rapid disc dispersal in \autoref{sec:sweeping}. 

The transition disc simulations can be used to test the dependency of the photoevaporative mass-loss rate on the inner hole radius. The results of this parameter study are presented in \autoref{fig:MRh} and \autoref{tab:massLossTD} (see Appendix \ref{Table}). In \autoref{fig:MRh} we plot the mass-loss rate as a function of the hole radius (black dots) which we fit with the following relation
\begin{equation}\label{eq:lorentzpeak}
\dot{M} (R_{\mathrm{H}}) = \frac{a}{1 + \left(\frac{\dot{M}-b}{c}\right)^2} + d
\end{equation}\noindent 
\noindent
(red solid lines). The primordial mass-loss rates (blue dots) are excluded from this fit but included in a second one (red dashed lines) for which we applied different functions. It is difficult to determine which curve would better match the mass-loss rate for small hole radii ($\textit{R}_{\textrm{H}}$ < 5\,au), as we do not have any hydrodynamical models for these transition discs. When a gap opens at very small disc radii, an inner disc is in general still present, shielding the outer disc from the star's direct radiation. By the time this inner disc will be accreted, the hole will however have developed to larger radii. It is therefore not realistic to model a transition disc with a very small hole radius, as such a disc would still behave like a primordial one. Thus, it is probably more appropriate to treat transition discs and primordial discs independent from each other and use the fit for which the primordial mass-loss rate is excluded. In a follow-up paper we will present a population synthesis model for which we will switch between our primordial and transition disc models, applying the fit for the hole radius dependency (red solid lines) down to 2$-$3\,au. 

From \autoref{fig:MRh} we notice a reversed (overall) trend similar to the one described in \autoref{sec:MassLoss}, as the mass-loss rate is decreasing with hole radius for no or moderate depletion and increasing with hole radius for strong depletion. With increasing hole radius the initial mass of the disc decreases as larger parts are cut compared to the primordial disc. For no or moderate depletion this means that more and more mass is removed from the disc regions where photoevaporation is effective, leading to weaker photoevaporative winds. For strong depletion on the other hand, the radiation can penetrate into disc regions that are not affected by the cut in the inner disc parts. Even though the disc mass is still lower for a larger hole radius in this case, the mass-loss rates increase with hole radius, as more disc layers and especially the midplane are irradiated directly. In principle, this effect of directly irradiated layers occurs as well for the moderate depletion, being however dominated by the opposite effect caused by the cut of the inner regions if the inner hole radius is large enough. Even though we can clearly identify the behaviour of the different curves, the absolute difference in the mass-loss rates for various hole radii is minimal. 

The behaviour explained above and in \autoref{sec:MassLoss} can indeed be seen when comparing the four disc masses for each carbon abundance case individually (along the rows) where we again find that the mass-loss rate is decreasing with disc mass if the depletion is low. For higher depletions on the other hand, the mass-loss rate is smaller for the lower-mass discs below a hole radius of $\textit{R}_{\textrm{H}} \approx \textrm{15\,au}$, while it is higher for larger radii. Moreover the slope of the curves is becoming steeper with increasing disc mass when the depletion is low and flatter when the depletion is high. 

Besides the data for the homogeneously depleted discs, we also included the mass-loss rates of the inhomogeneously depleted transition discs in \autoref{fig:MRh} (green dots). For the carbon depletion by a factor of 3, these values lie slightly below the ones for the homogeneously depleted discs, suggesting however a similar slope. In case of the carbon depletion by factor 10, the values lie very close to the ones for the homogeneously depleted discs for the two higher disc masses, but quite far off for the lower-mass discs. 
%Taking into account the mass loss rates of the primordial discs, we indeed find that the values are slightly lower than for smaller holes when the depletion is moderate and the disc mass low. Moreover we find, that for stronger depletion factors the primordial mass loss rates are not higher than the ones from the primordial discs, but they lie above the values predicted by the fit. This can be explained with the effect of the higher disc mass of the primordial discs becoming comparable to the effect of the directly hit layers. Comparing the primordial and transition disc simulations, we furthermore notice that the primordial mass loss rates lie closer to the values predicted by the fits when the disc mass is high. Due to the very large mass, some of the effects described above in the context of primordial and transition discs probably become washed out: As the disc mass is then high in any case, the effects caused my the difference in the disc mass become negligible.
\subsection{Mass-loss profiles \texorpdfstring{$\dot{\mathit{\Sigma}}$}{}}\label{sec:profiles}
The resulting mass-loss profiles from our primordial disc models are displayed in \autoref{fig:SigmaDotPrim1}, \autoref{fig:SigmaDotPrim2} and \autoref{fig:SigmaDotPrim3}. In \autoref{fig:SigmaDotPrim1} we present the profiles of the four different (homogeneous) carbon abundance set-ups for all disc masses. It strikes out that the profiles in general extend further with increasing depletion, whereas the difference between the high and the low carbon abundances is becoming more pronounced with increasing disc mass. Carbon depleted discs are thus experiencing a significant mass-loss at larger disc radii, which enables the formation of transition discs with large cavities that could still show an accretion signature. We will test the effect of our profiles on the disc evolution in a follow-up population synthesis model. 

As mentioned before, the total mass-loss rates of the homogeneously and inhomogeneously depleted discs are very similar, the corresponding profiles however show some substantial differences (compare examples in \autoref{fig:SigmaDotPrim2}). While the inhomogenously depleted discs experience a slightly enhanced mass-loss in the inner and outer part of the disc compared to the homogeneously depleted disc, the mass-loss is lower in the mid regions. In both cases, the profiles extend to a similarly large disc radius.

\autoref{fig:SigmaDotPrim3} displays a comparison of the profiles of the four different disc masses for each carbon abundance individually. Here it becomes evident that for solar metallicity and moderate carbon depletion the profiles are clearly different: While the mass-loss is similar for radii up to $\approx$ 50\,au, the profiles however extend to larger radii if the disc mass is low. In contrast to that these differences disappear with decreasing carbon abundance, as the X-ray opacity becomes low, resulting in very similar, disc mass independent profiles.   

Alongside the primordial mass-loss profiles we show some examples for transition disc profiles in \autoref{fig:SigmaDotTDlow} for the lowest-mass disc and in \autoref{fig:SigmaDotTDhigh} for the highest-mass disc. Regarding the solar metallicity, factor 3 depletion and inhomogeneously depleted discs, the overall shape of the profiles does not change for the transition discs compared to the primordial discs, with the peak however decreasing with increasing hole radius. Furthermore some of the features are becoming more pronounced for the transition disc profiles. In principle all profiles extend to a similar disc radius, which is however slightly below that for the primordial disc and increases slightly with hole radius, partly exceeding the profile for the primordial disc when the hole radius becomes very large. For the higher depletions (factor 10 and 100) on the other hand the profiles extend to smaller radii when the hole radius increases (but increase again for very large hole radii), with this effect being more pronounced for a lower-mass disc. One possible reason for this behaviour might be, that the strong wind in the inner part of the disc that occurs for large carbon depletion, shields the very outer part of the disc from the stars radiation. Therefore the photoevaporative wind significantly drops in these disc regions. With increasing hole radius, the effect becomes stronger, and thus the profiles shallower, as the wind will intensify with more layers being hit directly by high-energy stellar radiation. Being marked by a weaker disc wind, the higher carbon abundance simulations do not show this behaviour.   

Concerning the transition disc profiles, we note that the inner edge of the profiles should in principle be very sharp at the location of the hole radius, only beyond which the disc is present. As we applied a fit to our simulated data, which could not account for such an abrupt cut, this feature is not represented in the depicted profiles. For the purpose of this work and the following population synthesis this treatment is sufficient and won't influence the results. If however applied to other problems, a cut of the profile at the hole radius should be considered.   
\subsection{Rapid disc dispersal of carbon depleted discs}\label{sec:sweeping}
As mentioned in \autoref{sec:HoleM} some of the lower-mass transition discs are evolving extremely fast in the course of our simulations if the depletion is high and the hole radius relatively large. In \autoref{fig:Sweeping} we show an example for such a low-mass disc (0.005\,$\textit{M}_{*}$), depleted in carbon by a factor of 10 and harbouring a cavity with an initial hole radius of $\textit{R}_{\textrm{H}} \approx \textrm{30\,au}$. Here the disc is quickly moving outwards, thinning out rapidly and completely dispersed after about 500 orbits ($\approx$ 19000\,yrs). This represents the final stages of photoevaporation that can be observed directly in the course of the simulations for carbon depleted, lower-mass discs. Due to deeply penetrating X-rays (causing strong mass-loss rates) a metal depleted disc can thus experience a very rapid clearing of the order of 10$^4$ years, which inhibits any further planet formation in the disc and could furthermore prevent the formation of so-called relic discs. Relic discs are non-accreting transition discs, harbouring large holes, that are frequently predicted by current photoevaporation models, but not generally observed, thus representing one of the main open questions for these models. A full investigation of the impact of this rapid dispersal of (low-mass) carbon depleted discs is beyond the scope of this paper, but will be part of a forthcoming work on the demographics of transition discs. 
\section{Conclusions}\label{sec:Conclusions}
In this work, we performed radiation-hydrodynamical simulations of X-EUV driven photoevaporation in different solar metallicity and carbon depleted primordial and transition discs. We probed different carbon depletion factors (3, 10 and 100), disc masses between 0.005\,$\textit{M}_{*}$ and 0.1\,$\textit{M}_{*}$ as well as varying inner holes between 5\,au and 60\,au. Our models significantly improve on the previous  hydrostatic models of \citet{ErcolanoWeber2018}, by performing hydrodynamical calculations with new temperature prescriptions, based on tailored photoionisation and thermal calculations. The main results of our analysis are summarised in the following.
\\
\\
First, our new approach yields that carbon depletion results in higher gas temperatures compared to solar abundances, with the temperature increasing with degree of depletion (see \autoref{fig:TempCarbon}).   
\\
\\
From the hydrodynamical simulations, we determined new reliable total mass-loss rates of order 10$^{-8}$\,M$_{\odot}$\,yr$^{-1}$ to 10$^{-7}$\,M$_{\odot}$\,yr$^{-1}$ (compare \autoref{tab:massLoss} and \autoref{tab:massLossTD}) and found that the total mass-loss rate is about 2$-$6 times higher for carbon depleted discs compared to solar metallicity discs (depending on the depletion and the disc mass). The mass-loss in our calculations is dominated by the X-ray radiation. Even though we also included EUV in the irradiating spectrum, its contribution is negligible, as the EUV is already absorbed at small column densities and does not reach the high-density regions. FUV radiation is not included in our analysis. As FUV could in principle drive a significant mass-loss in the outer part of the disc, our results for the mass-loss rates represent a lower limit. Other authors like \citet{Gorti2009} or \citet{Nakatani2018} suggest that the effects of X-ray photoevaporation are minimal compared to FUV photoevaporation, thus a quantitative comparison of X-ray and FUV heating in low-metallicity discs is needed but outside the scope of this paper.
\\
\\
For each disc mass we found improved relations for the dependency of the total mass-loss rate on the carbon abundance, which predict a less extreme increase of the photoevaporative mass-loss with carbon abundance than the relation found by \citet{ErcolanoClarke2010} (see \autoref{fig:MZrel}) for the dependency of the mass-loss on the metallicity. These relations turn out to be weakly dependent on the disc mass. Moreover, we obtained scalings for the dependency of the total mass-loss rate on the disc mass for each carbon abundance set-up, which show a reversed behaviour depending on the degree of depletion (see \autoref{fig:DiscM}). In this context, we identified different effects being responsible for the opposite trends. 
\\
\\
Similar to the reversing behaviour of the disc mass dependencies we found opposing trends for the dependency of the total mass-loss rate on the hole radius, resulting from the fact that photoevaporation is effective in different disc regions for different carbon abundances and that a cut in the inner part of the disc is either affecting these regions or not (compare \autoref{fig:MRh}). Comparing the mass-loss rates for the homogeneously and inhomogeneously depleted discs, we found that the values are in principle very similar, including however some outliers in the case of the carbon depletion by a factor of 10 and the two lower disc masses. The according inhomogeneously depleted disc simulations behave less stable than the other simulations. Further tests (e.g. with higher resolution) could show if the mass-loss rates are resulting from numerical effects or if transition discs with solar abundances inside of 15\,au and strong carbon depletion outside of 15\,au are indeed experiencing an enhanced photoevaporative mass-loss due to the disc being less stable.  
\\
\\
In our analysis we calculated reasonable mass-loss profiles $\dot{\mathit{\Sigma}}$ for all simulated primordial and transition discs (compare \autoref{fig:SigmaDotPrim1} to \autoref{fig:SigmaDotTDhigh}). From the primordial disc profiles we can indeed conclude that the influence of X-ray photoevaporation is extended in carbon depleted discs, as the profiles extend to larger disc radii with increasing degree of depletion (\autoref{fig:SigmaDotPrim1}). In this context, the differences of the curves become more pronounced for higher disc masses, with the profiles for no or moderate depletion being clearly disc mass dependant while the profiles for higher depletions turn out to be very similar (\autoref{fig:SigmaDotPrim3}). Interestingly, even though the total mass-loss is comparable for the homogenously and inhomogeneously depleted discs, it is generated from different regions in the disc (\autoref{fig:SigmaDotPrim2}). While the corresponding $\dot{\mathit{\Sigma}}$ profiles extend to similar radii in both cases, the mass-loss is slightly enhanced close to the star and at larger disc radii and significantly lower for mid disc regions if the disc is depleted inhomogeneously.      
\\
\\
Some of our lower-mass transition discs are marked by a rapid disc dispersal, proceeding on a timescale of the order of 10$^{4}$\,yrs (compare \autoref{fig:Sweeping}). This will potentially help to prevent the formation of relic discs in a population synthesis model and will be studied in more detail in a follow-up paper.
\\
\\
The models of this work represent a detailed study of X-ray driven photoevaporation in carbon depleted discs and lay the foundation for a number of future investigations. Implementing the mass-loss profiles together with the total mass-loss rates into a population synthesis code could reveal the demographics of transition discs and show if carbon depletion can account for the majority of the observed diversity of transition discs and especially those discs that appear with large cavities and simultaneously strong accretion onto the central star. As we find a significant mass-loss at larger disc radii (up to $\approx$ 200\,au), we expect the formation of large cavities and even multiple holes, which we will test in a follow-up work. 
\section*{Acknowledgements}
We thank the anonymous referee for his/her detailed review, which significantly helped to improve this paper. We acknowledge the support from the DFG Research Unit "Planet Formation Witnesses and Probes: Transition Discs" (FOR 2634/1, ER 685/8-1 $\&$ ER 685/11-1). This work was performed partly on the computing facilities of the Computational Center for Particle and Astrophysics (C2PAP) and partly on the HPC system DRACO of the Max Planck Computing and $\&$ Data Facility (MPCDF).

%%%%%%%%%%%%%%%%%%%%%%%%%%%%%%%%%%%%%%%%%%%%%%%%%%



\bibliographystyle{mnras}
\bibliography{bibliography} % if your bibtex file is called example.bib

%%%%%%%%%%%%%%%%%%%%%%%%%%%%%%%%%%%%%%%%%%%%%%%%%%

%%%%%%%%%%%%%%%%% APPENDICES %%%%%%%%%%%%%%%%%%%%%
\appendix
\section{Temperature error}\label{appendix:TempError}
\begin{figure}
\centering
	\includegraphics[width=0.99\columnwidth]{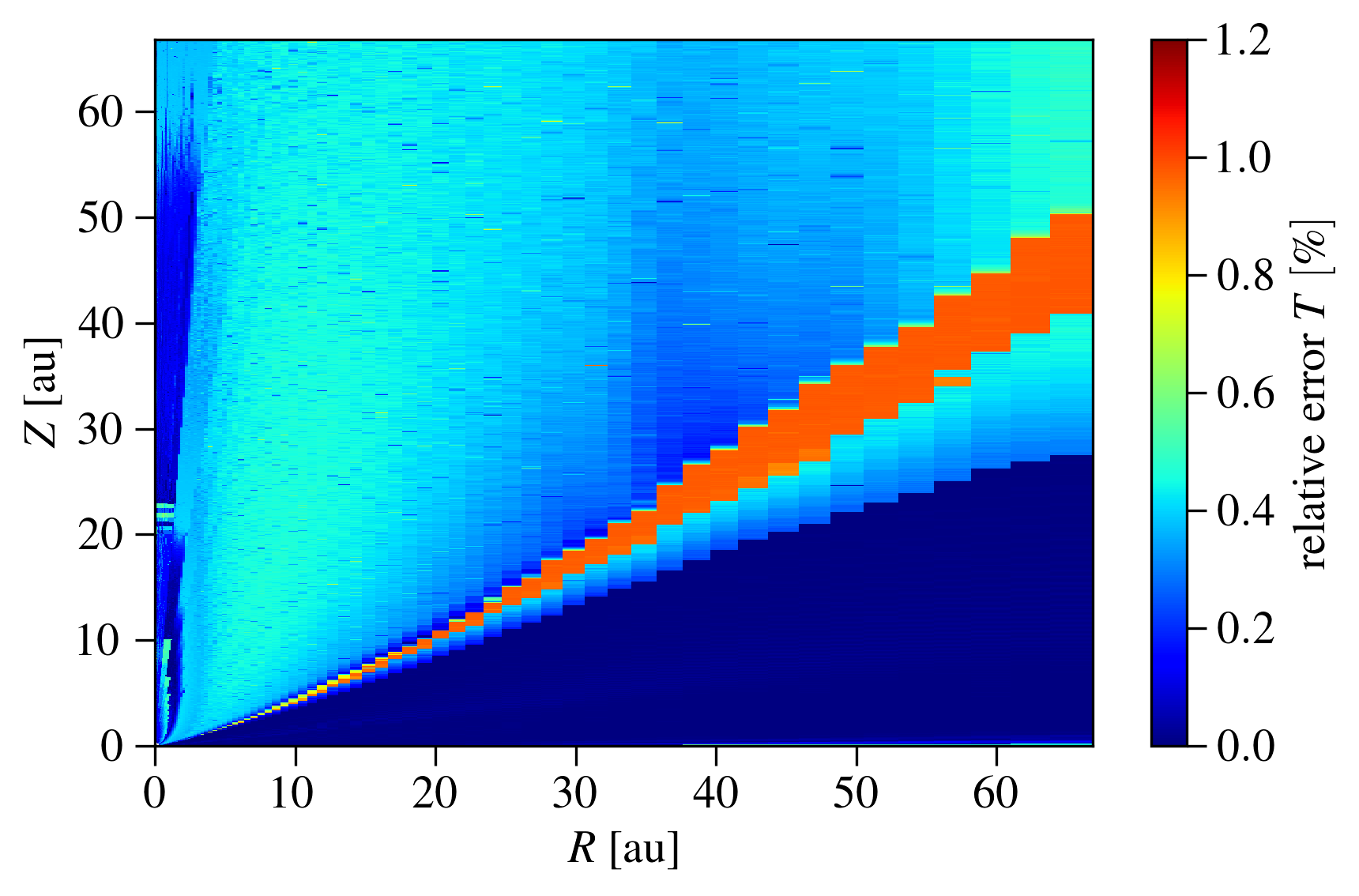}
	\includegraphics[width=0.99\columnwidth]{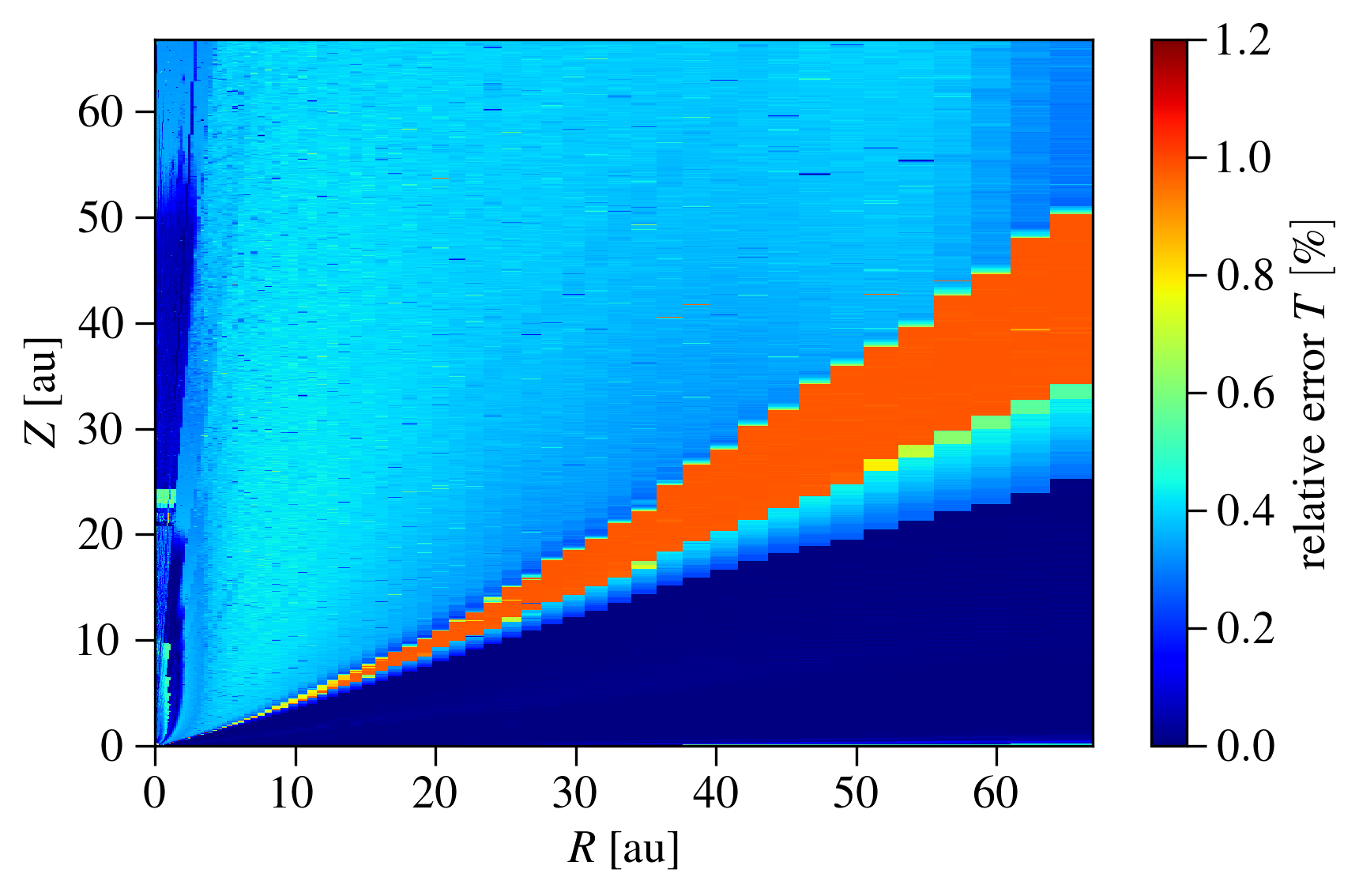}
    \caption{Relative error of the temperature determined in {\large \textsc{pluto}} with respect to the one post-processed with {\large \textsc{mocassin}} after a steady-state was reached in {\large \textsc{pluto}}. Shown are an example for the carbon depletion by a factor of 3 (top panel) and 100 (bottom panel) for the lowest-mass disc of $\textit{M}_{\textrm{disc}} = \textrm{0.005}\,\textit{M}_{*}$.}
    \label{fig:TempError}
\end{figure}

Using only a single-slab parametrisation for the column density, the models of \citet{Owen2010,Owen2011,Owen2012} can result in errors for the temperature of the order of 30\,\%. As shown in \autoref{fig:TempError} for carbon depletion by a factor of 3 and 100 respectively, this error is significantly reduced within our models. Even though the relative error is slightly increasing with degree of depletion, it is always less than 1\,\% for the whole computational domain in all simulations. The error was calculated by comparing the temperature coming directly from {\large \textsc{pluto}} to the temperature that is found from post-processing the steady-state from the {\large \textsc{pluto}} simulations in {\large \textsc{mocassin}}. 
\section{Choice of the (internal) disc radius of the streamlines calculation}\label{appendix:Radius}
\begin{figure}
\centering
\includegraphics[width=\columnwidth]{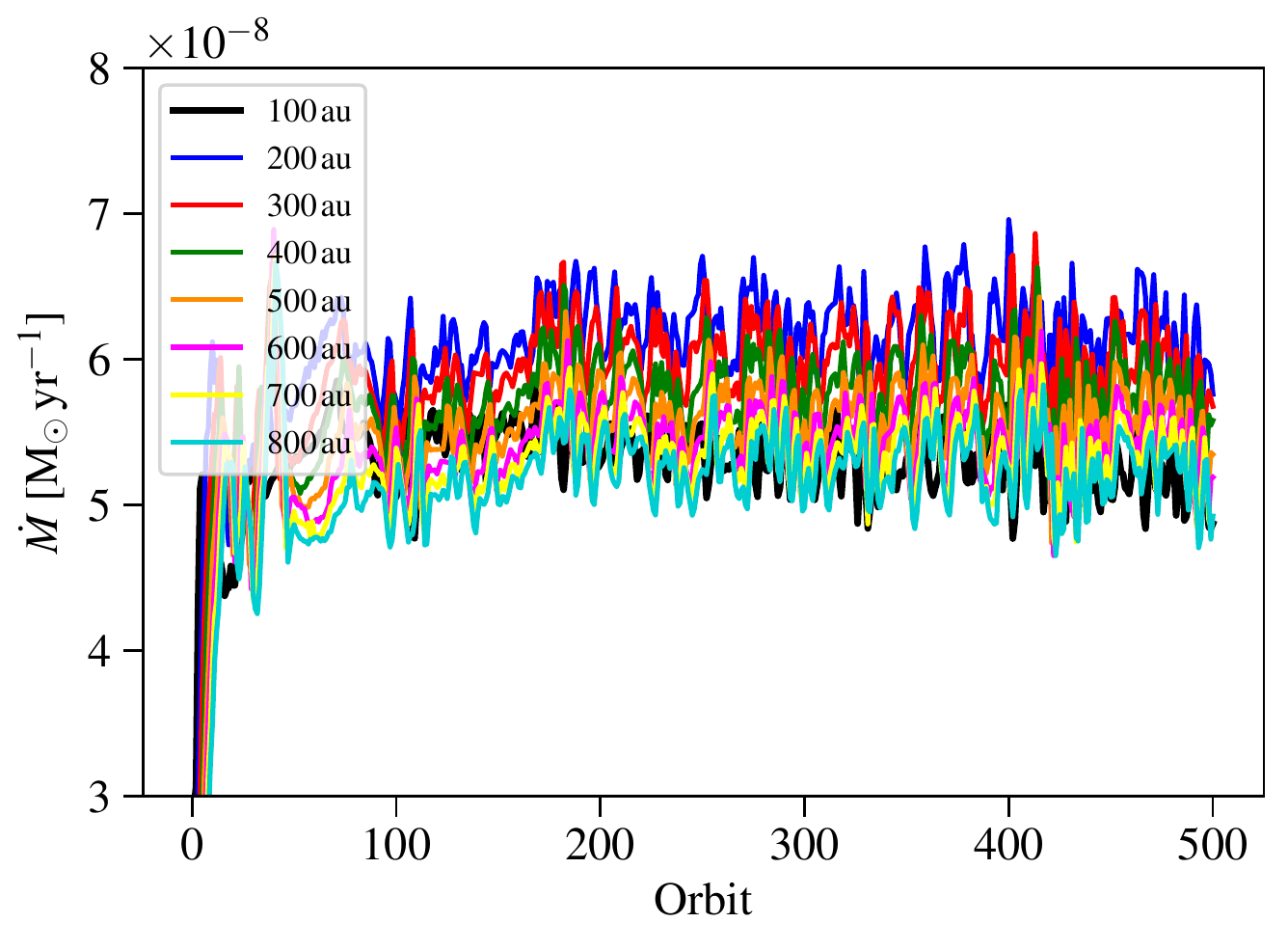}
\caption{Mass-loss rate as a function of orbits, shown for internal disc radii between 100\,au and 800\,au and the carbon depletion by a factor of 3 simulation of the 0.005\,$\textit{M}_{*}$ disc. The mass-loss rate is overall decreasing with increasing radius.}\label{fig:radius}
\end{figure}

While the outer radius of the computational domain is fixed at 1000\,au, the choice of the disc radius from which the streamlines are calculated is crucial for the distinction between the material that is actually removed from the disc and the material that is just redistributed within the disc. In this context, we tested different (internal) radii ranging from 100\,au to 800\,au for the primordial, low-mass disc simulations ($\textit{M}_{\textrm{disc}} \approx \textrm{0.005}\,\textit{M}_{*}$). The result of this test is shown in \autoref{fig:radius} for carbon depletion by a factor of 3 with the overall behaviour being representative for all simulations. From the plot we note that the lowest mass-loss rate is adjusted for the smallest radius of 100\,au, while the value is in general decreasing with increasing internal disc radius. The small value of the 100\,au radius, contradicting the overall trend, indicates that in this case important regions where photoevaporation was effective were cut out. The decrease in the mass-loss rate for larger radii is caused by the effect that some of the gas streamlines fall back below the sonic surface at larger disc radii. However we cannot fully trust those streamlines at large radii (\textit{r} > 200\,au) because the number of orbits they went through is limited and possibly they have not yet reached a stable state. 

Despite the variations, the mass-loss rate is comparable for all (internal) disc radii, possibly making them all suitable for the further calculations. Nevertheless, we decided to choose a radius of 200\,au, which yields the highest mass-loss rate. By doing so, we maximize the number of orbits at the given location, which is important for the streamlines stability, avoiding at the same time cutting too much of the outer disc regions. Moreover, we thus exclude the outermost regions which are possibly affected by the numerical oscillations and reflections from the outer boundary, that we described in \autoref{sec:Hydro}.

Even though a radius of 200\,au provides a good compromise for the purpose of this work, it would in principle be favourable to extend the hydrodynamical simulations in order to increase the number of orbits also for larger disc radii. As mentioned above we found that some streamlines in the beginning leave the disc but later fall back onto it. If the chosen radius is too small in these cases, streamlines, that are truly not contributing to the photoevaporative wind flow that leaves the disc, could be included in the mass-loss calculations. Performing additional hydrodynamical simulations could help to test the significance of this effect and yield detailed information about the influence of the different disc radii on the mass-loss rates and $\dot{\mathit{\Sigma}}$ profiles. 
\section{Test for radiative equilibrium}\label{appendix:equilib}
\begin{figure}
\centering
\includegraphics[width=\columnwidth]{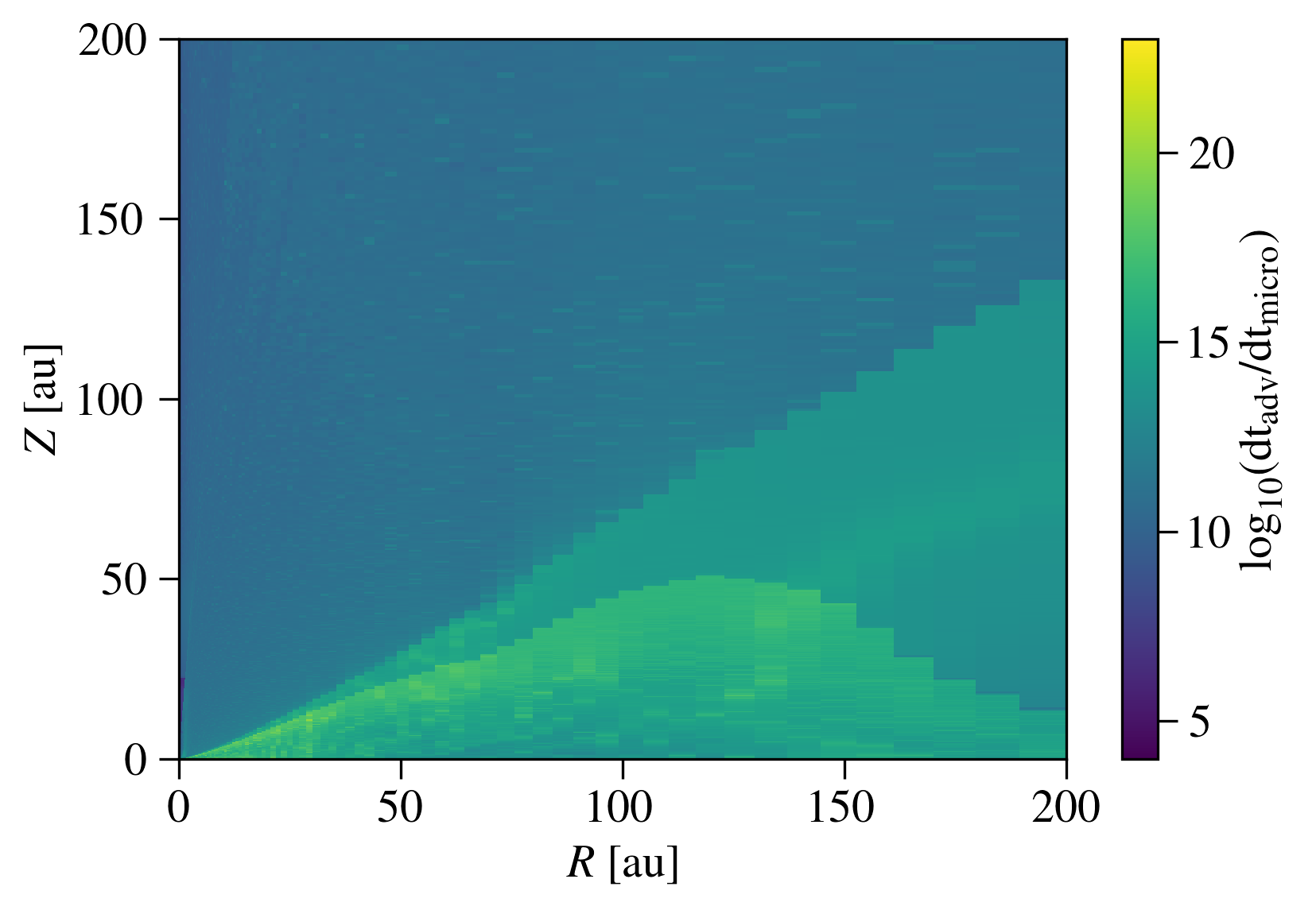}
\caption[Test for radiative equilibrium.]{Advection timescale $\tau_{\textrm{adv}}$ divided by the microphysical recombination timescale $\tau_{\textrm{rec}}$ in order to test for radiative equilibrium. Displayed is the test for the carbon depletion by a factor of 3 setup. The fraction is significantly larger than 1 for the whole computational domain.}\label{fig:eqTest}
\end{figure}

The approach we followed in this analysis is based on the assumption that the disc is in radiative equilibrium. This means that microphyiscal processes, which affect the temperature equilibrium, occur on timescales shorter than the hydrodynamical timescale. The most important microphysical process is hydrogen recombination, which proceeds on the longest timescale \citep{Ferland1979,Salz2015}
\begin{equation}
\tau_{\mathrm{rec}} = \frac{1}{\alpha_\mathrm{A}(T_{\mathrm{e}})n_{\mathrm{e}}} \simeq 1.5 \times 10^9\,\mathrm{s} \cdot \left(\frac{T_{\mathrm{e}}}{1\,\mathrm{K}}\right)^{0.8} \cdot \left(\frac{n_{\mathrm{e}}}{\mathrm{ptcls}/\mathrm{cm}^3}\right)^{-1} , 
\end{equation}  
with $\textit{T}_{\textrm{e}}$ as the electron temperature, $\textit{n}_{\textrm{e}}$ as the electron density and $\alpha_{\textrm{A}}$($\textit{T}_{\textrm{e}}$) as the temperature-dependent recombination rate. In order to check whether the hydrodynamical timescale is greater than this recombination timescale, we compared $\tau_{\textrm{rec}}$ to the advection timescale $\tau_{\textrm{adv}}$ for the regions that are are important for the wind dynamics. The result of this test is presented in \autoref{fig:eqTest}, where we plot the ratio of the advection and radiation timescale for the carbon depletion by a factor of 3 setup. In order to compute the advection time scale we have used the expression d\textit{x}/$\textit{v}_{\textrm{gas}}$ for each grid cell, which gives the time a gas parcel takes to cross a grid cell. \autoref{fig:eqTest} shows, that our assumption of radiative equilibrium is valid in the whole computational domain, as the hydrodynamical advection timescale is several orders of magnitude larger than the microphysical recombination timescale. Only very close to the \textit{Z}-axis there is a region that shows a smaller value of the fraction, although still considerably above 1. We performed this test for all carbon depletion setups, which yielded similar results as \autoref{fig:eqTest}. 
\section{Mass-loss rates of the transition disc simulations and additional \texorpdfstring{$\dot{\mathit{\Sigma}}$}{} profiles}\label{Table}
\begin{table*}
\caption{Average mass-loss rates and hole radii of the transition disc simulations.}\label{tab:massLossTD}

\centering
\begin{tabular}{l cccccccc}
\hline
\hline
& \multicolumn{2}{l}{\textit{disc mass 0.005\,M$_*$}} & \multicolumn{2}{l}{\textit{disc mass 0.01\,M$_*$}} & \multicolumn{2}{l}{\textit{disc mass 0.05\,M$_*$}} & \multicolumn{2}{l}{\textit{disc mass 0.1\,M$_*$}} \\
\hline
& $\textit{R}_{\textrm{H}}$ [au]& $\dot{M}$ [M$_{\odot}$\,yr$^{-1}$] & $\textit{R}_{\textrm{H}}$ [au]& $\dot{M}$ [M$_{\odot}$\,yr$^{-1}$] & $\textit{R}_{\textrm{H}}$ [au]& $\dot{M}$ [M$_{\odot}$\,yr$^{-1}$] & $\textit{R}_{\textrm{H}}$ [au]& $\dot{M}$ [M$_{\odot}$\,yr$^{-1}$] \\
\hline
\textit{solar} &  &  & $5.7 \pm 0.4 $ & $ (3.44 \pm 0.13) \times 10^{-8}$ & $5.4 \pm 0.4$ & $(2.49 \pm 0.17) \times 10^{-8}$ & $5.6 \pm 0.4$ &  $(2.05 \pm 0.13) \times 10^{-8}$\\
& / & / & $10.7 \pm 0.7 $ & $ (2.98 \pm 0.15) \times 10^{-8}$ & $10.2 \pm 0.7$ & $(2.1 \pm 0.14) \times 10^{-8}$ & $10.5 \pm 0.7$ & $(1.74 \pm 0.15) \times 10^{-8}$ \\ 
& / & / & $15.8 \pm 0.6 $ & $ (2.94 \pm 0.14) \times 10^{-8}$ & $15.1 \pm 0.6$ & $(1.94 \pm 0.14) \times 10^{-8}$ & $14.6 \pm 0.7$ & $(1.61 \pm 0.11) \times 10^{-8}$ \\
& / & / & $19.0 \pm 0.7 $ & $ (2.74 \pm 0.14) \times 10^{-8}$ & $18.1 \pm 0.6$  & $(1.89 \pm 0.11) \times 10^{-8}$ & $17.8 \pm 0.7$ & $(1.53 \pm 0.11) \times 10^{-8}$ \\
& / & / & $22.4 \pm 0.7 $ & $ (2.65 \pm 0.13 \times 10^{-8}$ & $21.0 \pm 0.7$ & $(1.81 \pm 0.12) \times 10^{-8}$ & $20.6 \pm 0.7$ &  $(1.47 \pm 0.1) \times 10^{-8}$\\
& / & / & $25.5 \pm 0.7 $ & $ (2.5 \pm 0.17) \times 10^{-8}$ & $23.6 \pm 0.5$ & $(1.75 \pm 0.12) \times 10^{-8}$ & $25.0 \pm 0.6$ & $(1.45 \pm 0.11) \times 10^{-8}$ \\
& / & / & / & / & $27.7 \pm 0.7$ & $(1.68 \pm 0.07) \times 10^{-8}$ & / & / \\
& / & / & $31.8 \pm 0.6 $ & $ (2.42 \pm 0.11) \times 10^{-8}$ & $31.9 \pm 0.6$ & $(1.64 \pm 0.14) \times 10^{-8}$ & $29.6 \pm 0.8$ & $(1.35 \pm 0.12) \times 10^{-8}$ \\
& / & / & $41.4 \pm 1.0 $ & $ (2.37 \pm 0.1) \times 10^{-8}$ & $38.9 \pm 0.7$ & $(1.57 \pm 0.12) \times 10^{-8}$ & $37.7 \pm 0.9$ & $(1.33 \pm 0.07) \times 10^{-8}$ \\
& / & / & $46.8 \pm 1.2 $ & $ (2.32 \pm 0.15) \times 10^{-8}$ & $47.1 \pm 2.1$ &  $(1.47 \pm 0.09) \times 10^{-8}$& $44.5 \pm 2.0$ &  $(1.26 \pm 0.11) \times 10^{-8}$\\
\\
\\
\textit{C/3} & $6.9 \pm 0.5$ & $ (7.24 \pm 0.28) \times 10^{-8}$ & $5.7 \pm 0.4$ & $ (7.19 \pm 0.23) \times 10^{-8}$ & $5.5 \pm 0.4$ & $(5.54 \pm 0.21) \times 10^{-8}$ & $6.8 \pm 0.5$ &  $(4.63 \pm 0.26) \times 10^{-8}$\\
& $11.0 \pm 0.7$ & $ (7.11 \pm 0.29) \times 10^{-8}$ & $10.8 \pm 0.7$ & $ (6.72 \pm 0.32) \times 10^{-8}$ & $11.3 \pm 0.7$ & $(4.92 \pm 0.26) \times 10^{-8}$ & $10.9 \pm 0.8$ &  $(4.13 \pm 0.3) \times 10^{-8}$\\ 
& $15.9 \pm 0.7$ & $ (6.8 \pm 0.27) \times 10^{-8}$ & $15.5 \pm 0.7$ & $ (6.25 \pm 0.33) \times 10^{-8}$ & $15.9 \pm 0.7$ & $(4.62 \pm 0.28) \times 10^{-8}$ & $14.9 \pm 0.6$ &  $(3.76 \pm 0.27) \times 10^{-8}$\\ 
& / & / & $18.9 \pm 0.7$ & $ (6.15 \pm 0.26) \times 10^{-8}$ & $18.8 \pm 0.6$ & $(4.45 \pm 0.3) \times 10^{-8}$ & / &  /\\
& $22.0 \pm 0.6$ & $ (6.48 \pm 0.44) \times 10^{-8}$ & $22.4 \pm 0.6$ & $ (5.91 \pm 0.27) \times 10^{-8}$ & $21.8 \pm 0.6$ & $(4.25 \pm 0.3) \times 10^{-8}$ & $21.2 \pm 0.8$ &  $(3.56 \pm 0.27) \times 10^{-8}$\\
& / & / & $25.9 \pm 0.6$ & $ (5.63 \pm 0.19) \times 10^{-8}$ & $25.2 \pm 0.6$ & $(4.09 \pm 0.25) \times 10^{-8}$ & / &  /\\
& $29.1 \pm 0.6$ & $ (6.11 \pm 0.28) \times 10^{-8}$ & / & / & $28.3 \pm 0.7$ & $(4.0 \pm 0.32) \times 10^{-8}$ & $27.3 \pm 1.7$ &  $(3.29 \pm 0.15) \times 10^{-8}$\\
& $35.9 \pm 0.5$ & $ (5.77 \pm 0.18) \times 10^{-8}$ & $33.8 \pm 0.8$ & $ (5.46 \pm 0.31) \times 10^{-8}$ & / & / & $33.0 \pm 0.6$ &  $(3.16 \pm 0.18) \times 10^{-8}$\\
& / & / & / & / & / & / & $36.8 \pm 1.1$ & $(3.15 \pm 0.11) \times 10^{-8}$\\
& $40.1 \pm 0.4$ & $ (5.65 \pm 0.25) \times 10^{-8}$ & $40.3 \pm 1.1$ & $ (5.35 \pm 0.26) \times 10^{-8}$ & $39.5 \pm 1.4$ & $(3.63 \pm 0.31) \times 10^{-8}$ & $39.7 \pm 1.1$ &  $(3.09 \pm 0.15) \times 10^{-8}$\\
& $43.9 \pm 0.7$ & $ (5.56 \pm 0.23) \times 10^{-8}$ & $46.7 \pm 1.4$ & $ (4.98 \pm 0.26) \times 10^{-8}$ & $46.9 \pm 2.0$ & $(3.32 \pm 0.27) \times 10^{-8}$ & $50.4 \pm 2.9$ &  $(2.89 \pm 0.21) \times 10^{-8}$\\
& $57.2 \pm 1.5$ & $ (5.05 \pm 0.29) \times 10^{-8}$ & / & / & / & / & / & / \\
\\
\\
\textit{solar} & $6.5 \pm 0.4$ & $ (6.89 \pm 0.25) \times 10^{-8}$ & $5.5 \pm 0.4$ & $ (6.69 \pm 0.33) \times 10^{-8}$ & $5.2 \pm 0.4$ & $(5.04 \pm 0.23) \times 10^{-8}$ & $6.5 \pm 0.5$ &  $(4.06 \pm 0.27) \times 10^{-8}$\\
\textit{+ C/3} & $10.8 \pm 0.7$ & $ (6.75 \pm 0.33) \times 10^{-8}$ & $10.8 \pm 0.7$ & $ (6.16 \pm 0.27) \times 10^{-8}$ & $10.2 \pm 0.6$ & $(4.35 \pm 0.26) \times 10^{-8}$ & $10.4 \pm 0.6$ &  $(3.87 \pm 0.28) \times 10^{-8}$\\
& $13.6 \pm 0.8$ & $ (6.61 \pm 0.33) \times 10^{-8}$ & $13.3 \pm 0.7$ & $ (6.02 \pm 0.3) \times 10^{-8}$ & $13.4 \pm 0.7$ & $(4.17 \pm 0.27) \times 10^{-8}$ & $13.3 \pm 0.7$ &  $(3.54 \pm 0.27) \times 10^{-8}$\\
\\
\\
\textit{C/10} & $ 6.3 \pm 0.4$ & $ (9.39 \pm 0.26) \times 10^{-8}$ & $5.4 \pm 0.3$ & $ (1.07 \pm 0.03) \times 10^{-7}$ & $5.4 \pm 0.3$ & $(1.31 \pm 0.04) \times 10^{-7}$ & $6.3 \pm 0.4$ &  $(1.29 \pm 0.04) \times 10^{-7}$\\
& $ 10.4 \pm 0.7$ & $ (1.1 \pm 0.04) \times 10^{-7}$ & $11.3 \pm 0.7$ & $ (1.17 \pm 0.03) \times 10^{-7}$ & $11.3 \pm 0.7$ & $(1.29 \pm 0.03) \times 10^{-7}$ & $11.1 \pm 0.7$ &  $(1.28 \pm 0.04) \times 10^{-7}$\\
& $ 15.1 \pm 1.1 $ & $ (1.32 \pm 0.06) \times 10^{-7}$ & $16.0 \pm 0.6$ & $ (1.36 \pm 0.03) \times 10^{-7}$ & $15.1 \pm 0.7$ & $(1.37 \pm 0.05) \times 10^{-7}$ & $15.1 \pm 0.7$ &  $(1.3 \pm 0.04) \times 10^{-7}$\\
& $ 19.3 \pm 1.3$ & $ (1.57 \pm 0.07) \times 10^{-7}$ & $19.2 \pm 0.5$ & $ (1.55 \pm 0.03) \times 10^{-7}$ & $18.5 \pm 0.6$ & $(1.44 \pm 0.06) \times 10^{-7}$ & / &  /\\
& / & / & $22.9 \pm 0.5$ & $ (1.73 \pm 0.05) \times 10^{-7}$ & $22.5 \pm 0.5$ & $(1.52 \pm 0.05) \times 10^{-7}$ & $21.7 \pm 0.6$ &  $(1.4 \pm 0.05) \times 10^{-7}$\\
& $ 24.0 \pm 1.6$ & $ (1.8 \pm 0.09) \times 10^{-7}$ & $26.7 \pm 0.5$ & $ (1.93 \pm 0.07) \times 10^{-7}$ & $26.0 \pm 0.6$ & $(1.6 \pm 0.05) \times 10^{-7}$ & / &  /\\
& / & / & $31.0 \pm 1.0$ & $ (2.22 \pm 0.08) \times 10^{-7}$ & $29.6 \pm 0.5$ & $(1.66 \pm 0.08) \times 10^{-7}$ & $28.8 \pm 0.8$ &  $(1.52 \pm 0.07) \times 10^{-7}$\\
& / & / & $34.4 \pm 1.2$ & $ (2.38 \pm 0.08) \times 10^{-7}$ & $33.0 \pm 0.8$ & $(1.82 \pm 0.09) \times 10^{-7}$ & $35.0 \pm 0.9$ &  $(1.67 \pm 0.05) \times 10^{-7}$\\
& / & / & / & / & $40.2 \pm 0.7$ & $(1.95 \pm 0.08) \times 10^{-7}$ & $39.4 \pm 1.0$ &  $(1.73 \pm 0.05) \times 10^{-7}$\\
& / & / & / & / & / & / & $42.9 \pm 0.8$ &  $(1.79 \pm 0.07) \times 10^{-7}$\\
& / & / & / & / & $49.5 \pm 2.3$ & $(2.14 \pm 0.22) \times 10^{-7}$ & $49.7 \pm 1.9$ &  $(1.93 \pm 0.06) \times 10^{-7}$\\
\\
\\
\textit{solar} & $6.5 \pm 0.4$ & $ (1.42 \pm 0.1) \times 10^{-7}$ & $5.3 \pm 0.3$ & $ (1.62 \pm 0.07) \times 10^{-7}$ & $5.2 \pm 0.3$ & $(1.35 \pm 0.04) \times 10^{-7}$ & $6.1 \pm 0.4$ &  $(1.26 \pm 0.05) \times 10^{-7}$\\
\textit{+ C/10} & $10.7 \pm 0.7$ & $ (1.11 \pm 0.03) \times 10^{-7}$ & $10.6 \pm 0.7$ & $ (1.87 \pm 0.08) \times 10^{-7}$ & $10.7 \pm 0.7$ & $(1.28 \pm 0.1) \times 10^{-7}$ & $10.0 \pm 0.6$ &  $(1.19 \pm 0.07) \times 10^{-7}$\\
\\
\\
\textit{C/100} & $6.4 \pm 0.4$ & $ (9.66 \pm 0.3) \times 10^{-8}$ & $5.3 \pm 0.3$ & $ (1.12 \pm 0.03) \times 10^{-7}$ & $5.2 \pm 0.4$ & $(1.51 \pm 0.07) \times 10^{-7}$ & $6.3 \pm 0.4$ &  $(1.6 \pm 0.06) \times 10^{-7}$\\
& $11.0 \pm 0.7$ & $ (1.12 \pm 0.04) \times 10^{-7}$ & $11.7 \pm 0.7$ & $ (1.18 \pm 0.03) \times 10^{-7}$ & $10.7 \pm 0.7$ & $(1.51 \pm 0.06) \times 10^{-7}$ & $11.1 \pm 0.7$ &  $(1.52 \pm 0.05) \times 10^{-7}$\\ 
& $15.5 \pm 1.1$ & $ (1.32 \pm 0.05) \times 10^{-7}$ & $15.5 \pm 0.6$ & $ (1.34 \pm 0.04) \times 10^{-7}$ & $15.6 \pm 0.6$ & $(1.50 \pm 0.04) \times 10^{-7}$ & $15.2 \pm 0.8$ &  $(1.56 \pm 0.06) \times 10^{-7}$\\
& $20.3 \pm 1.6$ & $ (1.59 \pm 0.07) \times 10^{-7}$ & $18.7 \pm 0.6$ & $ (1.57 \pm 0.05) \times 10^{-7}$ & $19.0 \pm 0.7$ & $(1.60 \pm 0.04) \times 10^{-7}$ & $21.7 \pm 0.6$ &  $(1.69 \pm 0.07) \times 10^{-7}$\\
& / & / & $23.1 \pm 1.3$ & $ (1.79 \pm 0.06) \times 10^{-7}$ & $22.2 \pm 0.8$ & $(1.69 \pm 0.06) \times 10^{-7}$ & / &  / \\
& $26.1 \pm 2.2$ & $ (1.85 \pm 0.12) \times 10^{-7}$ & / & / & $25.0 \pm 0.7$ & $(1.79 \pm 0.09) \times 10^{-7}$ & / &  /\\
& / & / & $28.4 \pm 1.2$ & $ (2.03 \pm 0.08) \times 10^{-7}$ & $29.8 \pm 0.7$ & $(1.95 \pm 0.09) \times 10^{-7}$ & $28.5 \pm 0.7$ & $(1.85 \pm 0.1) \times 10^{-7}$ \\
& / & / & $34.4 \pm 1.7$ & $ (2.38 \pm 0.13) \times 10^{-7}$ & $33.5 \pm 0.8$ & $(2.19 \pm 0.1) \times 10^{-7}$ & $35.9 \pm 0.7$ &  $(2.14 \pm 0.1) \times 10^{-7}$\\
& / & / & / & / & $42.8 \pm 1.0$ & $(2.51 \pm 0.12) \times 10^{-7}$ & $40.3 \pm 1.2$ &  $(2.29 \pm 0.11) \times 10^{-7}$\\
& / & / & / & / & / & / & $45.5 \pm 1.0$ &  $(2.48 \pm 0.1) \times 10^{-7}$\\
& / & / & / & / & $52.3 \pm 2.4$ & $(3.0 \pm 0.16) \times 10^{-7}$ & $52.2 \pm 2.1$ &  $(2.78 \pm 0.15) \times 10^{-7}$\\

\\
 \hline 
\end{tabular}
\end{table*} 

%\section{Plots for the mass-loss profiles $\dot{\Sigma}$}\label{appendix:Sigma}

\begin{figure*}
	\includegraphics[width=\columnwidth]{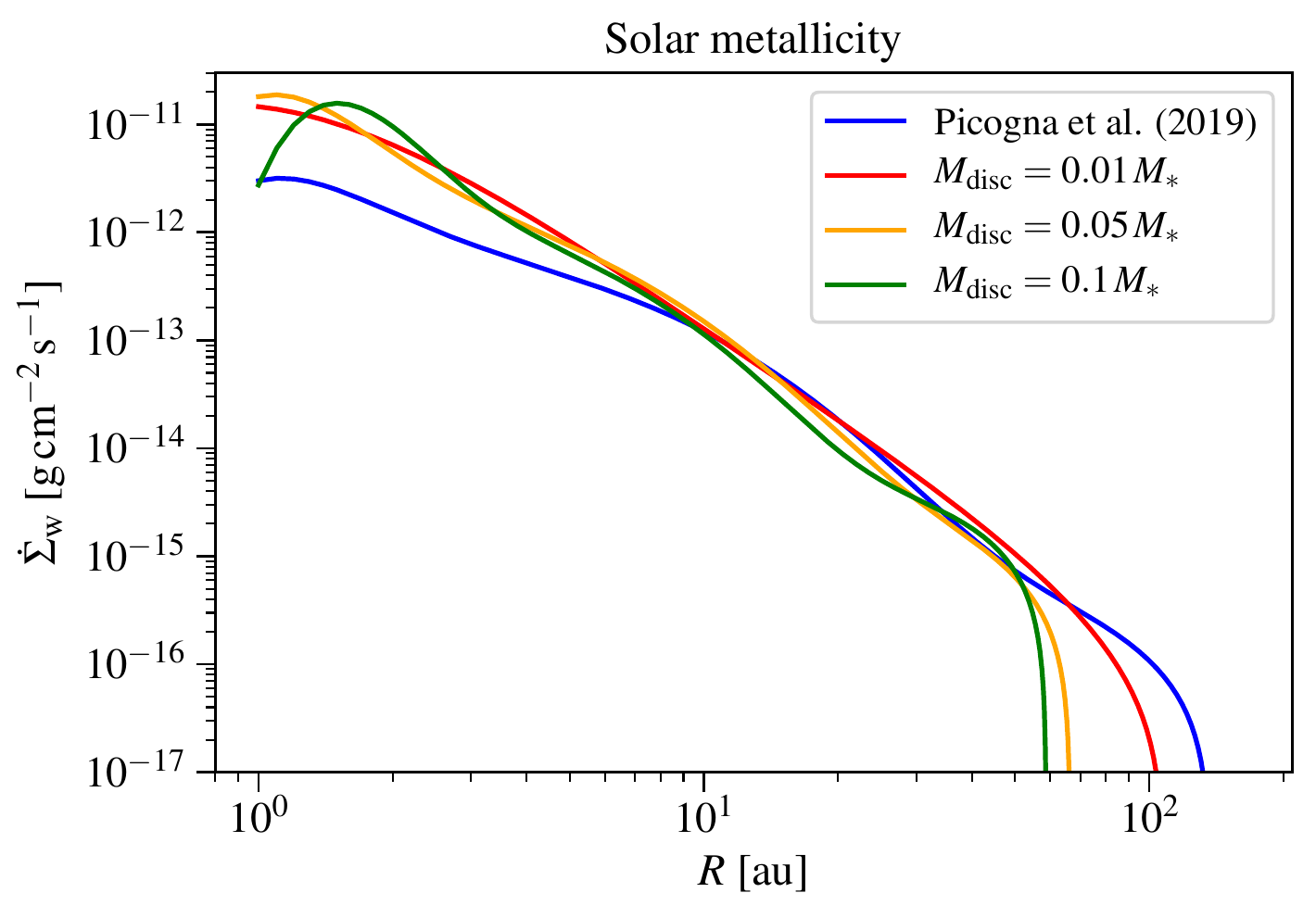}
	\includegraphics[width=\columnwidth]{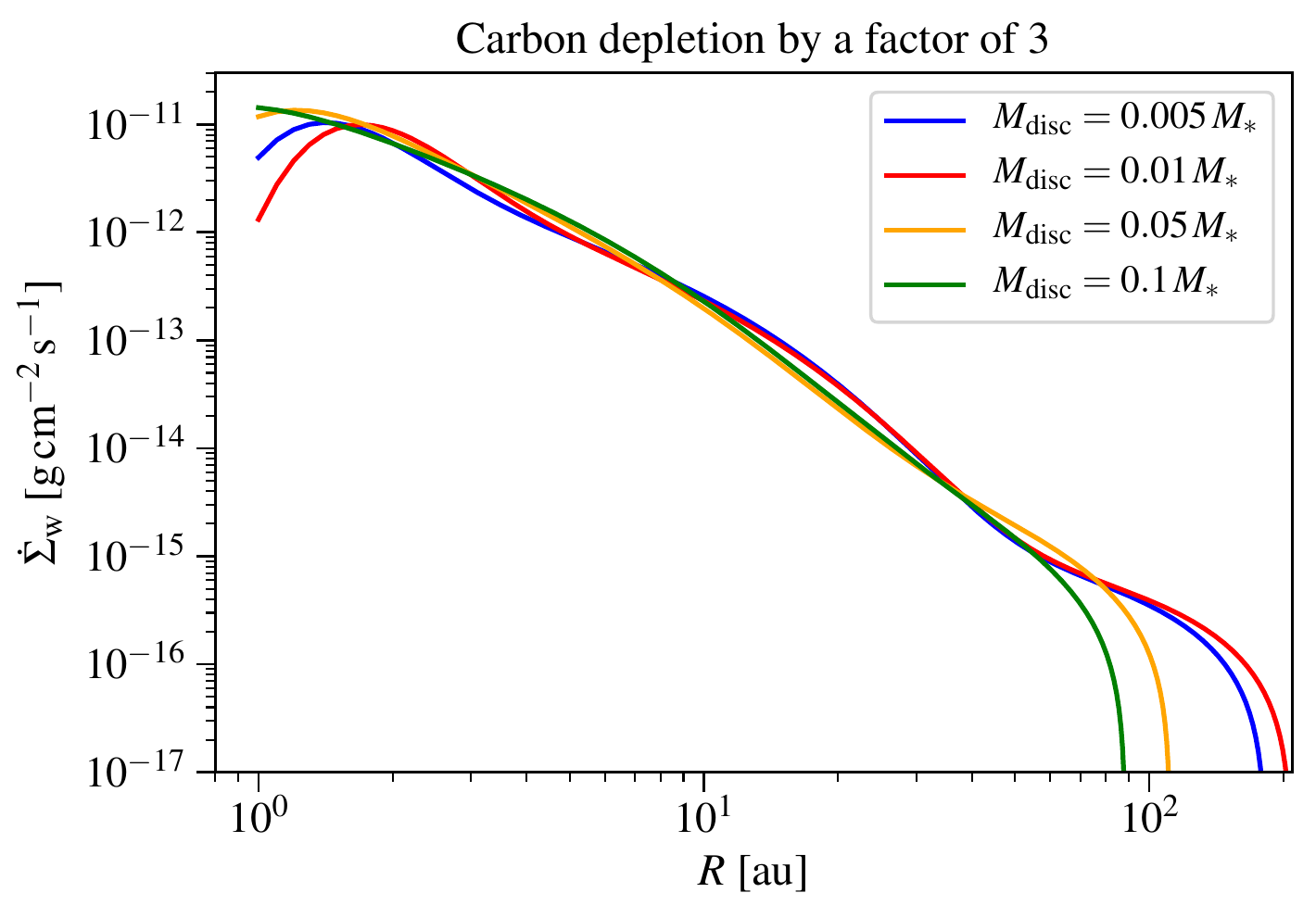}
	\includegraphics[width=\columnwidth]{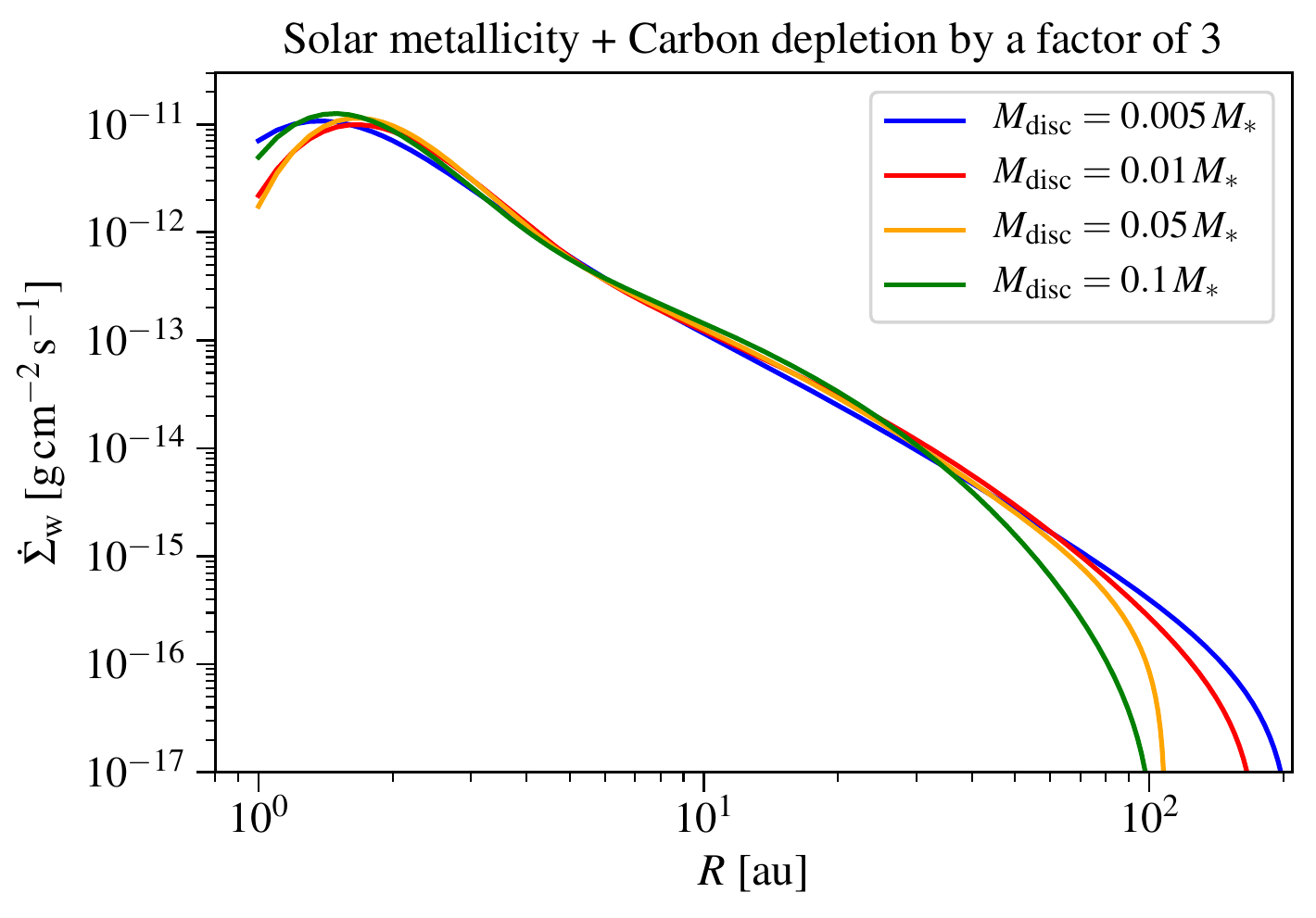}
	\includegraphics[width=\columnwidth]{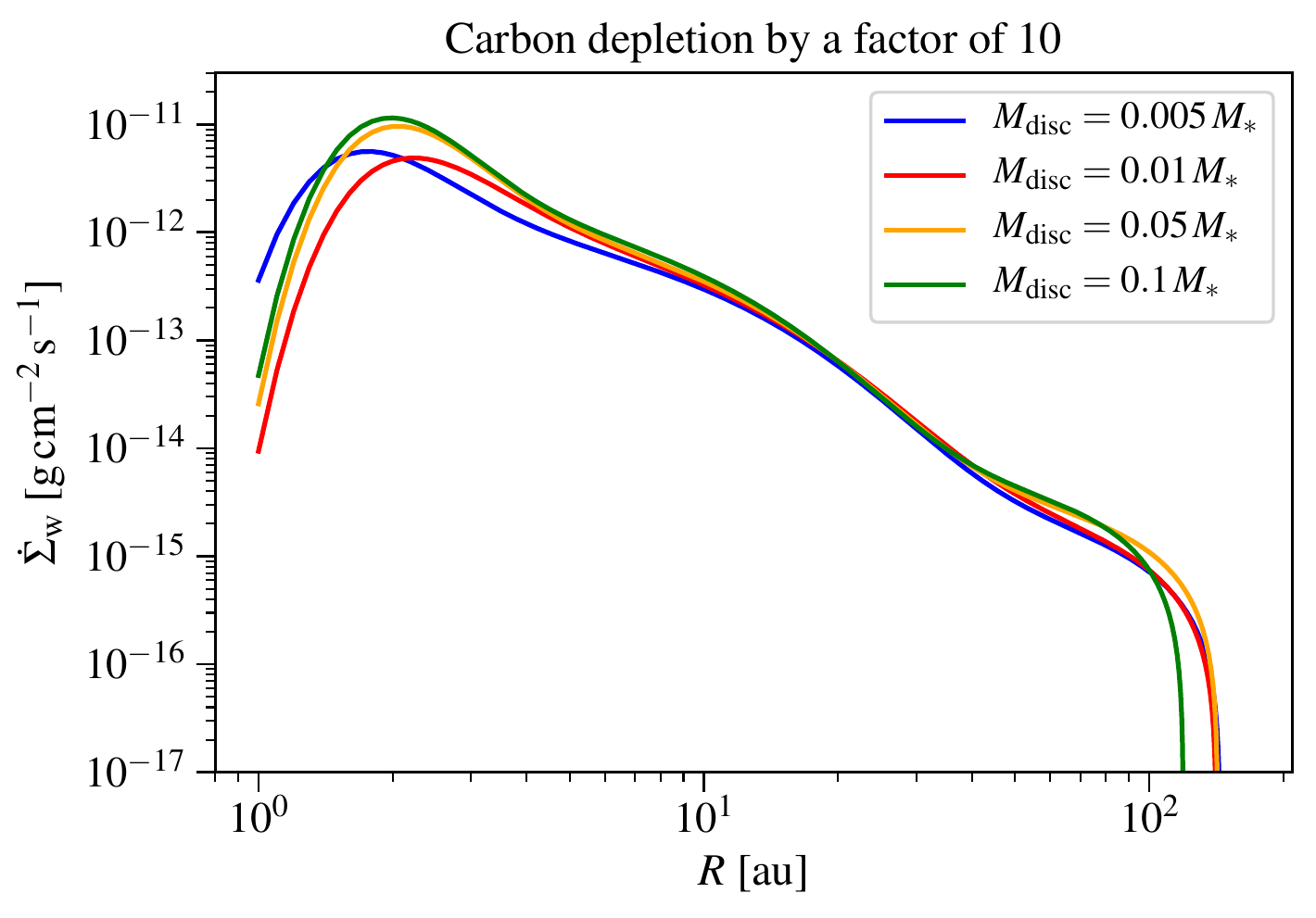}
	\includegraphics[width=\columnwidth]{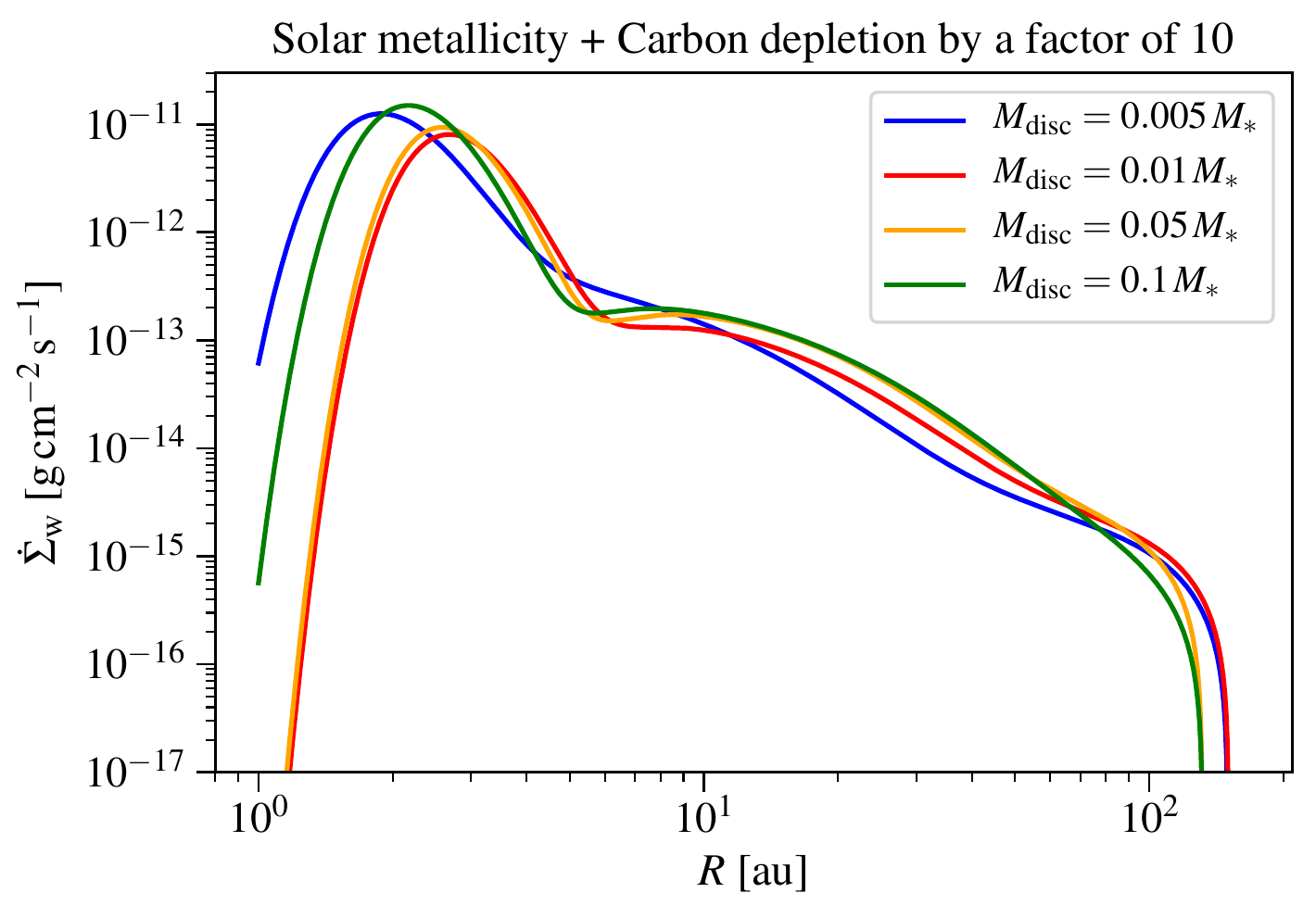}
	\includegraphics[width=\columnwidth]{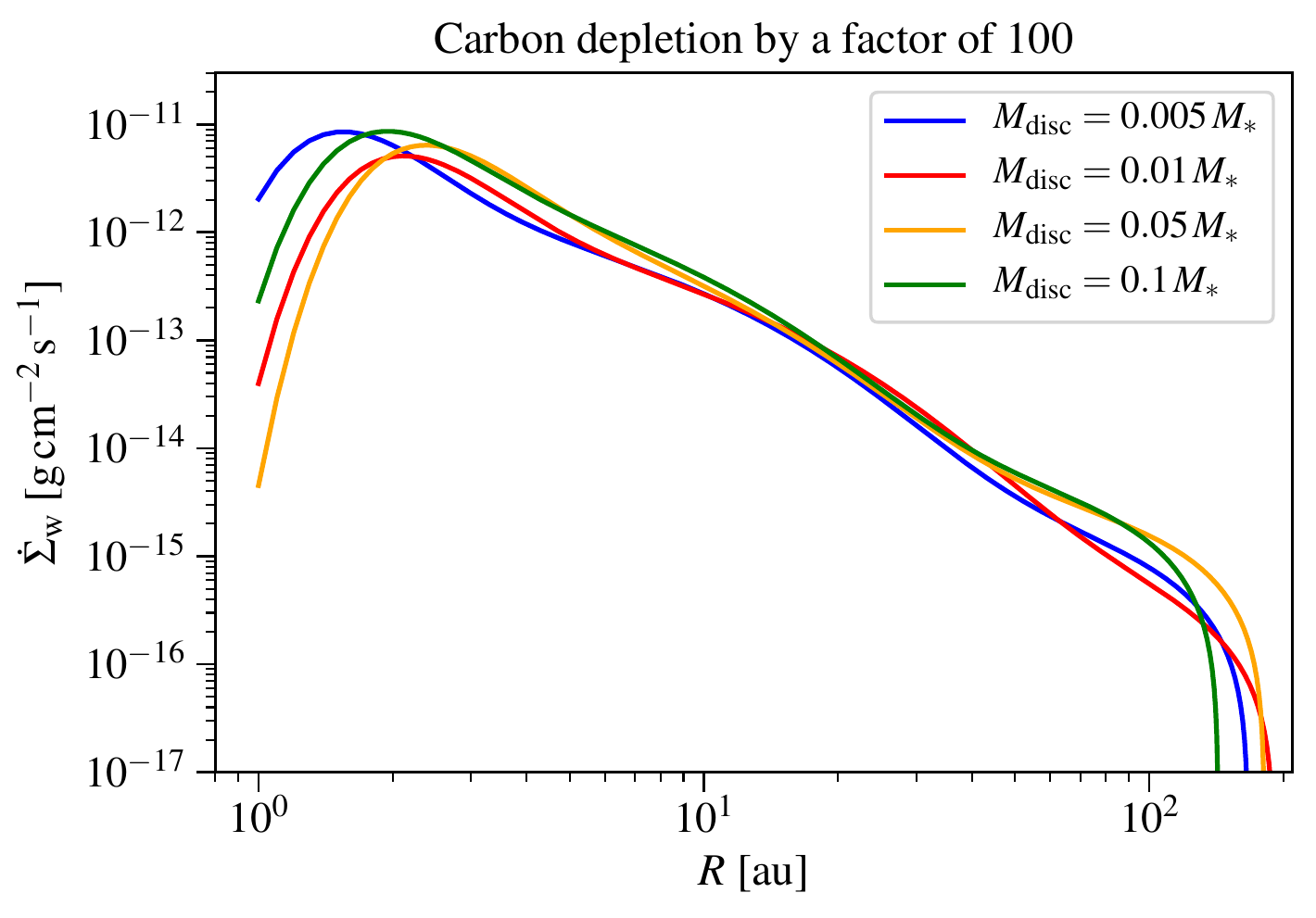}
    \caption{Comparison of the mass-loss profiles $\dot{\mathit{\Sigma}}$ of the four disc masses for each carbon abundance set-up. With increasing depletion the profiles become more similar.}
    \label{fig:SigmaDotPrim3}
\end{figure*}

\begin{figure*}
	\includegraphics[width=\columnwidth]{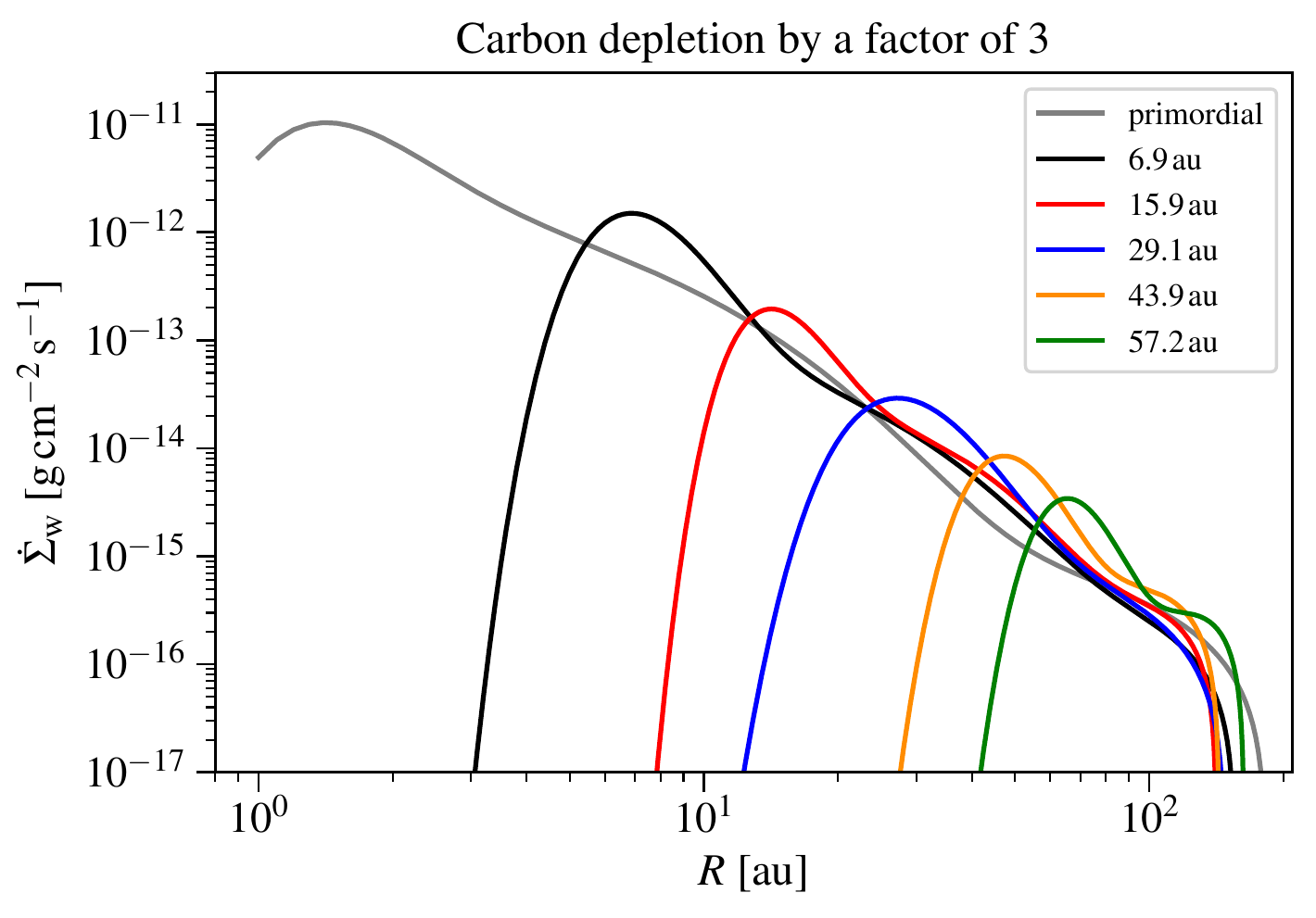}
	\includegraphics[width=\columnwidth]{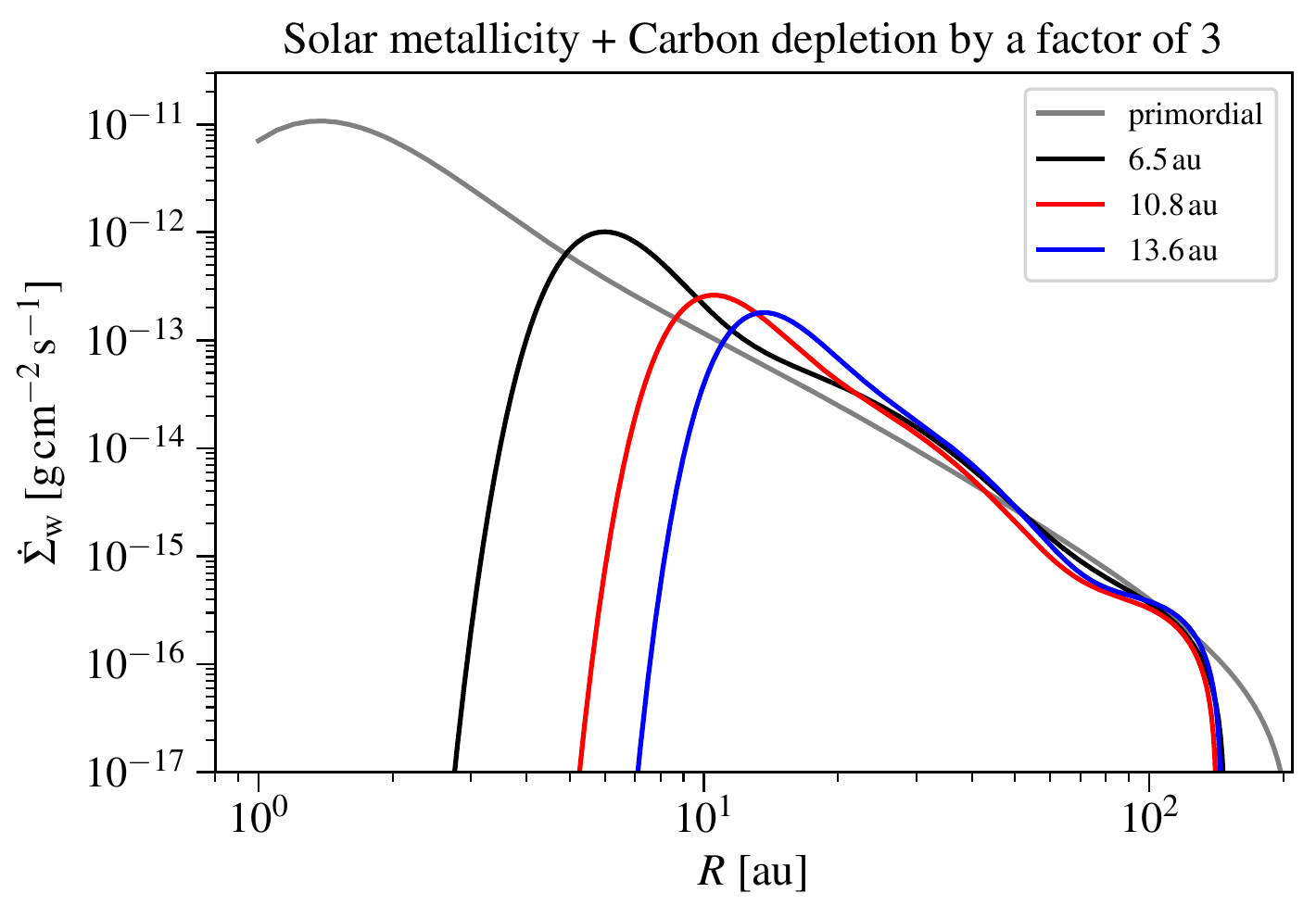}
	\includegraphics[width=\columnwidth]{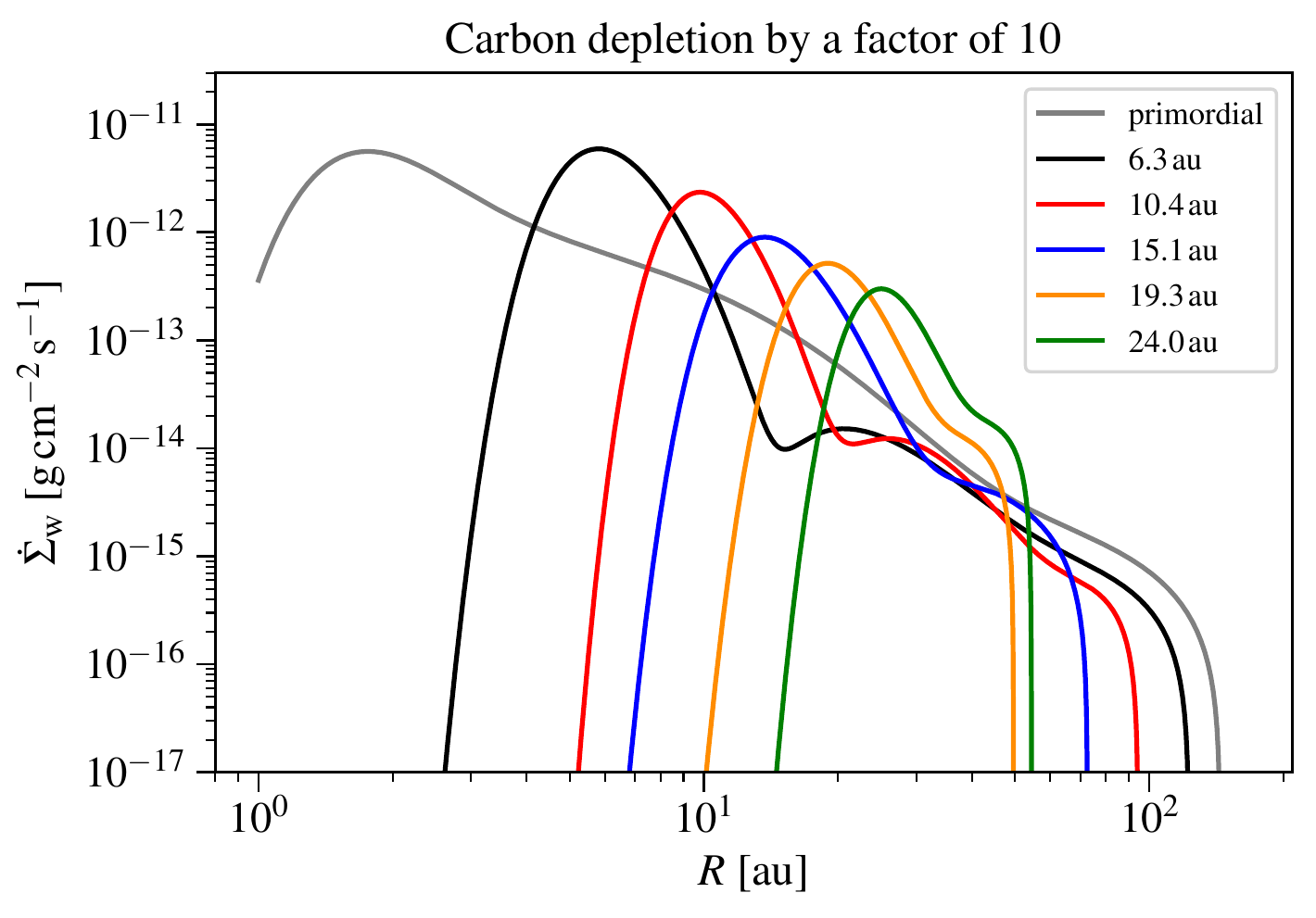}
	\includegraphics[width=\columnwidth]{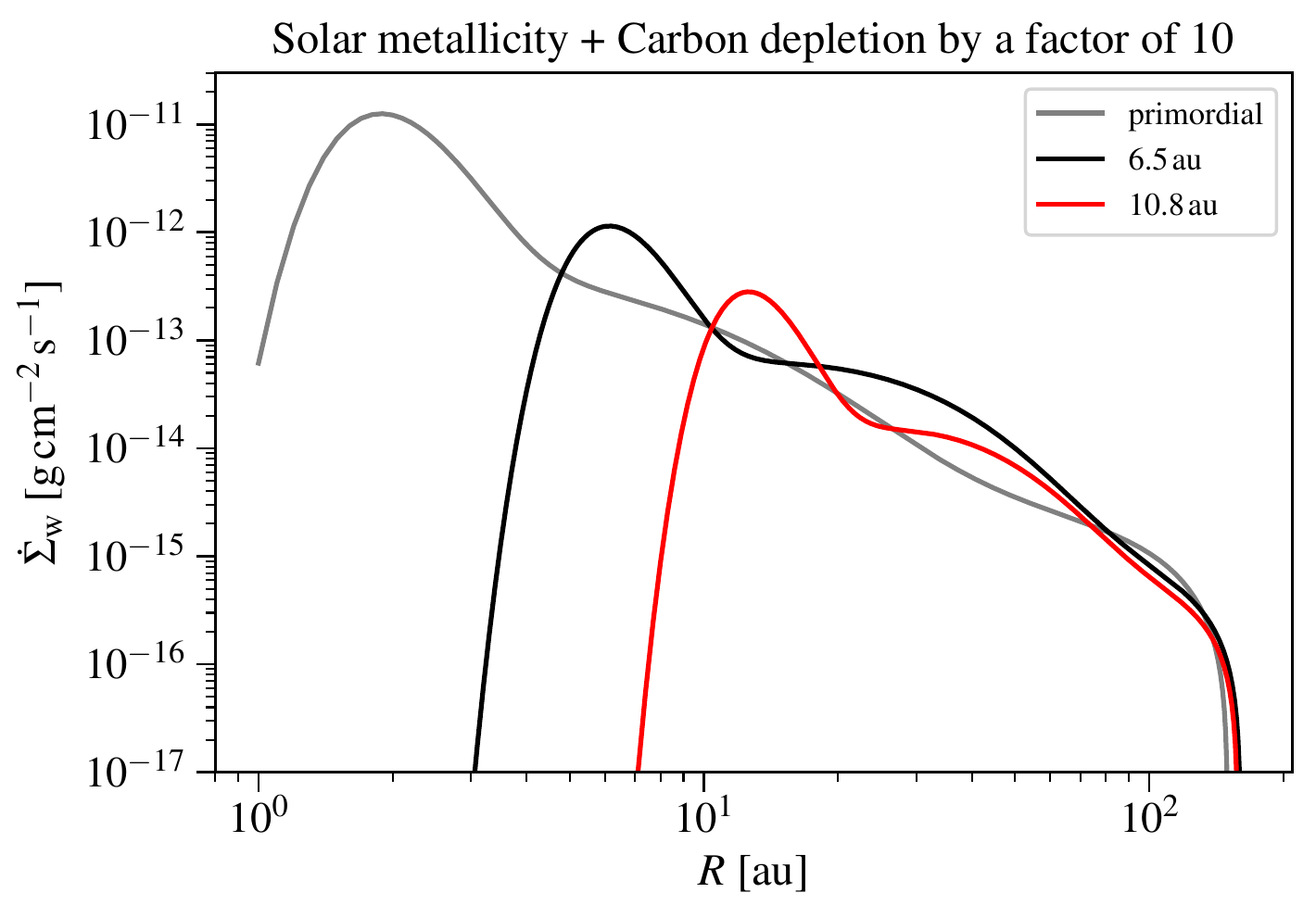}
	\includegraphics[width=\columnwidth]{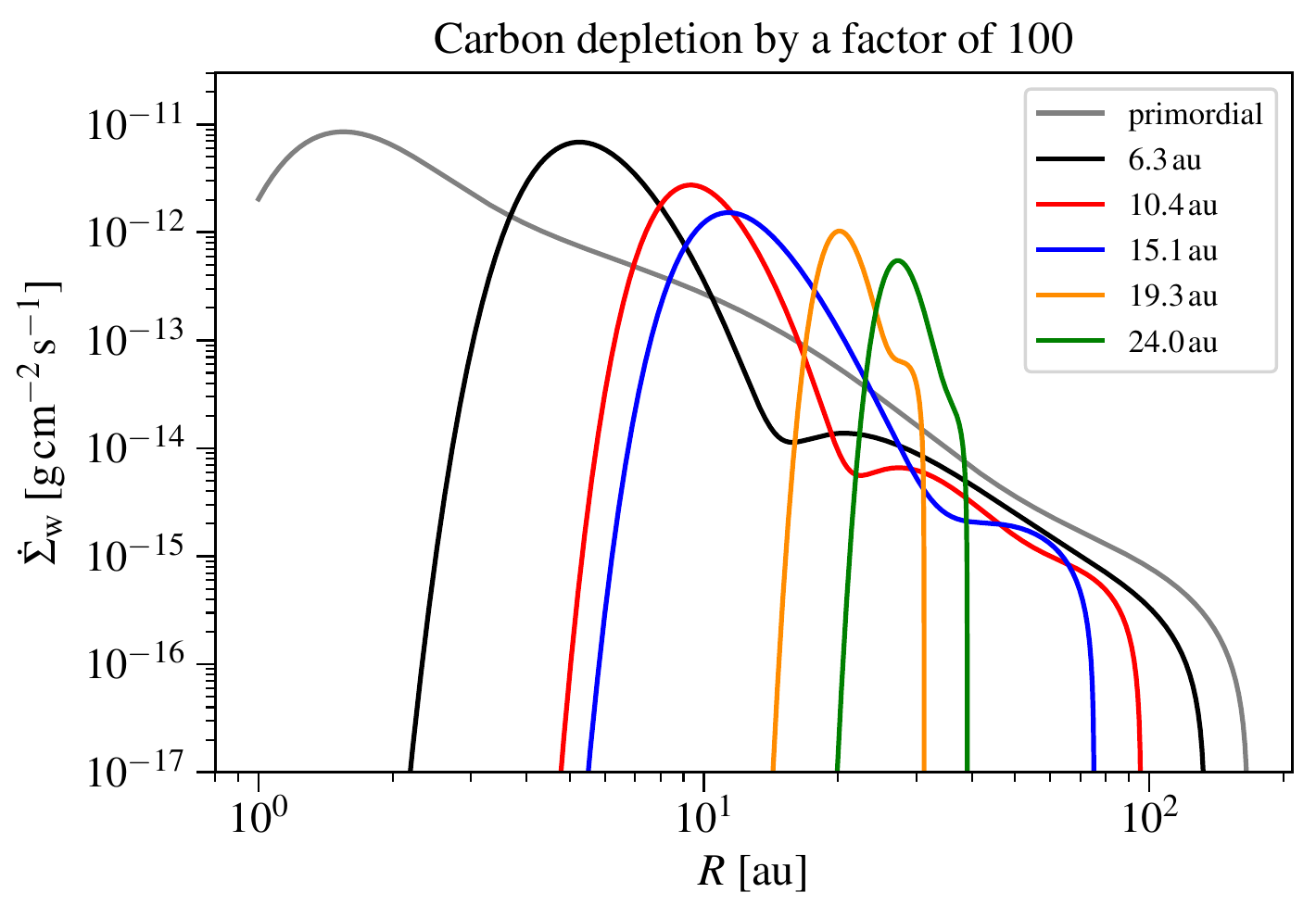}
    \caption{Examples of mass-loss profiles $\dot{\mathit{\Sigma}}$ for low-mass transition discs (0.005\,$\textit{M}_{*}$) with various hole radii, shown for the different carbon abundance set-ups.}
    \label{fig:SigmaDotTDlow}
\end{figure*}
\begin{figure*}
	\includegraphics[width=\columnwidth]{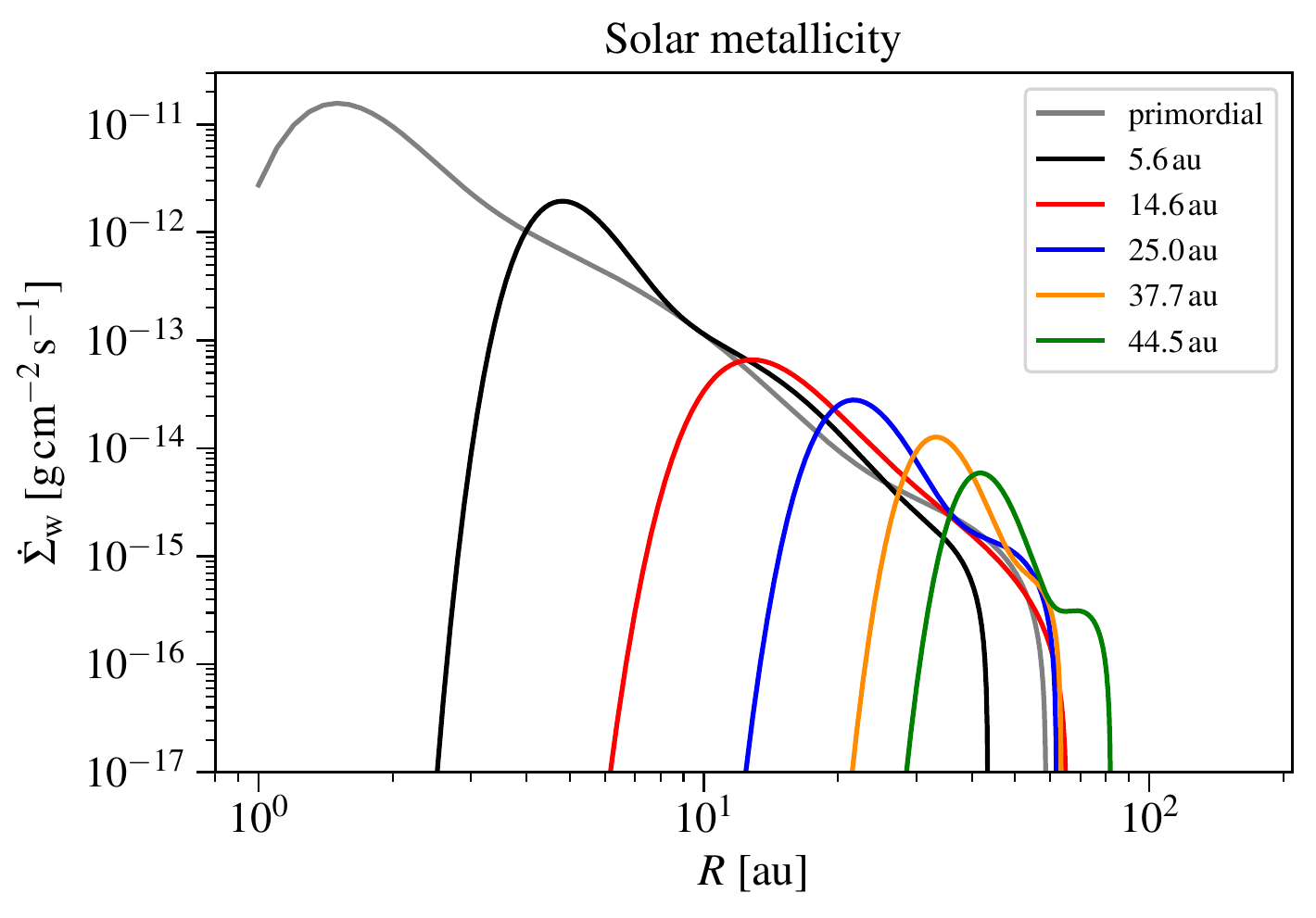}
	\includegraphics[width=\columnwidth]{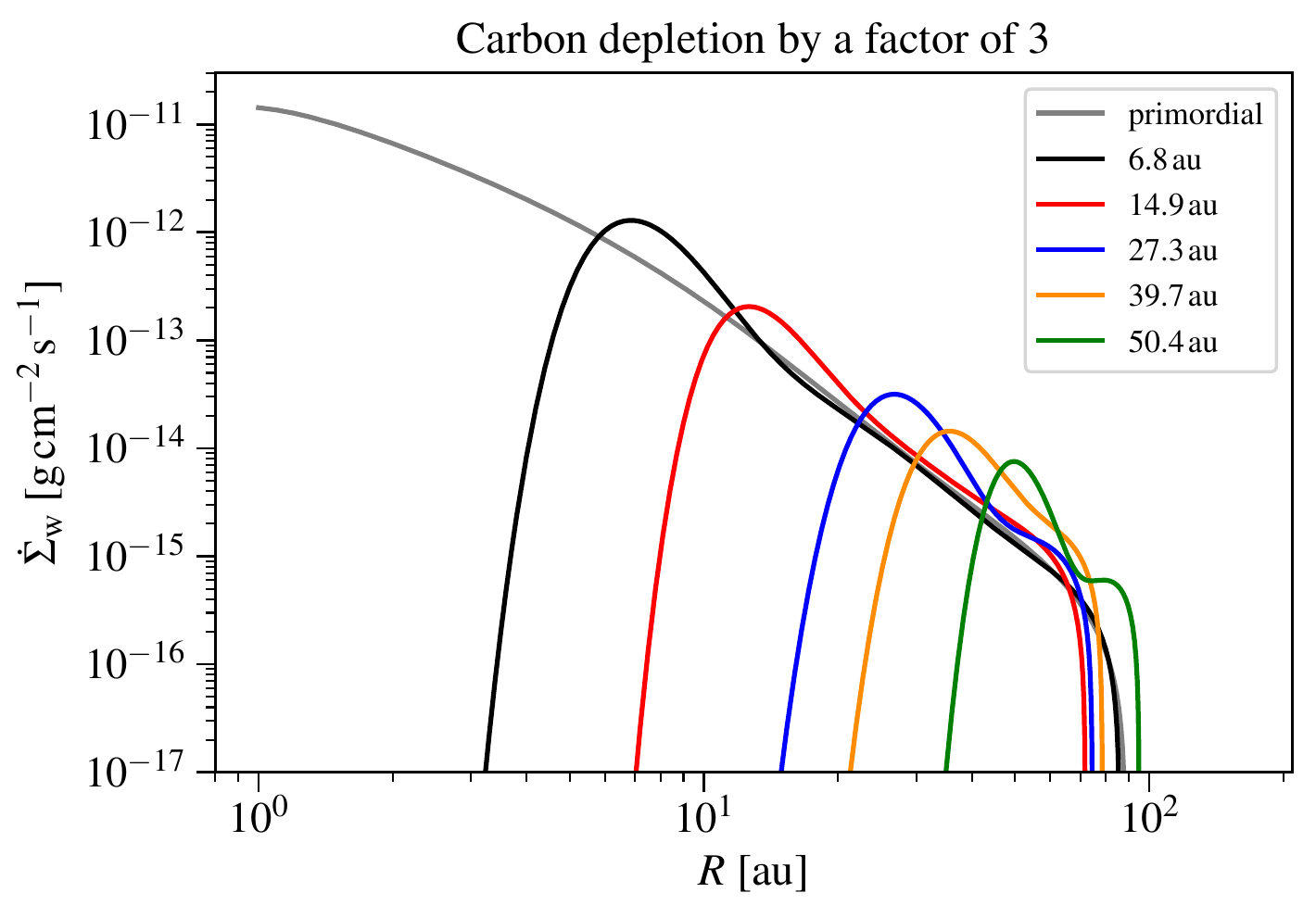}
	\includegraphics[width=\columnwidth]{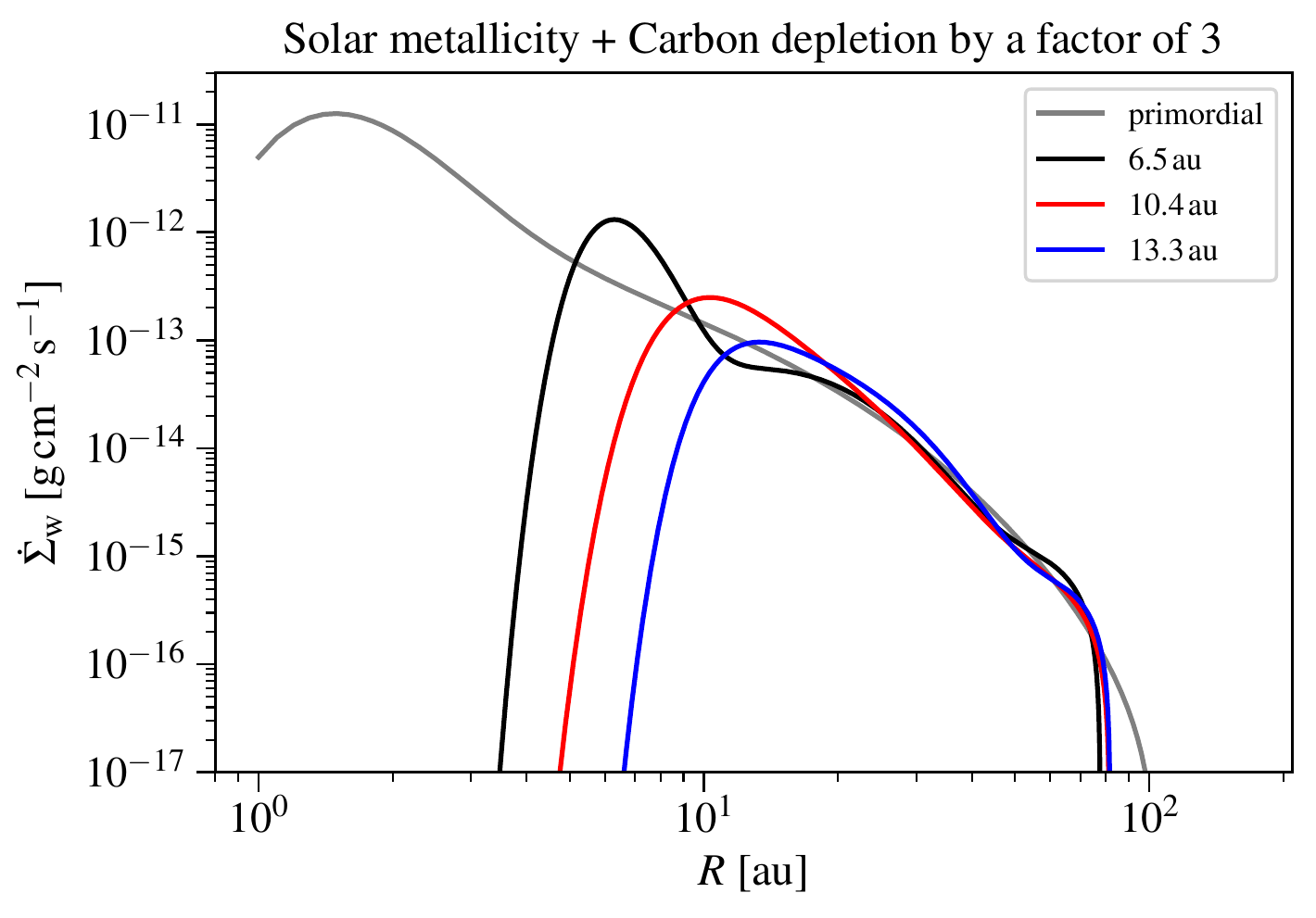}
	\includegraphics[width=\columnwidth]{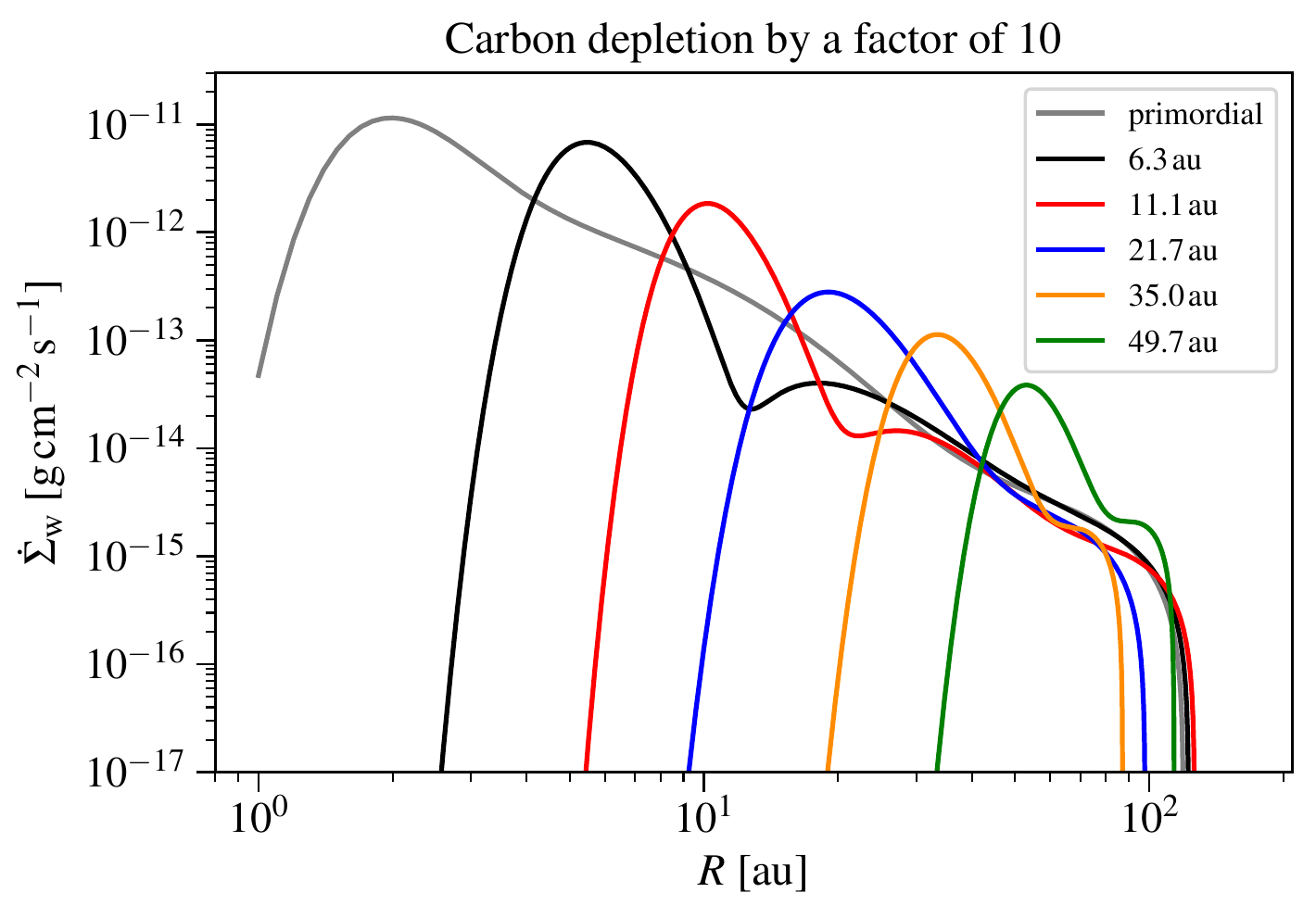}
	\includegraphics[width=\columnwidth]{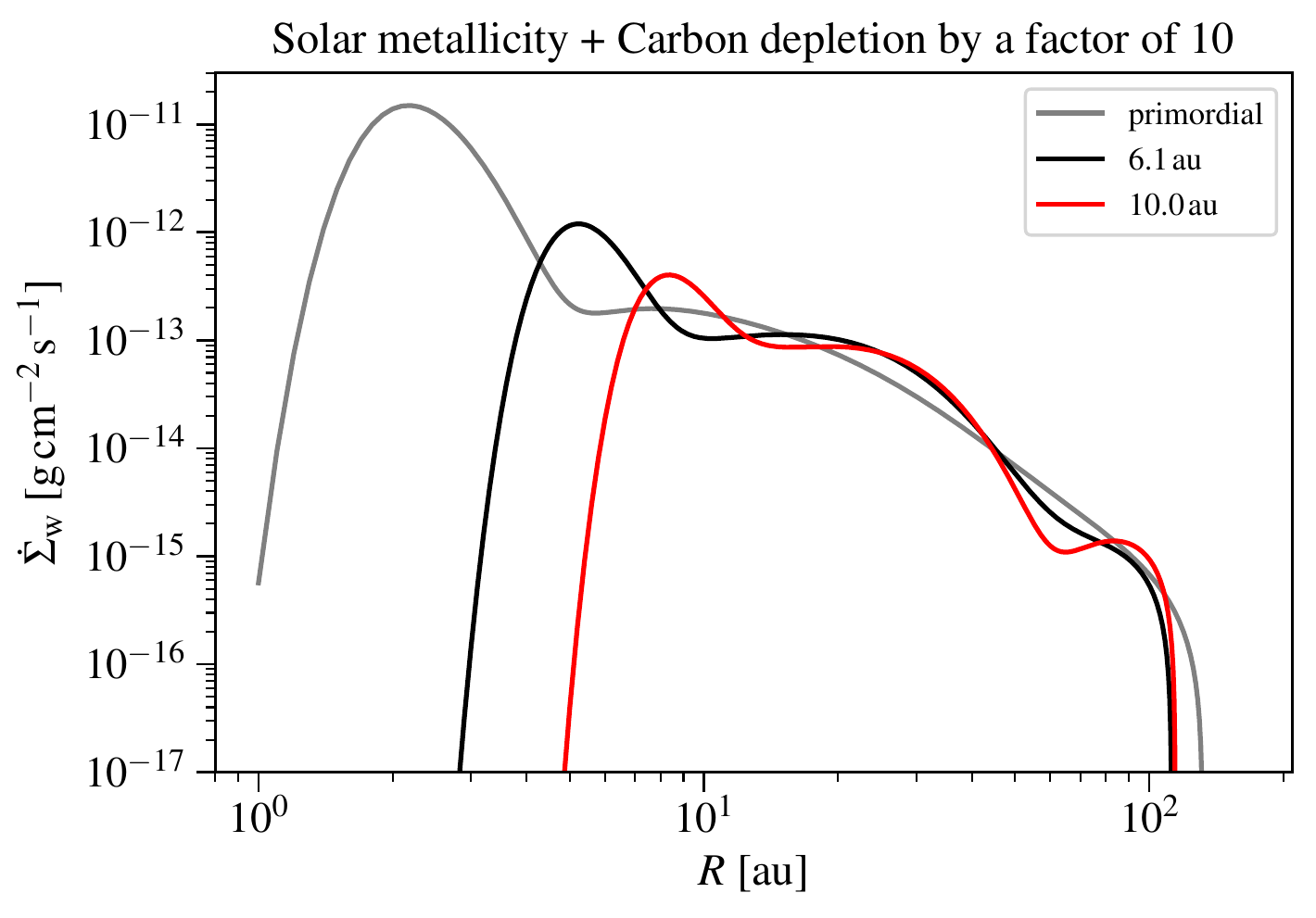}
    \includegraphics[width=\columnwidth]{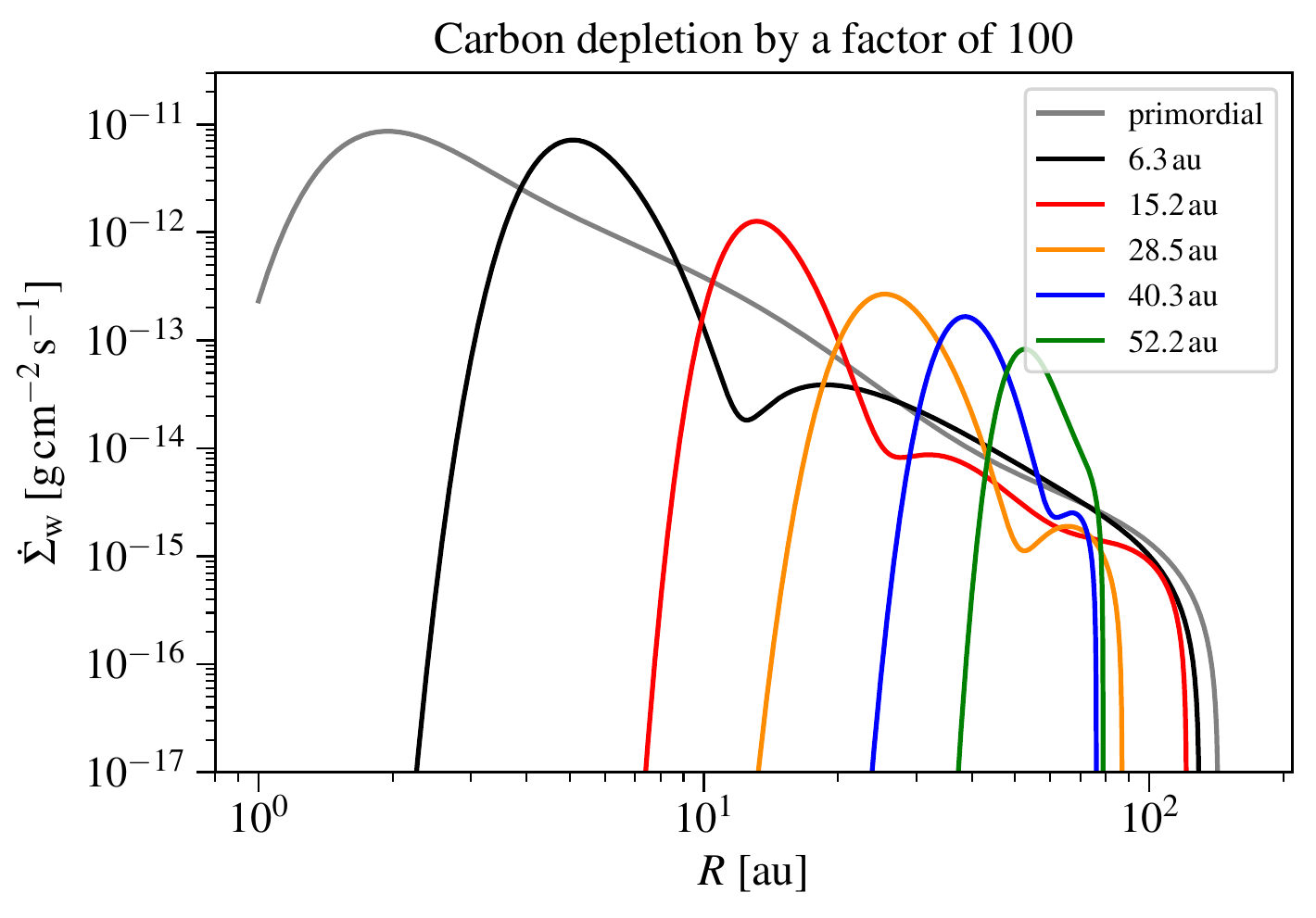}
    \caption{Examples of mass-loss profiles $\dot{\mathit{\Sigma}}$ for high-mass transition discs (0.1\,$\textit{M}_{*}$) with various hole radii, shown for the different carbon abundance set-ups.}
    \label{fig:SigmaDotTDhigh}
\end{figure*}
%%%%%%%%%%%%%%%%%%%%%%%%%%%%%%%%%%%%%%%%%%%%%%%%%%

% Don't change these lines
\bsp	% typesetting comment
\label{lastpage}
\end{document}